\documentclass[12pt]{article}
\title{    }

\global\arraycolsep=1pt
\usepackage[colorlinks=true,backref=true,linkcolor=black,anchorcolor=black,citecolor=black,filecolor=black,menucolor=black,pagecolor=black,urlcolor=black]{hyperref}

\usepackage{amsmath}
\usepackage{mathrsfs}
\usepackage[Symbol]{upgreek}
\usepackage{bbm}
\usepackage{dsfont}
\usepackage{amssymb}
\usepackage{textcomp}
\usepackage{youngtab}
\usepackage{caption}

\usepackage{graphicx}
\usepackage{wasysym}

\newcommand{\Scal}[1]{\Bigl ({#1} \Bigr )}
\newcommand{\scal}[1]{\bigl ({#1} \bigr )}
\def\bea{\begin{eqnarray}}
\def\eea{\end{eqnarray}}
\def\be{\begin{equation}}
\def\ee{\end{equation}}
\def\ie{{\it i.e.}\ }
\newcommand{\CR}{\nonumber \\*}

\newcommand{\trace}{\hbox {Tr}~}

\def\un{{\mathpzc{1}}}
\def\deux{{\mathpzc{2}}}
\def\trois{{\mathpzc{3}}}

\def\x{{\mathpzc{x}}}
\def\y{{\mathpzc{y}}}
\def\z{{\mathpzc{z}}}
\def\J{{J}}

\newcommand{\stfrac}[2]{{\textstyle \frac{#1}{#2}}}

\def\DEVIII#1#2#3#4#5#6#7#8{{\tiny $ { \left[ \begin{array}{ccccccc}  & & \mathfrak{#2} \hspace{-0.7mm}&&&& \vspace{ -1.5mm} \\ \mathfrak{#1}\hspace{-0.7mm} &  \mathfrak{#3} \hspace{-0.7mm}& \mathfrak{#4} \hspace{-0.7mm} & \mathfrak{#5}\hspace{-0.7mm}&\mathfrak{#6}\hspace{-0.7mm}&\mathfrak{#7}\hspace{-0.7mm}&\mathfrak{#8} \end{array}\right] }$}}
\def\DSOXVI#1#2#3#4#5#6#7#8{{\tiny $ {  \vspace{-2mm} \left[ \begin{array}{ccccccccc}  && \mathfrak{#8} \hspace{-0.7mm}&&&&&& \vspace{ -1.5mm} \\ \cdot \hspace{-0.5mm}& \mathfrak{#7}\hspace{-0.7mm} &\mathfrak{#6}\hspace{-0.7mm} &  \mathfrak{#5} \hspace{-0.7mm}& \mathfrak{#4} \hspace{-0.7mm} & \mathfrak{#3}\hspace{-0.7mm}&\mathfrak{#2}\hspace{-0.7mm}&\mathfrak{#1} \end{array}\right] }$}}

\def\DlacedLeft{{\fontsize{0.004pt}{0.0005pt}\selectfont  \scriptscriptstyle \mbox{$= \hspace{-2.2mm}  \langle$} \fontsize{12pt}{14.5pt}\selectfont }}

\def\DSpIV#1#2#3#4{{\tiny $ { \left[   \mathfrak{#1}\,  \mbox{-} \mathfrak{#2} \, \mbox{-} \mathfrak{#3} \hspace{-0.2mm}\DlacedLeft \hspace{0.2mm} \mathfrak{#4} \right] }$}}

\def\ie{{\it i.e.}\ }

\def\pA{{\scriptscriptstyle A}}
\def\pB{{\scriptscriptstyle B}}
\def\pC{{\scriptscriptstyle C}}

\def\asym{{\scriptscriptstyle 0}}
\def\invo{{\APLstar}}

\def\C{{\bf Q}}

\def\K{\mathcal{K}}

\def\V{{\mathcal{V}}}

\def\SU{SU_{\scriptscriptstyle \rm c}(8)}
\def\Sp{USp_{\scriptscriptstyle \rm c}(8)}

\def\Spin{Spin_{\scriptscriptstyle \rm c}}
\def\Sp{Sp_{\scriptscriptstyle \rm c}}

\def\IZ{{\mathds{Z}}}
\def\IR{{\mathds{R}}}

\def\IH{{\mathds{H}}}

\def\gl{\mathfrak{gl}}
\def\sl{\mathfrak{sl}}
\def\so{\mathfrak{so}}
\def\su{\mathfrak{su}}
\def\sp{\mathfrak{sp}}
\def\e{\mathfrak{e}}

\def\nn{\nonumber}
\def\DJo{$\;$\kern-.4em \hbox{D\kern-.8em\raise.15ex\hbox{--}\kern.35em okovi\'c}}

\DeclareMathAlphabet{\mathpzc}{OT1}{pzc}{m}{it}

\DeclareMathOperator{\ad}{ad}

\newcommand{\ord}[1]{{\scriptscriptstyle (#1)}}

\def\L{{\cal L}}

\def\nn{\nonumber}

\def\N{\mathcal{N}}

\def\C{{\mathscr{C}}}
\def\V{{\mathcal{V}}}

\def\p{{\mathpzc{p}}}

\def\invo{\APLstar}

\def\N{\mathcal{N}}
\newcommand{\eprint}[1]{{\href{http://arxiv.org/abs/#1}{\texttt{[#1}]}}}
\newcommand{\eprintN}[1]{{\href{http://arxiv.org/abs/#1}{\texttt{#1 [hep-th]}}}}

\def\DJo{$\;$\kern-.4em \hbox{D\kern-.8em\raise.15ex\hbox{--}\kern.35em okovi\'c}}

\begin{document}
\allowdisplaybreaks[1]
\renewcommand{\thefootnote}{\fnsymbol{footnote}}
\def\corr{$\spadesuit $}
\def\trefle{ $\clubsuit$}
\begin{titlepage}
\begin{flushright}
CPHT-RR011.0212\\
\end{flushright}

\bigskip
\bigskip
\centerline{\Large \bf Octonionic black holes}
\centerline{\Large \bf }
\bigskip
\bigskip
\centerline{{\bf Guillaume Bossard}\footnote{email: bossard@cpht.polytechnique.fr}}
\bigskip
\centerline{Centre de Physique Th\'eorique, Ecole Polytechnique, CNRS}
\centerline{91128 Palaiseau cedex, France}
\bigskip
\bigskip

\begin{abstract}
Using algebraic tools inspired by the study of nilpotent orbits in simple Lie algebras, we obtain a large class of solutions describing interacting non-BPS black holes in $\N=8$ supergravity, which depend on 44 harmonic functions. For this purpose, we consider a truncation $E_{6(6)} / Sp_{\scriptscriptstyle \rm c}(8,\mathds{R}) \subset E_{8(8)} / \Spin^*(16)$  of the non-linear sigma model describing stationary solutions  of the theory, which permits a reduction of algebraic computations to the multiplication of 27 by 27 matrices. The lift to $\N=8$ supergravity is then carried out without loss of information by using a pertinent representation of the moduli parametrizing $E_{7(7)} / \SU$ in terms of complex valued Hermitian matrices over the split octonions, which generalise the projective coordinates of exceptional special K\"{a}hler manifolds. We extract the electromagnetic charges, mass and angular momenta of the solutions, and exhibit the duality invariance of the black holes  distance separations. We discuss in particular a new type of interaction which appears when interacting non-BPS black holes are not aligned. Finally we will explain the possible generalisations toward the description of the most general stationary black hole solutions of $\N=8$ supergravity. 

\end{abstract}

\end{titlepage}

\renewcommand{\thefootnote}{\arabic{footnote}}
\setcounter{footnote}{0}

\tableofcontents

\section{Introduction}

The property that the weakly coupled calculation of the  Bekenstein--Hawking entropy of BPS black holes \cite{Strominger:1996sh} seems to generalise to non-BPS extremal black holes \cite{Emparan,DabhoSen,Dimitru}, suggested an increasing interest in characterising the entire moduli space of such solutions in various supergravity theories.  The understanding of non-BPS extremal black holes is an important step toward the study of black holes carrying a non-zero temperature. The classification of supersymmetric composite black hole solutions has permitted to understand the mismatch between the enumeration of spherically symmetric BPS black holes in $\N=2$ supergravity and the counting of BPS states within weakly coupled string theory \cite{Denef,DenefMoore}. The associated wall crossing formula can also be understood from the moduli space of regular solutions \cite{Manschot,Sen}. Because the weakly coupled computations for non-BPS black holes are not based on well established non-renormalisation theorems, it is important to extract the maximum of information from the classical solutions in the supergravity limit. These solutions have been studied extensively in the literature, see \cite{Ceresole:2007wx}--\cite{BossardRuef} for a non-exhaustive list of progress that have been achieved in the recent years. 

\vskip 2mm

The approach we will follow relies on the well established result that the stationary solutions of `irreducible' supergravity theories with abelian vector fields are described within a non-linear sigma model over a pseudo-Riemannian symmetric space $G/ K^*$ coupled to Euclidean gravity in three dimensions \cite{Breitenlohner:1987dg}. Here we will only discuss $\N=8$ supergravity, therefore we will fix notations accordingly and consider the coset space $E_{8(8)} / \Spin^*(16)$; although note that the generalisation of the content of this paper to exceptional $\N=2$ supergravity theories will be completely straightforward in our notations. The scalar momentum $P$ is defined as the component of the Maurer--Cartan form $\V^{-1} d \V$ in the coset component ${\bf 128} \cong \mathfrak{e}_{8(8)} \ominus \mathfrak{so}^*(16)$ of the Lie algebra. Solutions describing spherically symmetric black holes are then determined by the associated Noether charge in the Lie algebra $\mathfrak{e}_{8(8)}$ of $E_{8(8)}$, and can therefore be classified in terms of $E_{8(8)}$-orbits \cite{Breitenlohner:1987dg,Josef,Bossard:2009at}. In the extremal limit, the Noether charge is nilpotent and the spherically symmetric extremal black hole solutions are classified in terms of the class of nilpotent orbits of $E_{8(8)}$ in $\e_{8(8)}$ which lie in the closure of the minimal semi-simple orbit of $E_{8(8)}$ in $\e_{8(8)}$ \cite{Bossard:2009at,Bossard:2009my}. Then $P$ automatically lies in the same nilpotent orbit, and more precisely in its intersection with the coset component ${\bf 128}$. 

Relying on these properties, it has been proposed that the most general stationary solutions which geometry is fibered over a flat three-dimensional base are described by fields $\V$ valued in specific nilpotent subgroups \cite{BossardRuef}, such that $P$ is automatically nilpotent. Indeed, a vanishing three-dimensional Riemann tensor implies the equation 
\be \trace  P_\mu P_\nu = 0 \ee
and non-nilpotent solutions would then involve imaginary eigen values, which have been shown to lead to singularities in the spherically symmetric case \cite{Josef}. Moreover, all known under-rotating extremal solutions (as opposed to over-rotating extremal solutions generalising the extremal Kerr black hole) satisfy to this criterion, as it has been shown for the BPS solutions  in \cite{BossardBPS}, and for both the almost BPS solutions introduced in \cite{Goldstein:2008fq} and the composite non-BPS ones in \cite{BossardRuef}. 

In this paper we will exploit these ideas to derive generalisations of these solvable systems of differential equations to compute and study a large class of solutions in $\N=8$ supergravity. The generalisation to $\N=8$ is pertinent because this is very probably the simplest such supergravity theory at the quantum level, and on the other hand, it is complicated enough classically such that its solutions exhaust the possible classes of solutions one can have in supergravity coupled to abelian vector fields and scalars parametrizing a symmetric space. We will study the solvable systems of equations that are associated to the nilpotent orbits of $\e_{8(8)}$ which can be realised within $\so(4,4)$, and  which are therefore the direct generalisations of the systems of equations studied in the STU model in \cite{Bena:2011zq,Bena:2009en,BossardRuef} to the full $\N=8$ supergravity. However, as opposed to the BPS solutions, the non-BPS solutions do not in general sit in a given $\N=2$ truncation of the theory, and our solutions will include new harmonic functions. Although these functions will not dramatically   change the physical properties of the solutions, they modify the systems of differential equations in a non-trivial manner.

\vskip 2mm

Within the maximal $\N=2$ truncation of $\N=8$ supergravity, the relevant nilpotent subalgebras defining these solvable systems admit an $SU(2) \times_{\mathds{Z}_2} SL(3,\mathds{H})$ automorphism, and they decompose accordingly into a graded algebra; where $SL(3,\mathds{H}) \cong SU^*(6)$ is the group linearly realised on the special coordinates of the special K\"{a}hler space $SO^*(12) / U(6)$ of the maximal $\N=2$ truncation of the theory. The components of these nilpotent subalgebras in the coset component (which are associated to the harmonic functions defining the solutions) are
\bea \mathfrak{n}_{ \scriptscriptstyle \rm Almost \, BPS} &\cong& ({\bf 1}\oplus {\bf 15})^\ord{1} \oplus  \overline{\bf 15}^\ord{3} \oplus {\bf 1}^\ord{5} \CR
\mathfrak{n}_{ \scriptscriptstyle \rm non-BPS} &\cong& ({\bf 2}\otimes{\bf 15})^\ord{1} \oplus {\bf 2}^\ord{3} \CR
\mathfrak{n}_{ \scriptscriptstyle \rm BPS} &\cong& {\bf 32}^\ord{1}  
\eea
whereas for $\N=8$ supergravity one has
\bea
\mathfrak{n}_{ \scriptscriptstyle \rm Almost \, BPS} &\cong& ({\bf 1}\oplus {\bf 15})^\ord{1} \oplus ({\bf 2} \otimes \overline{\bf 6})^\ord{2} \oplus  \overline{\bf 15}^\ord{3} \oplus {\bf 1}^\ord{5} \CR
\mathfrak{n}_{ \scriptscriptstyle \rm non-BPS} &\cong& ({\bf 2}\otimes{\bf 15})^\ord{1} \oplus ({\bf 2} \otimes {\bf 6})^\ord{2} \oplus {\bf 2}^\ord{3} \CR
\mathfrak{n}_{ \scriptscriptstyle \rm BPS} &\cong&({\bf 2} \otimes {\bf 6})^\ord{1}\oplus  ({\bf 1} \oplus \overline{\bf 15})^\ord{2}  \oplus  ({\bf 1} \oplus{\bf 15})^\ord{3}  
\eea
such that $SU(2) \times_{\mathds{Z}_2}  SL(3,\mathds{H}) $ is also an automorphism of these nilpotent subalgebras. The integer superscripts indicate the grading preserved by these nilpotent algebras. It is such that the new generators of grade 2 are simply abelian in the generalised non-BPS nilpotent algebra. In this sense, the generalisation of the non-BPS system is the simplest. The almost BPS algebra is the next, with new generators of grade 2 that only commute to source the grade 5 scalar function. The BPS algebra is the most stringent generalisation, and the new generators of grade 1 then source half of the old ones.  We will see that in the associated, say `locally BPS system', these new generators are associated  to axion fields, and would source strings instead of black holes. The more general nilpotent subalgebras, which will be discussed in the last section, always have the property that new generators appear at grade 1, and modify strongly the algebra. It is not yet clear if they can describe composite black hole solutions. 

\vskip 2mm

We will make progress in the physical understanding of these solutions, which are also pertinent within the $\N=2$ truncations. In particular we will generalise the two non-BPS centres solution of \cite{BossardRuef} to an N centres solution. We will not assume the solution to be axisymmetric, although we will only be able to derive a closed form formula in the case in which the centres are all aligned. The missing explicit function will be defined as a convergent integral for which we will derive the asymptotic expansion in the near horizon and the asymptotic regions in Appendix \ref{Laplace}. Therefore we will not prove the regularity of the solution everywhere, but only in these different regions. We will nevertheless be able to prove the absence of closed time-like curve due to Dirac--Misner string singularities. These criteria are known to be enough for regularity by experience, but a complete proof would require a careful analysis of the solution everywhere. We can nevertheless rely on numerical simulations that have been carried out in \cite{Bena:2011zq,BossardRuef} to exhibit that these solutions are indeed generically regular everywhere.

In the non-BPS composite case the local properties of the non-BPS black holes at the horizon are as expected. In the case in which they are not aligned, they produce nevertheless an additional angular momentum which is related to the interactions of the centres three by three. Studying the equations for the distance separations between the black holes, we will see that they are not linear in their inverse distance separations, such that the interactions between the black holes do not only depend on their distance separations pairwise, but also on the geometry of the triangles joining them in triplets. 

We will study the composite non-BPS solutions in much more details, and in particular we will have an additional section devoted to the study of the duality invariance of their physical properties. This will permit to get some information on the general case, not associated to a specifically simple duality frame in which the solution can be written in closed form. In this section we will exhibit the general form of the momentum $P\in{\bf 128} $, which will permit to introduce a generalised fake superpotential for $\N=8$ supergravity. The introduction of auxiliary tensors that can be determined in terms of the electromagnetic charges of the black holes and the asymptotic moduli will permit to show the duality invariance of the equations determining the distance separations of the centres.

In the almost BPS system we will study the restricted case of one BPS centre and arbitrary many non-BPS ones. This solution will be complementary of the one derived in \cite{Bena:2009en}, where they considered one non-BPS centre and arbitrary many aligned BPS ones. We will see that even if this system does not allow for interactions between the non-BPS black holes in the absence of BPS black hole, the presence of the latter permits to consider charge configurations which do not locally commute for the various non-BPS black holes. The local properties of the BPS black hole in the near horizon region are somehow unexpected, because we will see that there is no enhancement of supersymmetry, and the BPS horizon is replaced by a surrounding under-rotating horizon,\footnote{By under-rotating horizon we mean an ergo-free horizon with zero angular velocity, but which is not spherically symmetric such that it would lead to an ADM angular momentum for a  single-centre black hole \cite{Rasheed:1995zv}.} which area is larger than the one of the original BPS centre.  In this case also we will find additional contributions to the angular momentum which appear when two non-BPS black holes are not aligned with the BPS centre. In general one should expect to have additional corrections when three BPS centres are not aligned with two non-BPS ones coming from the solution of the Laplace equation with five distinct point sources. 

\vskip 2mm

In order to carry out this program we will not consider directly the non-linear sigma model defined on $E_{8(8)} / \Spin^*(16)$. It turns out that the smallest linear representation of $E_{8(8)}$ is the adjoint which is 248-dimensional. Instead, we will consider the truncation to the $E_{6(6)}/Sp_{\scriptscriptstyle \rm c}(8,\mathds{R})$ coset space. This truncation contains all the information we need and the 27 dimensional fundamental representation of $E_{6(6)}$ can be managed with the help of a computer. It will appear that the $SL(3,\mathds{R})$ linearly realised symmetry of the solutions in the truncation is promoted to a linearly realised $SU(2) \times SL(3,\mathds{H}) \cong SU(2) \times SU^*(6)$ symmetry in $\N=8$ supergravity. Relying on this property, we will be able to generalise the solutions to $\N=8$ supergravity without ambiguities. Doing so, we will define particularly convenient coordinates for the symmetric space $E_{7(7)} / \SU$ which generalise somehow the special coordinates in very special K\"ahler geometry. The basic idea is to realise $E_{7(7)} / \SU$ as a $E_{7(7)} / ( U(1) \times E_{6(2)})$ fibered over the quaternionic space $SO(4,4)/(SO(4)\times SO(4)) $
\be \begin{array}{ccc} E_{7(7)} / ( U(1) \times E_{6(2)}) \hspace{10mm}  & \hookrightarrow & E_{7(7)} / \SU \\ && \downarrow \\ && SO(4,4)/(SO(4)\times SO(4))  \end{array} \ee
 The complex fibre $ E_{7(7)} / ( U(1) \times E_{6(2)})$ is a pseudo-Riemannian special K\"{a}hler space, and can be coordinatized in terms of Hermitian matrices over the split octonions, similarly as special coordinates for the exceptional special K\"{a}hler space $E_{7(-25)}/ ( U(1) \times E_{6(-78)})$. However, the fibre-bundle structure does not preserve the complete $E_{7(7)}$ isometry of the fibre. Therefore the associated special coordinates will not admit a linearly realised $E_{6(6)}$ symmetry preserving the equations of motion, as the special coordinates admit a linearly realised $E_{6(-26)}$ symmetry in the exceptional special  K\"{a}hler geometry \cite{GunaydinMagic}. 

In this paper we will parametrize the scalar fields  in terms of a complex exceptional Jordan algebra element $\mathpzc{t}$, \ie a 3 by 3 Hermitian matrix over the split octonions, which we will decompose into a 3 by 3 Hermitian matrix over the quaternions $\mathpzc{t}_\un$ and an antisymmetric matrix over the quaternions $\mathpzc{t}_2$ (equivalently a 3-vector of quaternions)
\be \mathpzc{t} = \mathpzc{t}_\un + \ell \mathpzc{t}_\deux \ee
and similarly for the electromagnetic charges $q_0, \mathpzc{Q}\, , \mathpzc{P}, p^0$ and the electromagnetic fields. This notation will be rather convenient in order to render straightforward the truncation of our solutions to $\N=2$  supergravity. In $\N=2$ supergravity, the components $\mathpzc{t}_\deux$ linear in the split imaginary unit $\ell$ are simply set to zero. The generalisation to any $\N=2$ supergravity coupled to vector multiplets which scalar fields parametrize a symmetric special K\"{a}hler space of cubic prepotential can be obtained by considering $\mathpzc{t}_\un$ (and the other Jordan algebra elements) to lie in the appropriate Jordan algebra \cite{GunaydinMagic,GunaydinJordan}. In particular, one can obtain the solutions of the STU model by considering the Jordan algebra of diagonal 3 by 3 real matrices, such that the three diagonal components of $\mathpzc{t}$ give the three STU moduli, and respectively for the charges. 

Similarly, the $SL(6) / SO(6)$ moduli of the truncated theory will be described in terms of special coordinates associated to the fibration
\be \begin{array}{ccc} SL(6,\mathds{R}) / ( U(1) \times SL(3,\mathds{C}))   & \hspace{5mm}  \hookrightarrow \hspace{5mm} &  SL(6,\mathds{R})/ SO(6) \\ && \downarrow \\ && \mathds{R}_+^* \times \mathds{R}_+^*   \end{array} \ee
which fibre will be coordinatized by a complex element of the Jordan algebra of Hermitian 3 by 3 matrices over the split complex. The solutions described in this paper will admit a linearly realised $SL(3,\mathds{R})$ symmetry, such that the two base space scalars are not sourced, and the solutions can be written in terms of symmetric 3 by 3 real matrices and a real 3 vector. The enhancement of this symmetry to $SU(2) \times SL(3,\mathds{H})$ in $\N=8$ supergravity is very constraining, and implies that the corresponding explicit solutions can be determined without ambiguities in terms of 3 by 3 Hermitian matrices over the quaternions, and a 3 vector of quaternions, such that the 16 $SO(4,4) / ( SO(4) \times SO(4))$ base space scalars are not sourced. For the most general ergo-free extremal solutions of $\N=8$ supergravity discussed in the last section, the linearly realised symmetry is reduced to $SO(4,4)$, which is a priori not big enough to determine without ambiguities the generalisation of a solution of the truncation to $\N=8$ supergravity. Nevertheless, combining the knowledge inherited from the most general solutions of the truncation with the explicit form of the first order linear system in $\N=8$ supergravity, one might be able to determine the most general solutions without considering directly the $E_{8(8)} / \Spin^*(16)$ representative. 

\section{Exceptional truncations}
In order to find solutions of $\N=8$ supergravity in four dimensions, and more generally of eleven-dimensional supergravity, it is often extremely useful to consider pertinent consistent truncations of the theory. Because one has been mainly interested in BPS solutions preserving a certain amount of supersymmetry, one was used to consider supersymmetric truncations of the theory. However, we will be studying non-BPS solutions in this paper, and there is no particular reason to consider truncations which are themselves supersymmetric. On the contrary, it will turn out that the most appropriate consistent truncations of $\N=8$ supergravity which are pertinent in the discussion of composite black hole solutions do not define the bosonic sector of supersymmetric theories. Rather than maintaining a certain amount of supersymmetry, these theories preserve somehow  ``exceptionality". 

$\N=8$ supergravity can be obtained as the compactification of eleven-dimensional supergravity on a seven-torus \cite{CremmerJulia}, or alternatively as the compactification of type IIB supergravity on a six-torus. The second theory admits a non-supersymmetric truncation which consists in setting to zero the scalar fields and the doublet of 2-form fields. The remaining theory consists in a self-dual 5-form field strength coupled to gravity. 

A further truncation can been carried out in eight dimensions, by considering gravity coupled to an $SL(2)/SO(2)$ non-linear sigma model and a 3-form, which 4-form field strength transforms in the fundamental of $SL(2)$ (by complex selfduality). 

These theories have the common property that after dimensional reduction down to four dimensions, the corresponding stationary solutions admit a duality group of type $E_{n(n)}$. Similarly as the exceptional $\N=2$ supergravity theories, their respective duality groups in 3, 4 and 5 dimensions define the magic square 
\be  \begin{array}{l} {\rm 5D} \\  {\rm 4D} \\ {\rm 3D} \end{array}  \left[ \begin{array}{ccc} \hspace{1mm} SL(3) \times SL(3) \hspace{1mm} & \hspace{2mm} SL(6) \hspace{2mm} & \hspace{2mm} E_{6(6)} \hspace{2mm} \\
 \hspace{2mm} SL(6) \hspace{2mm} & \hspace{2mm} SO(6,6) \hspace{2mm} & \hspace{2mm} E_{7(7)} \hspace{2mm} \\
 \hspace{2mm} E_{6(6)}  \hspace{2mm} & \hspace{2mm} E_{7(7)}  \hspace{2mm} & \hspace{2mm} E_{8(8)} \hspace{2mm} \end{array}\right] \ee
associated to the three exceptional Jordan algebras of Hermitian $3\times 3$ matrices over the three spit composition algebras. The realisation of split real form of exceptional Lie algebra as quasiconformal groups associated to exceptional Jordan algebras defined over split composition algebras has been defined and studied in \cite{Gunaydin:2000xr,Gunaydin:2009zza}. 

The stationary solutions of the corresponding four-dimensional theories are described by three-dimensional Euclidean gravity coupled to the non-linear sigma model over the pseudo-Riemannian symmetric spaces  
\be E_{6(6)} / \Sp(8,\IR) \subset E_{7(7)} / SU_{\scriptscriptstyle \rm c}(4,4) \subset E_{8(8)} / \Spin^*(16) \ee
respectively (where the index $\rm c$ states for the quotient by the $\mathds{Z}_2$ subgroup which leaves invariant the real ${\bf 27}$, the complexe ${\bf 28}$ and the real ${\bf 128}$ representations of $Sp(8,\IR), SU(4,4)$ and $Spin^*(16)$, respectively).

The main advantage in considering these truncations is that the smallest linear representation of $E_{8(8)}$ is of dimension 248, whereas $E_{6(6)}$ admits a 27-dimensional linear representation. Moreover, one will be able to lift the solutions associated to the $E_{6(6)}$ model to the general solutions of $\N=8$ supergravity by promoting real functions to functions defined over the quaternions.

For this purpose, let us first define the coordinates over $E_{6(6)} / \Sp(8,\IR) $.
\subsection{Conventions for $E_{6(6)}$}

In four dimensions, this theory consists of Einstein gravity coupled to 20 scalar fields parametrizing the symmetric space $SL(6)/SO(6)$ and 10 vector fields which field strengths transform altogether in the 3-form representation $\bf 20$ of $SL(6)$. The stationary metric will take the standard form 
\be \label{4Dmet}
 ds^2 = - e^{2U} \bigl( dt + \omega \bigr)^2 + e^{-2U}\delta_{\mu\nu} dx^\mu dx^\nu  \; ,
\ee
as well as the vector fields 
\be \label{4DEM}
 \sqrt{8}\, A_{ABC} = \zeta_{ABC}  \bigl( dt + \omega \bigr) +  w_{ABC} \; , 
\ee
where the tensors are antisymmetric with respect to the permutations of the three $SL(6)$  indices $ABC$. 

We will decompose the $\bf 27$ representation in terms of the four-dimensional duality group as
\bea \e_{6(6)} &\cong& {\bf 1}^\ord{-2} \oplus {\bf 20}^\ord{-1} \oplus \scal{\gl_1 \oplus \sl_6}^\ord{0} \oplus {\bf 20}^\ord{1} \oplus {\bf 1}^\ord{2}  \CR
{\bf 27} &\cong&  \overline{\bf 6}^\ord{-1} \oplus {\bf 15}^\ord{0} \oplus \overline{\bf 6}^\ord{1} \eea
such that a general element of $\e_{6(6)}$ acting on a vector $(R^\ord{1},S^\ord{-1},T^\ord{0})$ admits the following matrix form in terms of two antisymmetric tensors $E_{abc}$ and $F_{abc}$, and the traceless tensor $G^a{}_b$
\be \delta  \left(\begin{array}{c} R^a \\ S^a \\ T_{bc} \end{array}\right) =    \left(\begin{array}{ccc} 
\hspace{2mm} \delta^a_d H + G^a{}_d  \hspace{2mm}& \hspace{2mm} \delta^a_d E \hspace{2mm}  & \hspace{2mm} E^{aef}  \hspace{2mm}  \\
\hspace{2mm} \delta^a_d F  \hspace{2mm}& \hspace{2mm}-  \delta^a_d H  + G^a{}_d \hspace{2mm}  & \hspace{2mm} F^{aef}  \hspace{2mm}  \\
\hspace{2mm}  - F_{bcd}   \hspace{2mm}& \hspace{2mm} E_{bcd}  \hspace{2mm}  & \hspace{2mm} - 2 \delta_{[a}^{[e} G^{f]}{}_{b]}  \hspace{2mm} \end{array} \right)  \left(\begin{array}{c} R^d \\ S^d \\ T_{ef} \end{array}\right)  \ee 
where $E_{abc} \equiv \frac{1}{6} \varepsilon_{abcdef} E^{def}$ and respectively for $F_{abc}$. The elements of $\sp(8,\mathds{R})$ are defined from the involution associated to the metric 
\be \upeta =   \left(\begin{array}{ccc} 
\hspace{2mm} \delta^a_d   \hspace{2mm}& \hspace{2mm} 0 \hspace{2mm}  & \hspace{2mm} 0   \hspace{2mm}  \\
\hspace{2mm} 0 \hspace{2mm}& \hspace{2mm} \delta^a_d  \hspace{2mm}  & \hspace{2mm} 0 \hspace{2mm}  \\
\hspace{2mm}  0 \hspace{2mm}& \hspace{2mm} 0 \hspace{2mm}  & \hspace{2mm} -  \delta_{ab}^{ef}  \hspace{2mm} \end{array} \right)   \ee 
as ${\bf \rm x}^\ddagger \equiv \upeta {\bf \rm x}^t \upeta = - {\bf \rm x}$,  and they read
\be    \left(\begin{array}{ccc} 
\hspace{2mm} A^a{}_d  \hspace{2mm}& \hspace{2mm} \delta^a_d a \hspace{2mm}  & \hspace{2mm} X^{aef}  \hspace{2mm}  \\
\hspace{2mm} - \delta^a_d a  \hspace{2mm}& \hspace{2mm} A^a{}_d \hspace{2mm}  & \hspace{2mm} X_{aef}  \hspace{2mm}  \\
\hspace{2mm}  X^{bcd}   \hspace{2mm}& \hspace{2mm} X_{bcd}  \hspace{2mm}  & \hspace{2mm} - 2 \delta_{[a}^{[e} A^{f]}{}_{b]}  \hspace{2mm} \end{array} \right)\label{Sp} \ee 
where $A^a{}_b$ is antisymmetric. 

Using the standard convention that capital indices correspond to rigid $SL(6)$ (acting on the left) whereas small ones correspond to local $SO(6)$ (acting on the right), one has the representative
\bea   \V &=&  \exp\bigl[ \zeta^{ABC} {\bf E}_{ABC} + \sigma {\bf E} \bigr] \exp[ U {\bf H}] \left(\begin{array}{ccc} 
\hspace{2mm} v_a{}^{A^\prime}  \hspace{2mm}& \hspace{2mm} 0  \hspace{2mm}  & \hspace{2mm}0 \hspace{2mm}  \\
\hspace{2mm}0  \hspace{2mm}& \hspace{2mm} v_a{}^{A^\prime}  \hspace{2mm}  & \hspace{2mm} 0  \hspace{2mm}  \\
\hspace{2mm}  0    \hspace{2mm}& \hspace{2mm}  0 \hspace{2mm}  & \hspace{2mm} (v^{-1})_{[B^\prime}{}^b (v^{-1})_{C^\prime]}{}^c \hspace{2mm} \end{array} \right)  \begin{array}{c} \\ \\ \vspace{4mm} \end{array} \CR
&=& \left(\begin{array}{ccc} 
\hspace{2mm} e^U v_a{}^{A}  \hspace{2mm}& \hspace{2mm} e^{-U} v_a{}^{A} \sigma + \frac{1}{2} e^{-U} v_a{}^{B}{}  \zeta^{ADE} \zeta_{BDE}     \hspace{2mm}  & \hspace{2mm} (v^{-1})_D{}^b (v^{-1})_E{}^c  \zeta^{ADE} \hspace{2mm}  \\
\hspace{2mm}0  \hspace{2mm}& \hspace{2mm}e^{-U} v_a{}^{A}  \hspace{2mm}  & \hspace{2mm} 0  \hspace{2mm}  \\
\hspace{2mm}  0    \hspace{2mm}& \hspace{2mm}  e^{-U} v_a{}^{D}  \zeta_{BCD} \hspace{2mm}  & \hspace{2mm} (v^{-1})_{[B}{}^b (v^{-1})_{C]}{}^c \hspace{2mm} \end{array} \right) \eea
where one must note that $ v_a{}^{A}$ should be understood as the transverse $v^t$ acting on the right through the contraction with the index $a$. 

It will be convenient to make use of a non-triangular representative in $SL(6)/SO(6)$ as 
\begin{multline} v^t =   \exp\left(\begin{array}{cccccc} 
\hspace{1.5mm} 0 \hspace{1.5mm}& \hspace{1.5mm} a_1  \hspace{1.5mm}  & \hspace{1.5mm} 0  \hspace{1.5mm} & \hspace{1.5mm} \xi_0  \hspace{1.5mm} & \hspace{1.5mm} 0  \hspace{1.5mm} & \hspace{1.5mm} \zeta^2  \hspace{1.5mm} \\
\hspace{1.5mm} 0 \hspace{1.5mm}& \hspace{1.5mm}0 \hspace{1.5mm}  & \hspace{1.5mm} 0  \hspace{1.5mm} & \hspace{1.5mm} 0 \hspace{1.5mm} & \hspace{1.5mm} 0 \hspace{1.5mm} & \hspace{1.5mm} 0   \hspace{1.5mm} \\\hspace{1.5mm} 0 \hspace{1.5mm}& \hspace{1.5mm} \xi^3 \hspace{1.5mm}  & \hspace{1.5mm}0  \hspace{1.5mm} & \hspace{1.5mm} a_2  \hspace{1.5mm} & \hspace{1.5mm} 0  \hspace{1.5mm} & \hspace{1.5mm} \eta_0    \hspace{1.5mm} \\\hspace{1.5mm} 0 \hspace{1.5mm}& \hspace{1.5mm} 0 \hspace{1.5mm}  & \hspace{1.5mm} 0 \hspace{1.5mm} & \hspace{1.5mm} 0  \hspace{1.5mm} & \hspace{1.5mm} 0 \hspace{1.5mm} & \hspace{1.5mm} 0   \hspace{1.5mm} \\\hspace{1.5mm} 0 \hspace{1.5mm}& \hspace{1.5mm} \zeta_0 \hspace{1.5mm}  & \hspace{1.5mm} 0 \hspace{1.5mm} & \hspace{1.5mm} \eta^1  \hspace{1.5mm} & \hspace{1.5mm}0 \hspace{1.5mm} & \hspace{1.5mm}  a_3   \hspace{1.5mm} \\
\hspace{1.5mm} 0 \hspace{1.5mm}& \hspace{1.5mm} 0 \hspace{1.5mm}  & \hspace{1.5mm} 0 \hspace{1.5mm} & \hspace{1.5mm} 0 \hspace{1.5mm} & \hspace{1.5mm} 0 \hspace{1.5mm} & \hspace{1.5mm}0   \hspace{1.5mm} 
\end{array} \right)  \exp \left(\begin{array}{cccccc} 
\hspace{1.5mm}0  \hspace{1.5mm}& \hspace{1.5mm} 0  \hspace{1.5mm}  & \hspace{1.5mm} 0  \hspace{1.5mm} & \hspace{1.5mm} 0   \hspace{1.5mm} &\hspace{1.5mm}   0  \hspace{1.5mm} & \hspace{1.5mm} 0  \hspace{1.5mm} \\
\hspace{1.5mm} 0 \hspace{1.5mm}& \hspace{1.5mm}0  \hspace{1.5mm}  & \hspace{1.5mm} 0  \hspace{1.5mm} & \hspace{1.5mm} \xi^1  \hspace{1.5mm} & \hspace{1.5mm} 0 \hspace{1.5mm} & \hspace{1.5mm}  \zeta_3  \hspace{1.5mm} \\\hspace{1.5mm} \xi_2  \hspace{1.5mm}& \hspace{1.5mm} 0 \hspace{1.5mm}  & \hspace{1.5mm}0  \hspace{1.5mm} & \hspace{1.5mm} 0\hspace{1.5mm} & \hspace{1.5mm} 0  \hspace{1.5mm} & \hspace{1.5mm} 0    \hspace{1.5mm} \\
\hspace{1.5mm} 0 \hspace{1.5mm}& \hspace{1.5mm} 0 \hspace{1.5mm}  & \hspace{1.5mm} 0 \hspace{1.5mm} & \hspace{1.5mm}0 \hspace{1.5mm} & \hspace{1.5mm}0 \hspace{1.5mm} & \hspace{1.5mm} \eta^2    \hspace{1.5mm} \\
\hspace{1.5mm} \zeta^1  \hspace{1.5mm}& \hspace{1.5mm} 0 \hspace{1.5mm}  & \hspace{1.5mm} \eta_3  \hspace{1.5mm} & \hspace{1.5mm} 0 \hspace{1.5mm} & \hspace{1.5mm}0 \hspace{1.5mm} & \hspace{1.5mm} 0   \hspace{1.5mm} \\
\hspace{1.5mm} 0 \hspace{1.5mm}& \hspace{1.5mm} 0 \hspace{1.5mm}  & \hspace{1.5mm} 0 \hspace{1.5mm} & \hspace{1.5mm} 0 \hspace{1.5mm} & \hspace{1.5mm} 0 \hspace{1.5mm} & \hspace{1.5mm}0   \hspace{1.5mm} 
\end{array} \right) \times  \\*  \exp \left(\begin{array}{cccccc} 
\hspace{0.5mm}  {-\phi_1+\varsigma_+} \hspace{0.5mm}& \hspace{1.5mm} 0 \hspace{1.5mm}  & \hspace{1.5mm} 0  \hspace{1.5mm} & \hspace{1.5mm} 0 \hspace{1.5mm} & \hspace{1.5mm} 0 \hspace{1.5mm} & \hspace{1.5mm} 0\hspace{1.5mm} \\
\hspace{1.5mm} 0 \hspace{1.5mm}& \hspace{0.5mm}  {\phi_1 + \varsigma_+ } \hspace{0.5mm}  & \hspace{1.5mm} 0  \hspace{1.5mm} & \hspace{1.5mm} 0 \hspace{1.5mm} & \hspace{1.5mm} 0 \hspace{1.5mm} & \hspace{1.5mm} 0 \hspace{1.5mm} \\\hspace{1.5mm} 0 \hspace{1.5mm}& \hspace{1.5mm} 0 \hspace{1.5mm}  & \hspace{0.5mm}  {-\phi_2 - \varsigma_+ - \varsigma_- }  \hspace{0.5mm} & \hspace{1.5mm} 0  \hspace{1.5mm} & \hspace{1.5mm} 0 \hspace{1.5mm} & \hspace{1.5mm} 0   \hspace{1.5mm} \\\hspace{1.5mm} 0 \hspace{1.5mm}& \hspace{1.5mm} 0 \hspace{1.5mm}  & \hspace{1.5mm} 0 \hspace{1.5mm} & \hspace{0.5mm}  {\phi_2 - \varsigma_+ - \varsigma_- } \hspace{0.5mm} & \hspace{1.5mm} 0 \hspace{1.5mm} & \hspace{1.5mm} 0    \hspace{1.5mm} \\\hspace{1.5mm} 0 \hspace{1.5mm}& \hspace{1.5mm} 0 \hspace{1.5mm}  & \hspace{1.5mm} 0 \hspace{1.5mm} & \hspace{1.5mm} 0 \hspace{1.5mm} & \hspace{0.5mm}  {-\phi_3 + \varsigma_-} \hspace{0.5mm} & \hspace{1.5mm}  0   \hspace{1.5mm} \\
\hspace{1.5mm} 0 \hspace{1.5mm}& \hspace{1.5mm} 0 \hspace{1.5mm}  & \hspace{1.5mm} 0 \hspace{1.5mm} & \hspace{1.5mm} 0 \hspace{1.5mm} & \hspace{1.5mm} 0 \hspace{1.5mm} & \hspace{0.5mm}  {\phi_3+ \varsigma_-}   \hspace{0.5mm} 
\end{array} \right) \end{multline}
which gives 
\begin{multline} v^t =  \left(\begin{array}{cccccc} 
\hspace{1.5mm} 1 \hspace{1.5mm}& \hspace{1.5mm} a_1  \hspace{1.5mm}  & \hspace{1.5mm} 0  \hspace{1.5mm} & \hspace{1.5mm} \xi_0  \hspace{1.5mm} & \hspace{1.5mm} 0  \hspace{1.5mm} & \hspace{1.5mm} \zeta^2  \hspace{1.5mm} \\ \hspace{1.5mm} 0 \hspace{1.5mm}& \hspace{1.5mm} 1  \hspace{1.5mm}  & \hspace{1.5mm} 0  \hspace{1.5mm} & \hspace{1.5mm} 0 \hspace{1.5mm} & \hspace{1.5mm} 0 \hspace{1.5mm} & \hspace{1.5mm} 0   \hspace{1.5mm} \\\hspace{1.5mm} 0 \hspace{1.5mm}& \hspace{1.5mm} \xi^3 \hspace{1.5mm}  & \hspace{1.5mm} 1  \hspace{1.5mm} & \hspace{1.5mm} a_2  \hspace{1.5mm} & \hspace{1.5mm} 0  \hspace{1.5mm} & \hspace{1.5mm} \eta_0    \hspace{1.5mm} \\\hspace{1.5mm} 0 \hspace{1.5mm}& \hspace{1.5mm} 0 \hspace{1.5mm}  & \hspace{1.5mm} 0 \hspace{1.5mm} & \hspace{1.5mm}1  \hspace{1.5mm} & \hspace{1.5mm} 0 \hspace{1.5mm} & \hspace{1.5mm} 0   \hspace{1.5mm} \\\hspace{1.5mm} 0 \hspace{1.5mm}& \hspace{1.5mm} \zeta_0 \hspace{1.5mm}  & \hspace{1.5mm} 0 \hspace{1.5mm} & \hspace{1.5mm} \eta^1  \hspace{1.5mm} & \hspace{1.5mm} 1 \hspace{1.5mm} & \hspace{1.5mm}  a_3   \hspace{1.5mm} \\
\hspace{1.5mm} 0 \hspace{1.5mm}& \hspace{1.5mm} 0 \hspace{1.5mm}  & \hspace{1.5mm} 0 \hspace{1.5mm} & \hspace{1.5mm} 0 \hspace{1.5mm} & \hspace{1.5mm} 0 \hspace{1.5mm} & \hspace{1.5mm} 1   \hspace{1.5mm} 
\end{array} \right)\times \\*   \left(\begin{array}{cccccc} 
 \hspace{0.0mm} e^{-\phi_1+ \varsigma_+ }   \hspace{0.0mm}&  \hspace{0.0mm} 0   \hspace{0.0mm}  &  \hspace{0.0mm} 0   \hspace{0.0mm} &  \hspace{0.0mm} 0    \hspace{0.0mm} &  \hspace{0.0mm} 0   \hspace{0.0mm} &  \hspace{0.0mm} 0   \hspace{0.0mm} \\
 \hspace{0.0mm} 0  \hspace{0.0mm}&  \hspace{0.0mm}  e^{\phi_1+ \varsigma_+ }  \hspace{0.0mm}  &  \hspace{0.0mm} 0   \hspace{0.0mm} &  \hspace{0.0mm}  e^{\phi_2  -\varsigma_+ -  \varsigma_- }  \xi^1   \hspace{0.0mm} &  \hspace{0.0mm} 0  \hspace{0.0mm} & \hspace{-0.8mm} e^{\phi_3  + \varsigma_- }  ({ \scriptstyle  \zeta_3 + \stfrac{1}{2} \xi^1 \eta^2  } ) \hspace{-0.8mm} \\ \hspace{0.0mm} e^{-\phi_1+ \varsigma_+ }  \xi_2   \hspace{0.0mm}&  \hspace{0.0mm} 0  \hspace{0.0mm}  &  \hspace{0.0mm} e^{-\phi_2 -  \varsigma_+ - \varsigma_- }   \hspace{0.0mm} &  \hspace{0.0mm} 0 \hspace{0.0mm} &  \hspace{0.0mm} 0   \hspace{0.0mm} &  \hspace{0.0mm} 0     \hspace{0.0mm} \\ \hspace{0.0mm} 0  \hspace{0.0mm}&  \hspace{0.0mm} 0  \hspace{0.0mm}  &  \hspace{0.0mm} 0  \hspace{0.0mm} &  \hspace{0.0mm}  e^{\phi_2  -\varsigma_+ -  \varsigma_- }   \hspace{0.0mm} &  \hspace{0.0mm}0  \hspace{0.0mm} &  \hspace{0.0mm} e^{\phi_3  + \varsigma_- } \eta^2     \hspace{0.0mm} \\\hspace{-0.8mm}  e^{-\phi_1+ \varsigma_+ } ( {\scriptstyle  \zeta^1  + \stfrac{1}{2} \xi_2 \eta_3 } ) \hspace{-0.8mm}  &  \hspace{0.0mm} 0  \hspace{0.0mm}  &  \hspace{0.0mm}  e^{-\phi_2 -  \varsigma_+ - \varsigma_- } \eta_3  \hspace{0.0mm} &  \hspace{0.0mm} 0  \hspace{0.0mm} &  \hspace{0.0mm} e^{-\phi_3  +  \varsigma_- }  \hspace{0.0mm} &  \hspace{0.0mm} 0    \hspace{0.0mm} \\
 \hspace{0.0mm} 0  \hspace{0.0mm}&  \hspace{0.0mm} 0  \hspace{0.0mm}  &  \hspace{0.0mm} 0  \hspace{0.0mm} &  \hspace{0.0mm} 0  \hspace{0.0mm} &  \hspace{0.0mm} 0  \hspace{0.0mm} &  \hspace{0.0mm} e^{\phi_3  + \varsigma_- }   \hspace{0.0mm} 
\end{array} \right) \label{vtExpression} \end{multline} 
where $t^i \equiv a_i + i e^{-2\phi_i}$ are the $STU$ moduli. The reason for this choice is that the matrix on the right corresponds to the $GL(1) \times SL(3) /SO(3)  \times SL(3)/ SO(3)$ representative (with the two $SL(3)$ factors respectively in upper and lower triangular form), which parametrizes the five-dimensional scalars and the dilaton $\phi \equiv \sum_i \phi_i$ that appears in the dimensional reduction from five to four dimensions, whereas the left matrix corresponds to the axion fields associated to the five dimensional electromagnetic fields, which transform in the ${\bf 3} \otimes {\bf 3}$ with respect to the five-dimensional duality group $SL(3) \times SL(3)$. 

\subsubsection*{Exceptional Jordan algebra}

This representation will permit to use the homomorphism $SL(3) \times SL(3) \cong SL(3,\mathds{C}^*)$ where $\mathds{C}^*$ is the split complex composition algebra, with imaginary unite $\ell$ satisfying $\ell^2 = 1$. In this way one can rewrite the axion fields as a (Jordan algebra) Hermitian matrix over $\mathds{C}^*$ 
\be  \mathpzc{a} = \left(\begin{array}{ccc} 
\hspace{2mm} a_1   \hspace{2mm}& \hspace{2mm} \alpha_3   \hspace{2mm}  & \hspace{2mm} \alpha^*_2   \hspace{2mm} \\
\hspace{2mm} \alpha^*_3   \hspace{2mm}& \hspace{2mm} a_2   \hspace{2mm}  & \hspace{2mm} \alpha_1   \hspace{2mm} \\
\hspace{2mm}\alpha_2   \hspace{2mm}& \hspace{2mm} \alpha^*_1  \hspace{2mm}  & \hspace{2mm} a_3    \hspace{2mm} 
\end{array} \right) \ee
with 
\be \alpha_1 = \frac{1}{2} ( \xi_0 + \xi^3 ) + \frac{\ell}{2} ( \xi^3 - \xi_0 ) \; , \quad  \alpha_2 = \frac{1}{2} ( \zeta^2 + \zeta_0 ) + \frac{\ell}{2} ( \zeta^2 - \zeta_0 ) \; ,\quad \alpha_3 =  \frac{1}{2} ( \eta_0 + \eta^1 ) + \frac{\ell}{2} ( \eta^1 - \eta_0 ) \ee
Using the property  (which defines the homomorphism $\mathds{C}^* \cong \mathds{R} \oplus \mathds{R}$)  
\be \Scal{ \frac{ 1\pm \ell }{2}}^2 = \frac{ 1\pm \ell }{2} \ee
one computes that the components of the Maurer--Cartan form $v^{t\, -1} dv^t$ associated to the axions take the form
\be \rho(\varsigma)  \left(\begin{array}{ccc} 
\hspace{2mm} e^{\phi_1}   \hspace{2mm}& \hspace{2mm} 0  \hspace{2mm}  & \hspace{2mm} 0  \hspace{2mm} \\
\hspace{2mm} e^{\phi_2} \xi^*   \hspace{2mm}& \hspace{2mm} e^{\phi_2}  \hspace{2mm}  & \hspace{2mm} 0  \hspace{2mm} \\
 \hspace{1.0mm} e^{\phi_3} ( {\scriptstyle \zeta +\stfrac{1}{2} \xi^* \eta^*})    \hspace{1.0mm}& \hspace{2mm} e^{\phi_3} \eta^* \hspace{2mm}  & \hspace{2mm} e^{\phi_3}   \hspace{2mm} 
\end{array} \right) d  \left(\begin{array}{ccc} 
\hspace{2mm} a_1   \hspace{2mm}& \hspace{2mm} \alpha_3   \hspace{2mm}  & \hspace{2mm} \alpha^*_2   \hspace{2mm} \\
\hspace{2mm} \alpha^*_3   \hspace{2mm}& \hspace{2mm} a_2   \hspace{2mm}  & \hspace{2mm} \alpha_1   \hspace{2mm} \\
\hspace{2mm}\alpha_2   \hspace{2mm}& \hspace{2mm} \alpha^*_1  \hspace{2mm}  & \hspace{2mm} a_3    \hspace{2mm} 
\end{array} \right) 
 \left(\begin{array}{ccc} 
\hspace{2mm} e^{\phi_1}   \hspace{2mm}& \hspace{2mm} e^{\phi_2} \xi   \hspace{2mm}  & \hspace{1.0mm} e^{\phi_3} ( {\scriptstyle \zeta^* +\stfrac{1}{2} \xi\eta})    \hspace{1.0mm} \\
\hspace{2mm} 0   \hspace{2mm}& \hspace{2mm} e^{\phi_2}  \hspace{2mm}  & \hspace{2mm} e^{\phi_3} \eta  \hspace{2mm} \\
 \hspace{2mm} 0  \hspace{2mm} & \hspace{2mm} 0  \hspace{2mm}  & \hspace{2mm} e^{\phi_3}   \hspace{2mm} 
\end{array} \right) \label{MaurAxions} \ee 
where 
\be \xi =  \frac{1}{2} ( \xi^1 -\xi_2 ) - \frac{\ell}{2} ( \xi^1 + \xi_2 ) \; , \quad \zeta = \frac{1}{2} ( \zeta_3 -\zeta^1 ) + \frac{\ell}{2} ( \zeta_3 + \zeta^1 ) \; ,\quad  \eta =  \frac{1}{2} ( \eta^2 - \eta_3 ) - \frac{\ell}{2} ( \eta^2 + \eta_3 ) \ee
and $\rho(\varsigma)$ acts on the resulting matrix such that on any Hermitian matrix
\be \rho(\varsigma)   \left(\begin{array}{ccc} 
\hspace{2mm} b_1   \hspace{2mm}& \hspace{2mm} \beta_3   \hspace{2mm}  & \hspace{2mm} \beta^*_2   \hspace{2mm} \\
\hspace{2mm} \beta^*_3   \hspace{2mm}& \hspace{2mm} b_2   \hspace{2mm}  & \hspace{2mm} \beta_1   \hspace{2mm} \\
\hspace{2mm}\beta_2   \hspace{2mm}& \hspace{2mm} \beta^*_1  \hspace{2mm}  & \hspace{2mm} b_3    \hspace{2mm} 
\end{array} \right) =  \left(\begin{array}{ccc} 
\hspace{2mm} b_1   \hspace{2mm}& \hspace{2mm} \rho_3(\varsigma) \beta_3   \hspace{2mm}  & \hspace{2mm} \rho_2^*(\varsigma) \beta^*_2   \hspace{2mm} \\
\hspace{2mm} \rho^*_3(\varsigma) \beta^*_3   \hspace{2mm}& \hspace{2mm} b_2   \hspace{2mm}  & \hspace{2mm}\rho_1 (\varsigma) \beta_1   \hspace{2mm} \\
\hspace{2mm} \rho_2(\varsigma)\beta_2   \hspace{2mm}& \hspace{2mm}\rho_1^*(\varsigma) \beta^*_1  \hspace{2mm}  & \hspace{2mm} b_3    \hspace{2mm} 
\end{array} \right)   \ee
with 
\be \rho_1(\varsigma) = e^{\ell( 2 \varsigma_+ + \varsigma_-)} \; , \quad  \rho_2(\varsigma) = e^{\ell( - \varsigma_+ + \varsigma_-)} \; , \quad  \rho_3(\varsigma) = e^{-\ell(  \varsigma_+ + 2\varsigma_-)} \ee 
the three phases satisfying the triality condition 
\be \rho^*_i(\varsigma) = \rho_{i+1}(\varsigma) \rho_{i+2}(\varsigma) \ee
The matrix 
\be \upsilon^{-1} =  \left(\begin{array}{ccc} 
\hspace{2mm} e^{\phi_1}   \hspace{2mm}& \hspace{2mm} e^{\phi_2} \xi   \hspace{2mm}  & \hspace{1.0mm} e^{\phi_3} ( {\scriptstyle \zeta^* +\stfrac{1}{2} \xi\eta})    \hspace{1.0mm} \\
\hspace{2mm} 0   \hspace{2mm}& \hspace{2mm} e^{\phi_2}  \hspace{2mm}  & \hspace{2mm} e^{\phi_3} \eta  \hspace{2mm} \\
 \hspace{2mm} 0  \hspace{2mm} & \hspace{2mm} 0  \hspace{2mm}  & \hspace{2mm} e^{\phi_3}   \hspace{2mm} 
\end{array} \right) \ee
on the right of (\ref{MaurAxions}) also defines partially the $GL(1) \times SL(3,\mathds{C}^*)$ coset representative. In analogy with the $\N=2$ case for which the $GL(1)\times SL(3,\mathds{C})$ Maurer--Cartan form can be written as
\be d \upsilon \, \upsilon^{-1} + \upsilon^{\dagger\, -1} d \upsilon^\dagger =   \upsilon^{\dagger\, -1} d (\upsilon^\dagger \upsilon ) \upsilon^{-1} \ee
we will define the complex coordinates 
\be \mathpzc{t} =  \mathpzc{a}  + i \upsilon^\dagger \upsilon \ee
with 
\begin{multline}    \upsilon^\dagger \upsilon
=  \\ \left(\begin{array}{ccc} 
\hspace{2mm}  e^{-2\phi_1}  \hspace{2mm}& \hspace{2mm} -  e^{-2\phi_1} \xi \hspace{2mm}  &  \hspace{1.0mm}  -  e^{-2\phi_1}( {\scriptstyle \zeta^* - \stfrac{1}{2} \xi \eta})    \hspace{1.0mm} \\
\hspace{2mm} -  e^{-2\phi_1} \xi^*   \hspace{2mm}& \hspace{2mm}   e^{-2\phi_2} + e^{-2\phi_1} |\xi|^2    \hspace{2mm}  &  \hspace{1.0mm} -  e^{-2\phi_2} \eta + e^{-2\phi_1 } {\scriptstyle \xi^*} ( {\scriptstyle \zeta^* - \stfrac{1}{2} \xi \eta})     \hspace{1.0mm} \\
 \hspace{1.0mm} -  e^{-2\phi_1}( {\scriptstyle \zeta - \stfrac{1}{2}  \eta^* \xi^*})   \hspace{1.0mm}&  \hspace{1.0mm} {\small -  e^{-2\phi_2} \eta^* +  e^{-2\phi_1 } ( {\scriptstyle \zeta - \stfrac{1}{2}  \eta^* \xi^* }){\scriptstyle  \xi} } \hspace{1.0mm}  &  \hspace{1.0mm}  {\scriptstyle e^{-2\phi_3} +  e^{-2\phi_2} |\eta|^2 + e^{-2 \phi_1}   |{\scriptstyle \zeta^* - \stfrac{1}{2} \xi \eta}|^2 }  \hspace{1.0mm} 
\end{array} \right) \end{multline} 
The metric is still defined in terms of the complex 1-form valued in the Jordan algebra 
\be D \mathpzc{t} \equiv \rho(\varsigma) \Scal{   \upsilon^{\dagger\, -1}  d \mathpzc{t} \, \upsilon^{-1} } \ee
but instead of being given by the $SL(3,\mathds{C}^*)$  invariant trace, as 
\be ds^2 \ne \trace  D \mathpzc{t} \star D \bar{\mathpzc{t}} \ee
with 
\be \bar{\mathpzc{t}} \equiv  \mathpzc{a} - i \upsilon^\dagger \upsilon \ee
the definition requires to split $\mathds{C}^*$ into $\mathds{R} \oplus \mathds{R}$ as
\be  D \mathpzc{t}_\un + \ell D \mathpzc{t}_\deux \ee
and the metric is given by the $\mathds{R}_+^* \times \mathds{R}_+^*$ invariant trace 
\be ds^2 =  \trace  D \mathpzc{t}_\un \star D \bar{\mathpzc{t}}_\un +   \trace  D \mathpzc{t}_\deux \star D \bar{\mathpzc{t}}_\deux + 4 d\varsigma_+^{\; 2} + 4 d\varsigma_-^{\; 2} \label{splitMetric} \ee
Therefore the metric is not directly determined in terms of the variables $\mathpzc{t}$ and is not invariant under the holomorphic action of the symmetry group of the determinant of the Jordan algebra $SL(3,\mathds{C}^*)$. Nevertheless, we will see that these coordinates are particularly convenient for describing black hole solutions. 

\vskip 2mm

In the same way, the components of $\zeta^{ABC} E_{ABC}$, rewritten as a $6\times 15$ matrix 
\be \left(\begin{array}{ccccccccccccccc} 
\hspace{1.5mm} 0 \hspace{1.5mm}& \hspace{1.5mm} P^0 \hspace{1.5mm}  & \hspace{1.5mm} P^3 \hspace{1.5mm} & \hspace{1.5mm} P^2  \hspace{1.5mm} & \hspace{1.5mm} Q_1  \hspace{1.5mm} & \hspace{1.5mm} p_1  \hspace{1.5mm} & \hspace{1.5mm} 0 \hspace{1.5mm} & \hspace{1.5mm} -p_3 \hspace{1.5mm} & \hspace{1.5mm} 0 \hspace{1.5mm} & \hspace{1.5mm} - \bar \x \hspace{1.5mm} & \hspace{1.5mm} \y \hspace{1.5mm} & \hspace{1.5mm} 0 \hspace{1.5mm}& \hspace{1.5mm} 0 \hspace{1.5mm} & \hspace{1.5mm} \z \hspace{1.5mm} & \hspace{1.5mm} q^2 \hspace{1.5mm} \\
\hspace{1.5mm} 0 \hspace{1.5mm}& \hspace{1.5mm} P^1 \hspace{1.5mm}  & \hspace{1.5mm} Q_2  \hspace{1.5mm} & \hspace{1.5mm} Q_3   \hspace{1.5mm} & \hspace{1.5mm} - Q_0 \hspace{1.5mm} & \hspace{1.5mm} \bar \y \hspace{1.5mm} & \hspace{1.5mm} p_3 \hspace{1.5mm} & \hspace{1.5mm} 0  \hspace{1.5mm} & \hspace{1.5mm} \bar \x \hspace{1.5mm} & \hspace{1.5mm} 0 \hspace{1.5mm} & \hspace{1.5mm} q^1 \hspace{1.5mm} & \hspace{1.5mm} -\z \hspace{1.5mm}& \hspace{1.5mm} -q^2  \hspace{1.5mm} & \hspace{1.5mm} 0 \hspace{1.5mm} & \hspace{1.5mm} 0 \hspace{1.5mm} \\

\hspace{1.5mm} \z \hspace{1.5mm}& \hspace{1.5mm} 0 \hspace{1.5mm}  & \hspace{1.5mm} 0  \hspace{1.5mm} & \hspace{1.5mm} \x  \hspace{1.5mm} & \hspace{1.5mm} q^3   \hspace{1.5mm} & \hspace{1.5mm} 0 \hspace{1.5mm} & \hspace{1.5mm} P^0 \hspace{1.5mm} & \hspace{1.5mm} P^1  \hspace{1.5mm} & \hspace{1.5mm} P^3 \hspace{1.5mm} & \hspace{1.5mm} Q_2  \hspace{1.5mm} & \hspace{1.5mm} p_2 \hspace{1.5mm} & \hspace{1.5mm} 0 \hspace{1.5mm}& \hspace{1.5mm} -p_1 \hspace{1.5mm} & \hspace{1.5mm} 0 \hspace{1.5mm} & \hspace{1.5mm} - \bar \y \hspace{1.5mm} \\
\hspace{1.5mm} q^2 \hspace{1.5mm}& \hspace{1.5mm} -\x \hspace{1.5mm}  & \hspace{1.5mm} -q^3  \hspace{1.5mm} & \hspace{1.5mm} 0  \hspace{1.5mm} & \hspace{1.5mm} 0   \hspace{1.5mm} & \hspace{1.5mm} 0 \hspace{1.5mm} & \hspace{1.5mm} P^2 \hspace{1.5mm} & \hspace{1.5mm} Q_3   \hspace{1.5mm} & \hspace{1.5mm} Q_1  \hspace{1.5mm} & \hspace{1.5mm} - Q_0   \hspace{1.5mm} & \hspace{1.5mm} \bar \z \hspace{1.5mm} & \hspace{1.5mm} p_1 \hspace{1.5mm}& \hspace{1.5mm} 0  \hspace{1.5mm} & \hspace{1.5mm} \bar \y \hspace{1.5mm} & \hspace{1.5mm} 0 \hspace{1.5mm} \\

\hspace{1.5mm} p_3  \hspace{1.5mm}& \hspace{1.5mm} 0 \hspace{1.5mm}  & \hspace{1.5mm} -p_2  \hspace{1.5mm} & \hspace{1.5mm} 0  \hspace{1.5mm} & \hspace{1.5mm} - \bar \z  \hspace{1.5mm} & \hspace{1.5mm} \bar \x \hspace{1.5mm} & \hspace{1.5mm} 0  \hspace{1.5mm} & \hspace{1.5mm} 0   \hspace{1.5mm} & \hspace{1.5mm} \y  \hspace{1.5mm} & \hspace{1.5mm} q^1    \hspace{1.5mm} & \hspace{1.5mm} 0  \hspace{1.5mm} & \hspace{1.5mm} P^0 \hspace{1.5mm}& \hspace{1.5mm} P^2  \hspace{1.5mm} & \hspace{1.5mm} P^1 \hspace{1.5mm} & \hspace{1.5mm} Q_3 \hspace{1.5mm} \\

\hspace{1.5mm} \bar \x  \hspace{1.5mm}& \hspace{1.5mm} p_2  \hspace{1.5mm}  & \hspace{1.5mm} 0  \hspace{1.5mm} & \hspace{1.5mm} \bar \z \hspace{1.5mm} & \hspace{1.5mm} 0   \hspace{1.5mm} & \hspace{1.5mm} q^3 \hspace{1.5mm} & \hspace{1.5mm} - \y   \hspace{1.5mm} & \hspace{1.5mm} - q^1   \hspace{1.5mm} & \hspace{1.5mm} 0  \hspace{1.5mm} & \hspace{1.5mm} 0    \hspace{1.5mm} & \hspace{1.5mm} 0  \hspace{1.5mm} & \hspace{1.5mm} P^3 \hspace{1.5mm}& \hspace{1.5mm} Q_1  \hspace{1.5mm} & \hspace{1.5mm} Q_2  \hspace{1.5mm} & \hspace{1.5mm} - Q_0  \hspace{1.5mm} 
\label{EQP} 

\end{array} \right) \ee 
can be recombined into two scalars $q_0 \equiv Q_0$ and $p^0 \equiv P^0$ and two Hermitian matrices $\mathpzc{Q}$
\be \mathpzc{Q} =  \left(\begin{array}{ccc} 
\hspace{2mm} Q_1   \hspace{2mm}& \hspace{2mm} \frac{q^3-\mathpzc{z}}{2} + \ell \frac{q^3+\mathpzc{z}}{2}  \hspace{2mm}  & \hspace{2mm}  \frac{q^2-\mathpzc{y}}{2} + \ell \frac{q^2+\mathpzc{y}}{2} \hspace{2mm} \\
\hspace{2mm}  \frac{q^3-\mathpzc{z}}{2} - \ell \frac{q^3+\mathpzc{z}}{2}    \hspace{2mm}& \hspace{2mm} Q_2   \hspace{2mm}  & \hspace{2mm} \frac{q^1-\mathpzc{x}}{2} + \ell \frac{q^1+\mathpzc{x}}{2}    \hspace{2mm} \\
\hspace{2mm} \frac{q^2-\mathpzc{y}}{2} - \ell \frac{q^2+\mathpzc{y}}{2}    \hspace{2mm}& \hspace{2mm} \frac{q^1-\mathpzc{x}}{2} - \ell \frac{q^1+\mathpzc{x}}{2}   \hspace{2mm}  & \hspace{2mm} Q_3    \hspace{2mm} 
\end{array} \right) \ee
and $\mathpzc{P}$
\be \mathpzc{P} =  \left(\begin{array}{ccc} 
\hspace{2mm} P_1   \hspace{2mm}& \hspace{2mm} \frac{p_3-\bar{\mathpzc{z}}}{2} + \ell \frac{p_3+\bar{\mathpzc{z}}}{2}  \hspace{2mm}  & \hspace{2mm}  \frac{p_2-\bar{\mathpzc{y}}}{2} + \ell \frac{p_2+\bar{\mathpzc{y}}}{2} \hspace{2mm} \\
\hspace{2mm}  \frac{p_3-\bar{\mathpzc{z}}}{2} - \ell \frac{p_3+\bar{\mathpzc{z}}}{2}    \hspace{2mm}& \hspace{2mm} P_2   \hspace{2mm}  & \hspace{2mm} \frac{p_1-\bar{\mathpzc{x}}}{2} + \ell \frac{p_1+\bar{\mathpzc{x}}}{2}    \hspace{2mm} \\
\hspace{2mm} \frac{p_2-\bar{\mathpzc{y}}}{2} - \ell \frac{p_2+\bar{\mathpzc{y}}}{2}    \hspace{2mm}& \hspace{2mm} \frac{p_1-\bar{\mathpzc{x}}}{2} - \ell \frac{p_1+\bar{\mathpzc{x}}}{2}   \hspace{2mm}  & \hspace{2mm} P_3    \hspace{2mm} 
\end{array} \right) \ee

These coordinates render the $\N=2$ truncation of the theory rather transparent, because the latter is simply defined by the real solutions (no $\ell$ components) with $\varsigma_\pm = 0 $. The coordinates then correspond to the conventional special coordinates associated to the special K\"{a}hler space  $Sp(6,\mathds{R}) / U(3) $ \cite{GunaydinMagic,GunaydinJordan}. Requiring all the symmetric 3 by 3 matrices to be diagonal corresponds then to the STU truncation. In these notations the two truncations of the solutions we will describe in the following will be transparent. The corresponding three-dimensional models are defined over the symmetric spaces
\be SO(4,4)/\scal{SO(2,2) \times SO(2,2) } \subset F_{4(4)} / \scal{SL(2) \times_{\mathds{Z}_2} Sp(6,\mathds{R})} \subset E_{6(6)} / \Sp(8,\mathds{R}) \ee

\subsection{Lift to $E_{7(7)}$ special coordinates}
\label{E7Special}
We will now describe how these coordinates can be generalised to the symmetric space  $E_{7(7)} / \SU $ of $\N=8$ supergravity moduli. Just like $SL(3,\mathds{R}) \times SL(3,\mathds{R}) \cong SL(3,\mathds{C}^*)$, it is  possible to realise $E_{6(6)}$ as $SL(3,\mathds{O^*})$. Recall that the split octonions are defined as doublets of quaternions
\be  k + \ell s \; , \quad e + \ell y \; \ee
which are multiplied according to the Cayley rule  \cite{Jordan}
\be ( e + \ell y ) ( k +  \ell s ) = e k + s y^* + \ell ( e^* s + ky ) \label{Cayley} \ee
and for which the complex conjugation reads
 \be ( k + \ell s )^* = k^* - \ell s \ee
For the case of the $E_{6(6)}$ truncation one considered the split complex, and these relations reduced to say that $\ell^2 = 1$ and $\ell^* = - \ell$. As opposed to the case of the associative composition algebra, there is no representation of $SL(3,\mathds{O^*})$ on 3-vectors of split octonions because of the lack of associativity. However, the associativity problem is resolved in the case of the representation on Hermitian 3 by 3 matrices over the split octonions. More precisely, the action of traceless 3 by 3 matrices over  $\mathds{O^*}$ on a Hermitian 3 by 3 matrix closes modulo $\mathfrak{g}_{2(2)}$ automorphism transformations. This Lie algebra representation exponentiates to a group representation of $E_{6(6)}$. 

Let us consider a 3 by 3 matrix over any composition algebra which diagonal components are all real
\be X =  \left(\begin{array}{ccc} 
\hspace{2mm} h_1   \hspace{2mm}& \hspace{2mm} e_3   \hspace{2mm}  & \hspace{2mm} e^*_2   \hspace{2mm} \\
\hspace{2mm} f^*_3   \hspace{2mm}& \hspace{2mm} h_2   \hspace{2mm}  & \hspace{2mm} e_1   \hspace{2mm} \\
\hspace{2mm}f_2   \hspace{2mm}& \hspace{2mm} f^*_1  \hspace{2mm}  & \hspace{2mm} h_3    \hspace{2mm} 
\end{array} \right) \ee
By definition, the infinitesimal transformation of a Hermitian matrix $\mathpzc{a} $
\be \delta \mathpzc{a} = X \mathpzc{a} + \mathpzc{a} X^\dagger \label{Repa}  \ee
preserves the condition that $\mathpzc{a}$ is Hermitian. In order to check that this transformation defines a Lie algebra representation, it will be convenient to define the algebra in the associated differential graded algebra $\bigwedge \sl^*_3(\mathds{O}^*)$, which can be understood as the complex of Grassmann variables valued in the co-algebra on which acts the differential $\delta$ as 
\be \delta X^i = \frac{1}{2} C^i{}_{jk} X^j X^k \label{CartanComplex} \ee
where $C^i{}_{jk}$ are the structure constants. The differential is also defined on vectors associated to the representations of the Lie algebra as in (\ref{Repa}). In this formalism, the Jacobi identity is then equivalent to the nilpotency of the differential $\delta$, as it is in the BRST formalism.  For anti-commuting parameters, one will use the convention that 
\be (e f)^* = - f^* e^* \ee
and therefore the identities on the algebra get minus signs associated to the grading. We define 
\be \delta X = X^2 - \frac{1}{2} \left(\begin{array}{ccc} 
{\scriptstyle e_3 f_3^* + e^*_2 f_2 + f_3 e_3 + f_2^* e_2 }  & 0   \hspace{2mm}  & \hspace{2mm} 0   \hspace{2mm} \\
\hspace{2mm} 0& {\scriptstyle f_3^* e_3 + e_1 f_1^* + e_3^* f_3 + f_1 e^*_1  } & 0  \hspace{2mm} \\
\hspace{2mm} 0  \hspace{2mm}& \hspace{2mm} 0    &{\scriptstyle    f_2 e_2^* + f_1^* e_1 + e_2 f_2^* + e_1 f_1^* }
\end{array} \right) \ee 
where the second term is required in order to ensure that the diagonal elements remain real.  One computes that $\delta^2$ does not vanish, but gives  
\be \delta^2 X = \sum_i \mathfrak{r}(e_i, f_i) X = \sum_i \left(\begin{array}{ccc} 
\hspace{2mm} 0   \hspace{2mm}& \hspace{2mm} \mathfrak{r}_{3-i}(e_i, f_i)  e_3   \hspace{2mm}  & \hspace{2mm} \scal{ \mathfrak{r}_{2-i}(e_i,f_i) e_2 }^*    \hspace{2mm} \\
\hspace{2mm}\scal{ \mathfrak{r}_{3-i}(e_i, f_i) f_3  }^* \hspace{2mm}& \hspace{2mm} 0   \hspace{2mm}  & \hspace{2mm} \mathfrak{r}_{1-i}(e_i,f_i) e_1   \hspace{2mm} \\
\hspace{2mm}\mathfrak{r}_{2-i}(e_i,f_i)f_2   \hspace{2mm}& \hspace{2mm}\scal{ \mathfrak{r}_{1-i}(e_i,f_i) f_1}^*  \hspace{2mm}  & \hspace{2mm} 0    \hspace{2mm} 
\end{array} \right) \label{X2Oct} \ee
where the indices of $ \mathfrak{r}_{i}(a,b)$ are defined modulo $3$ and 
\bea \mathfrak{r}_{1}(a,b) \alpha  &=& - \frac{1}{2} a^* ( b\alpha ) - \frac{1}{2}b^* ( a \alpha)  \CR
 \mathfrak{r}_{2}(a,b)\alpha &=& \frac{1}{2} ( \alpha a ) b^* + \frac{1}{2} ( \alpha b) a^* \CR
   \mathfrak{r}_{3}(a,b)\alpha &=& - a ( b^* \alpha + (-1)^\alpha \alpha^* b ) - b ( a^* \alpha + (-1)^\alpha \alpha^* a ) \eea
for anticommuting composition algebra elements $a$ and $b$ and where $(-)^\alpha$ is minus if $\alpha$ is anticommuting (\ie $(-)^\alpha=-1$ in (\ref{X2Oct}), whereas $(-)^\alpha=1$ in (\ref{X2alpha}) to come). These transformations generate the automorphisms of the algebra of cyclic products of three copies of the composition algebra. Indeed $ \mathfrak{r}_{3}(a,b)$ is manifestly an infinitesimal rotation whereas  $\mathfrak{r}_{1}(a,b)$ and  $\mathfrak{r}_{2}(a,b)$ correspond to spinor transformations. Their group analogue are the triality related fundamental representations of $Spin(4,4)$ satisfying 
\be  \scal{ \rho_i(\varsigma) \beta^* \alpha^* }^* = \scal{ \rho_{i+1}(\varsigma) \alpha} \scal{  \rho_{i+2}(\varsigma)  \beta } \ee
for the split octonions. We conclude that the algebra closes modulo $\so(4,4)$ transformations in this case. This representation  corresponds precisely to the graded decomposition of $\e_{6(6)}$
\be \e_{6(6)} \cong {\bf 8}_2^\ord{-2} \oplus ( {\bf 8}_1 \oplus {\bf 8}_3 )^\ord{-1} \oplus \scal{ \gl_1 \oplus \gl_1 \oplus \so(4,4) }^\ord{0}   \oplus ( {\bf 8}_1 \oplus {\bf 8}_3 )^\ord{1}\oplus {\bf 8}_2^\ord{2} \ee
One checks also that $E_{6(6)}$ is represented in this way on the Jordan algebra such that 
\be \delta^2 \mathpzc{a} =  \sum_i \mathfrak{r}(e_i, f_i) \mathpzc{a} =  \sum_i \left(\begin{array}{ccc} 
\hspace{2mm} 0   \hspace{2mm}& \hspace{2mm} \mathfrak{r}_{3-i}(e_i, f_i)  \alpha_3   \hspace{2mm}  & \hspace{2mm} \scal{ \mathfrak{r}_{2-i}(e_i,f_i) \alpha_2 }^*    \hspace{2mm} \\
\hspace{2mm}\scal{ \mathfrak{r}_{3-i}(e_i, f_i) \alpha_3  }^* \hspace{2mm}& \hspace{2mm} 0   \hspace{2mm}  & \hspace{2mm} \mathfrak{r}_{1-i}(e_i,f_i) \alpha_1   \hspace{2mm} \\
\hspace{2mm}\mathfrak{r}_{2-i}(e_i,f_i)\alpha_2   \hspace{2mm}& \hspace{2mm}\scal{ \mathfrak{r}_{1-i}(e_i,f_i) \alpha_1}^*  \hspace{2mm}  & \hspace{2mm} 0    \hspace{2mm} 
\end{array} \right) \label{X2alpha} \ee
The coordinates we have been using for $SL(6)$ will extend straightforwardly to $E_{7(7)}$ in this way. To take care of the associativity issue, we will define the ordered exponential
\begin{multline}  : \exp[X] \mathpzc{a} \exp[X^\dagger] : \; \equiv \mathpzc{a} + X \mathpzc{a} + \mathpzc{a} X^\dagger + \frac{1}{2} \scal{ X ( X \mathpzc{a}+ \mathpzc{a} X^\dagger) + ( X\mathpzc{a}+ \mathpzc{a}X^\dagger) X^\dagger } \\ + \frac{1}{6} \Scal{  X  \scal{ X ( X \mathpzc{a}+ \mathpzc{a} X^\dagger) + ( X\mathpzc{a}+ \mathpzc{a}X^\dagger) X^\dagger } +  \scal{ X ( X \mathpzc{a}+ \mathpzc{a} X^\dagger) + ( X\mathpzc{a}+ \mathpzc{a}X^\dagger) X^\dagger } X^\dagger } + \dots \label{NormalProd}  \end{multline}
\subsubsection*{Special coordinates for $E_{7(7)} / \SU$} 
The $GL(1) \times E_{6(6)}$ coset element will be represented by 16 scalar fields (written collectively $\varsigma^I$) parametrizing the quaternionic symmetric space $SO(4,4)/(SO(4) \times SO(4))$, three scalars $\phi_i$ and three $\mathds{O}^*$ valued scalars $\xi,\eta,\zeta$ such that the group is represented in the $\overline{\bf 27}$ as 
\be  \mathpzc{a}^\prime  =  \rho(\varsigma) \left[  \left(\begin{array}{ccc} 
\hspace{0.5mm} e^{\phi_1}   \hspace{0.5mm}& \hspace{0.5mm} 0  \hspace{0.5mm}  & \hspace{0.5mm} 0  \hspace{0.5mm} \\
\hspace{0.5mm} 0  \hspace{0.5mm}& \hspace{0.5mm} e^{\phi_2}  \hspace{0.5mm}  & \hspace{0.5mm} 0  \hspace{0.5mm} \\
 \hspace{0.5mm}  0     \hspace{0.5mm}& \hspace{0.5mm} 0 \hspace{0.5mm}  & \hspace{0.5mm} e^{\phi_3}   \hspace{0.5mm} 
\end{array} \right)    : \exp  \left(\begin{array}{ccc} 
\hspace{0.5mm} 0  \hspace{0.5mm}& \hspace{0.5mm} 0  \hspace{0.5mm}  & \hspace{0.5mm} 0  \hspace{0.5mm} \\
\hspace{0.5mm} \xi^*   \hspace{0.5mm}& \hspace{0.5mm} 0 \hspace{0.5mm}  & \hspace{0.5mm} 0  \hspace{0.5mm} \\
 \hspace{0.5mm}  \zeta   \hspace{0.5mm}& \hspace{0.5mm}  \eta^* \hspace{0.5mm}  & \hspace{0.5mm} 0  \hspace{0.5mm} 
\end{array} \right)  \mathpzc{a} \exp \left(\begin{array}{ccc} 
\hspace{0.5mm} 0  \hspace{0.5mm}& \hspace{0.5mm} \xi   \hspace{0.5mm}  & \hspace{0.5mm} \zeta^*     \hspace{0.5mm} \\
\hspace{0.5mm} 0   \hspace{0.5mm}& \hspace{0.5mm} 0\hspace{0.5mm}  & \hspace{0.5mm}  \eta  \hspace{0.5mm} \\
 \hspace{0.5mm} 0  \hspace{0.5mm} & \hspace{0.5mm} 0  \hspace{0.5mm}  & \hspace{0.5mm} 0   \hspace{0.5mm} 
\end{array} \right) :   \left(\begin{array}{ccc} 
\hspace{0.5mm} e^{\phi_1}   \hspace{0.5mm}& \hspace{0.5mm} 0  \hspace{0.5mm}  & \hspace{0.5mm}0    \hspace{0.5mm} \\
\hspace{0.5mm} 0   \hspace{0.5mm}& \hspace{0.5mm} e^{\phi_2}  \hspace{0.5mm}  & \hspace{0.5mm} 0  \hspace{0.5mm} \\
 \hspace{0.5mm} 0  \hspace{0.5mm} & \hspace{0.5mm} 0  \hspace{0.5mm}  & \hspace{0.5mm} e^{\phi_3}   \hspace{0.5mm} 
\end{array} \right) \right] \ee
One can still define the complex coordinates $\mathpzc{t}$ by acting on the identity of the Jordan algebra in the conjugate representation 
\bea \mathpzc{t} &=& \mathpzc{a} + i   : \exp  \left(\begin{array}{ccc} 
\hspace{0.0mm} 0  \hspace{0.0mm}& \hspace{0.0mm} 0  \hspace{0.0mm}  & \hspace{0.0mm} 0  \hspace{0.0mm} \\
\hspace{0.0mm} -\xi^*   \hspace{0.0mm}& \hspace{0.0mm} 0 \hspace{0.0mm}  & \hspace{0.0mm} 0  \hspace{0.0mm} \\
 \hspace{0.0mm}  -\zeta   \hspace{0.0mm}& \hspace{0.0mm} - \eta^* \hspace{0.0mm}  & \hspace{0.0mm} 0  \hspace{0.0mm} 
\end{array} \right)   \left(\begin{array}{ccc} 
\hspace{0.0mm} e^{-2\phi_1}   \hspace{0.0mm}& \hspace{0.0mm} 0  \hspace{0.0mm}  & \hspace{0.0mm} 0  \hspace{0.0mm} \\
\hspace{0.0mm} 0  \hspace{0.0mm}& \hspace{0.0mm} e^{-2\phi_2}  \hspace{0.0mm}  & \hspace{0.0mm} 0  \hspace{0.0mm} \\
 \hspace{0.0mm}  0     \hspace{0.0mm}& \hspace{0.0mm} 0 \hspace{0.0mm}  & \hspace{0.0mm} e^{-2\phi_3}   \hspace{0.0mm} 
\end{array} \right) \exp \left(\begin{array}{ccc} 
\hspace{0.0mm} 0  \hspace{0.0mm}& \hspace{0.0mm}- \xi   \hspace{0.0mm}  & \hspace{0.0mm}- \zeta^*     \hspace{0.0mm} \\
\hspace{0.0mm} 0   \hspace{0.0mm}& \hspace{0.0mm} 0\hspace{0.0mm}  & \hspace{0.0mm}  -\eta  \hspace{0.0mm} \\
 \hspace{0.0mm} 0  \hspace{0.0mm} & \hspace{0.0mm} 0  \hspace{0.0mm}  & \hspace{0.0mm} 0   \hspace{0.0mm} 
\end{array} \right) :  \\
&=& \mathpzc{a} + i \left(\begin{array}{ccc} 
\hspace{2mm}  e^{-2\phi_1}  \hspace{2mm}& \hspace{2mm} -  e^{-2\phi_1} \xi \hspace{2mm}  &  \hspace{1.0mm}  -  e^{-2\phi_1}( {\scriptstyle \zeta^* - \stfrac{1}{2} \xi \eta})    \hspace{1.0mm} \\
\hspace{2mm} -  e^{-2\phi_1} \xi^*   \hspace{2mm}& \hspace{2mm}   e^{-2\phi_2} + e^{-2\phi_1} |\xi|^2    \hspace{2mm}  &  \hspace{1.0mm} -  e^{-2\phi_2} \eta + e^{-2\phi_1 } {\scriptstyle \xi^*}( {\scriptstyle \zeta^* - \stfrac{1}{2} \xi \eta})     \hspace{1.0mm} \\
 \hspace{1.0mm} -  e^{-2\phi_1}( {\scriptstyle \zeta - \stfrac{1}{2}  \eta^* \xi^*})   \hspace{1.0mm}&  \hspace{1.0mm} {\small -  e^{-2\phi_2} \eta^* + i e^{-2\phi_1 } ( {\scriptstyle \zeta - \stfrac{1}{2}  \eta^* \xi^* }) {\scriptstyle \xi} } \hspace{1.0mm}  &  \hspace{1.0mm}  {\scriptstyle e^{-2\phi_3} +  e^{-2\phi_2} |\eta|^2 + e^{-2 \phi_1}   |{\scriptstyle \zeta^* - \stfrac{1}{2} \xi \eta}|^2 }  \hspace{1.0mm} 
\end{array} \right) \nn
\eea
The metric is defined in terms of 
\be D\mathpzc{t} = \rho(\varsigma) \left[  \left(\begin{array}{ccc} 
\hspace{0.5mm} e^{\phi_1}   \hspace{0.5mm}& \hspace{0.5mm} 0  \hspace{0.5mm}  & \hspace{0.5mm} 0  \hspace{0.5mm} \\
\hspace{0.5mm} 0  \hspace{0.5mm}& \hspace{0.5mm} e^{\phi_2}  \hspace{0.5mm}  & \hspace{0.5mm} 0  \hspace{0.5mm} \\
 \hspace{0.5mm}  0     \hspace{0.5mm}& \hspace{0.5mm} 0 \hspace{0.5mm}  & \hspace{0.5mm} e^{\phi_3}   \hspace{0.5mm} 
\end{array} \right)    : \exp  \left(\begin{array}{ccc} 
\hspace{0.5mm} 0  \hspace{0.5mm}& \hspace{0.5mm} 0  \hspace{0.5mm}  & \hspace{0.5mm} 0  \hspace{0.5mm} \\
\hspace{0.5mm} \xi^*   \hspace{0.5mm}& \hspace{0.5mm} 0 \hspace{0.5mm}  & \hspace{0.5mm} 0  \hspace{0.5mm} \\
 \hspace{0.5mm}  \zeta   \hspace{0.5mm}& \hspace{0.5mm}  \eta^* \hspace{0.5mm}  & \hspace{0.5mm} 0  \hspace{0.5mm} 
\end{array} \right)  d\mathpzc{t} \exp \left(\begin{array}{ccc} 
\hspace{0.5mm} 0  \hspace{0.5mm}& \hspace{0.5mm} \xi   \hspace{0.5mm}  & \hspace{0.5mm} \zeta^*     \hspace{0.5mm} \\
\hspace{0.5mm} 0   \hspace{0.5mm}& \hspace{0.5mm} 0\hspace{0.5mm}  & \hspace{0.5mm}  \eta  \hspace{0.5mm} \\
 \hspace{0.5mm} 0  \hspace{0.5mm} & \hspace{0.5mm} 0  \hspace{0.5mm}  & \hspace{0.5mm} 0   \hspace{0.5mm} 
\end{array} \right) :   \left(\begin{array}{ccc} 
\hspace{0.5mm} e^{\phi_1}   \hspace{0.5mm}& \hspace{0.5mm} 0  \hspace{0.5mm}  & \hspace{0.5mm}0    \hspace{0.5mm} \\
\hspace{0.5mm} 0   \hspace{0.5mm}& \hspace{0.5mm} e^{\phi_2}  \hspace{0.5mm}  & \hspace{0.5mm} 0  \hspace{0.5mm} \\
 \hspace{0.5mm} 0  \hspace{0.5mm} & \hspace{0.5mm} 0  \hspace{0.5mm}  & \hspace{0.5mm} e^{\phi_3}   \hspace{0.5mm} 
\end{array} \right) \right] \ee
similarly as in (\ref{splitMetric}) by
\be ds^2 = \trace  D \mathpzc{t}_\un \star D \bar{\mathpzc{t}}_\un +  \trace  D \mathpzc{t}_\deux \star D \bar{\mathpzc{t}}_\deux + g_{IJ}(\varsigma)  d\varsigma^I d \varsigma^J  \label{SOsplitmetric} \ee

The determinant is defined in the Jordan algebra by \cite{Jordan}
\be \det[\mathpzc{a}] = a_1 a_2 a_3 - \sum_i a_i |\alpha_i|^2 + \alpha_1 \alpha_2 \alpha_3 + \alpha_3^* \alpha_2^* \alpha_1^* \ee
and we have that 
\be i \det[ \mathpzc{t} -   \bar{\mathpzc{t}} ] = 8 \,  e^{-2\phi_1} e^{-2\phi_2} e^{-2\phi_3} \ee
This determinant defines the symmetric cross product such that 
\be \mathpzc{a} \times \mathpzc{a} = \mathpzc{a}^2 - \trace  \mathpzc{a} \cdot  \mathpzc{a}  - \frac{1}{2} \trace \mathpzc{a} ^2 + \frac{1}{2} \mbox{Tr}^2 \, \mathpzc{a}   \ee
The coordinates can be chosen independently of a specific parametrization by requiring that this determinant is the product of three strictly positive fields. Under this condition one can define the Jordan inverse of $(\mathpzc{t}- \bar{\mathpzc{t}})$ as
\be (\mathpzc{t}- \bar{\mathpzc{t}})^{-1} =   \det[ \mathpzc{t} -   \bar{\mathpzc{t}} ]^{-1} (\mathpzc{t}- \bar{\mathpzc{t}}) \times (\mathpzc{t}- \bar{\mathpzc{t}}) \ee
For a Jordan algebra defined over a field, like in $\N=2$ special geometry, the scalar field metric would be simply given by \cite{GunaydinJordan}
\be ds^2 = -\frac{4}{  \det[ \mathpzc{t} -   \bar{\mathpzc{t}} ]^{2}} \trace  (\mathpzc{t}- \bar{\mathpzc{t}}) \times (\mathpzc{t}- \bar{\mathpzc{t}})   d \mathpzc{t}\,     (\mathpzc{t}- \bar{\mathpzc{t}}) \times (\mathpzc{t}- \bar{\mathpzc{t}}) d\bar{\mathpzc{t}}  \ee
which implies that the metric is invariant with respect to a linear action of the symmetry group preserving the determinant. However in our case the metric is defined as in (\ref{SOsplitmetric}) and is not preserved by the complete $E_{6(6)}$ symmetry, but only by its $SO(4,4)$ subgroup. The solutions we will describe in this paper satisfy nevertheless that $ D \mathpzc{t}_\deux$ is real, such that the remaining coordinates admit an enhanced linearly realised symmetry $SU(2) \times_{\mathds{Z}_2} SL(3,\mathds{H})$, as for the solutions sitting in the maximal $\N=2$ truncation of the theory. \footnote{Note however that this symmetry is incompatible with $SO(4,4)$. Indeed, extracting $\mbox{Im}[\mathpzc{t}_\deux]$ from the special coordinates and combining it with the $\varsigma$'s to parametrize the symmetric space $F_{4(4)} / ( Sp(1) \times Sp(3))$ permits to compensate for the linear $E_{6(6)}$ realisation, however, only its $SU(2) \times_{\mathds{Z}_2} SL(3,\mathds{H})$ preserves the condition $\mbox{Im}[\mathpzc{t}_\deux] = 0 $.}

\subsubsection*{The electromagnetic charges in the ${\bf 56}$}

The electromagnetic charges will be described in the same way. We consider the decomposition of eletromagnetic charges according to the $E_{6(6)}\subset E_{7(7)}$ subgroup, such that they decompose into ${\bf 1} \oplus {\bf 27}$ electric charges, $q_0$ and the Jordan algebra element 
\be \mathpzc{Q} =  \left(\begin{array}{ccc} 
\hspace{2mm} Q_1   \hspace{2mm}& \hspace{2mm} \frac{q^3-\mathpzc{z}}{2} + \ell \frac{q^3+\mathpzc{z}}{2}  \hspace{2mm}  & \hspace{2mm}  \Scal{ \frac{q^2-\mathpzc{y}}{2}}^* + \ell \frac{q^2+\mathpzc{y}}{2} \hspace{2mm} \\
\hspace{2mm}  \Scal{ \frac{q^3-\mathpzc{z}}{2}}^* - \ell \frac{q^3+\mathpzc{z}}{2}    \hspace{2mm}& \hspace{2mm} Q_2   \hspace{2mm}  & \hspace{2mm} \frac{q^1-\mathpzc{x}}{2} + \ell \frac{q^1+\mathpzc{x}}{2}    \hspace{2mm} \\
\hspace{2mm} \frac{q^2-\mathpzc{y}}{2} - \ell \frac{q^2+\mathpzc{y}}{2}    \hspace{2mm}& \hspace{2mm} \Scal{ \frac{q^1-\mathpzc{x}}{2} }^* - \ell \frac{q^1+\mathpzc{x}}{2}   \hspace{2mm}  & \hspace{2mm} Q_3    \hspace{2mm} 
\end{array} \right) \ee
and ${\bf 1} \oplus \overline{\bf 27}$ magnetic charges,  $p^0$ and the Jordan algebra element 
\be \mathpzc{P} =  \left(\begin{array}{ccc} 
\hspace{2mm} P_1   \hspace{2mm}& \hspace{2mm} \frac{p_3-\bar{\mathpzc{z}}}{2} + \ell \frac{p_3+\bar{\mathpzc{z}}}{2}  \hspace{2mm}  & \hspace{2mm}  \Scal{ \frac{p_2-\bar{\mathpzc{y}}}{2}}^*  + \ell \frac{p_2+\bar{\mathpzc{y}}}{2} \hspace{2mm} \\
\hspace{2mm}  \Scal{ \frac{p_3-\bar{\mathpzc{z}}}{2}}^* - \ell \frac{p_3+\bar{\mathpzc{z}}}{2}    \hspace{2mm}& \hspace{2mm} P_2   \hspace{2mm}  & \hspace{2mm} \frac{p_1-\bar{\mathpzc{x}}}{2} + \ell \frac{p_1+\bar{\mathpzc{x}}}{2}    \hspace{2mm} \\
\hspace{2mm} \frac{p_2-\bar{\mathpzc{y}}}{2} - \ell \frac{p_2+\bar{\mathpzc{y}}}{2}    \hspace{2mm}& \hspace{2mm} \Scal{  \frac{p_1-\bar{\mathpzc{x}}}{2}}^*  - \ell \frac{p_1+\bar{\mathpzc{x}}}{2}   \hspace{2mm}  & \hspace{2mm} P_3    \hspace{2mm} 
\end{array} \right) \ee
where $q^i, p_i , \mathpzc{x}, \mathpzc{y} , \mathpzc{z}, \bar{\mathpzc{x}}, \bar{\mathpzc{y}}, \bar{\mathpzc{z}}$ are quaternions. 

It will often be useful to decompose a 3 by 3 Hermitian matrix over the split octonions into a 3 by 3 Hermitian matrix over the quaternions and a 3 vector over the quaternions. For this purpose we define the Hodge like operation $\invo$ from a 3 vector over the quaternions $Y$ to a 3 by 3 antisymmetric matrix over the quaternions 
\be \invo  \left(\begin{array}{c} 
\hspace{0.5mm} Y_1 \hspace{0.5mm} \\
\hspace{0.5mm} Y_2  \hspace{0.5mm}\\
 \hspace{0.5mm}  Y_3     \hspace{0.5mm}
\end{array} \right)   =  \left(\begin{array}{ccc} 
\hspace{0.5mm}0  \hspace{0.5mm}& \hspace{0.5mm} Y_3  \hspace{0.5mm}  & \hspace{0.5mm} -Y_2  \hspace{0.5mm} \\
\hspace{0.5mm} -Y_3  \hspace{0.5mm}& \hspace{0.5mm} 0  \hspace{0.5mm}  & \hspace{0.5mm} Y_1  \hspace{0.5mm} \\
 \hspace{0.5mm}  Y_2     \hspace{0.5mm}& \hspace{0.5mm} -Y_1 \hspace{0.5mm}  & \hspace{0.5mm}0   \hspace{0.5mm} 
\end{array} \right) \ee
Using this notation, we will write 
\be \mathpzc{Q} = \mathpzc{q} + \ell \invo q \; , \qquad \mathpzc{P} = \mathpzc{p} + \ell \invo p \ee
One has then \footnote{Whereas the Jordan product decomposes as
\be \mathpzc{Q}^2 = \mathpzc{q}^2 + q q^\dagger - ( q^\dagger q) \mathds{1} + \ell \invo ( \trace [ \mathpzc{q}] - \mathpzc{q} ) q \ee}
\be \mathpzc{Q} \times \mathpzc{Q} = \mathpzc{q} \times \mathpzc{q} + q q^\dagger - \ell \invo \mathpzc{q} q \ee
where $q q^\dagger$ is the Hermitian matrix defined from the tensor product of the vector $q$ with its Hermitian conjugate $q^\dagger$, and $\mathpzc{q} q$ is the ordinary action of a Hermitian matrix on a vector. In particular, we will consider $\mathpzc{p} \mathpzc{q} q$ as the repeated action of the matrix such that the factor $\mathpzc{p} \mathpzc{q}$ is the ordinary product of two Hermitian matrices, and not the symmetric Jordan product. In this way 
\be \det[ \mathpzc{Q}\, ] = \det[ \mathpzc{q}] + ( q^\dagger \mathpzc{q} q) \ee
and 
\be \trace [ \mathpzc{Q}\,  \mathpzc{P} ] = \trace[ \mathpzc{q}\,  \mathpzc{p} ] -  ( q^\dagger p) - ( p^\dagger q) \ee
Using these equations, the quartic invariant \cite{Ferrara:2006yb}
\be I_4 \equiv p^0 \det[ \mathpzc{Q}\, ] - q_0 \det[ \mathpzc{P}] + \trace[( \mathpzc{Q}\times \mathpzc{Q})( \mathpzc{P}\times \mathpzc{P})] - \frac{1}{4} \scal{ q_0 p^0 + \trace [  \mathpzc{Q}\,  \mathpzc{P} ]}^2  \ee
decomposes as 
\begin{multline}  I_4 = p^0 \scal{ \det[ \mathpzc{q}] + ( q^\dagger \mathpzc{q} q) } - q_0 \scal{ \det[ \mathpzc{p}] + ( p^\dagger \mathpzc{p} \, p)} + \trace[ ( \mathpzc{q}\times \mathpzc{q})( \mathpzc{p}\times \mathpzc{p})] \\ + ( q^\dagger ( \mathpzc{p}\times \mathpzc{p}) q) +  ( p^\dagger ( \mathpzc{q}\times \mathpzc{q}) p) + ( q^\dagger p)(p^\dagger q) - ( p^\dagger \mathpzc{p} \, \mathpzc{q}  q) - (q^\dagger \mathpzc{q} \, \mathpzc{p}\,  p ) \\  - \frac{1}{4} \scal{ q_0 p^0 + \trace [ \mathpzc{q} \, \mathpzc{p}] - ( q^\dagger p) - (p^\dagger q) }^2 \end{multline}  

\vskip 2mm

Let us mention that all these algebraic relations can be written in terms of  $SU(2)\times SU^*(6)$ tensors ($SU^*(6) \cong SL(3,\mathds{H})$). The charge $\mathpzc{q}$ can be written in term of an antisymmetric tensor of $SU^*(6)$ $Q_{ab}$, and $q$ as an $SU(2)\times SU^*(6)$ tensor $Q^a_\alpha$, while $\mathpzc{P}$ is defined in terms of tensors transforming in the conjugate representations $P^{ab}$ and $P_a^\alpha$. In this way one has 
\bea \trace[ \mathpzc{Q}\, \mathpzc{P}] &=& \frac{1}{2} Q_{ab} P^{ab} - \frac{1}{2} Q_\alpha^a P_a^\alpha  \CR
( \mathpzc{Q}\times \mathpzc{Q}\, )^{ab} &=& \frac{1}{8} \varepsilon^{abcdef} Q_{cd} Q_{ef} + \frac{1}{2} \varepsilon^{\alpha\beta} Q^a_\alpha Q^b_\beta \CR
(\mathpzc{Q}\times \mathpzc{Q}\, )_a^\alpha &=& - \varepsilon^{\alpha\beta} Q_{ab} Q^b_\beta \CR
\det[ \mathpzc{Q}\, ] &=& \frac{1}{48} \varepsilon^{abcdef} Q_{ab} Q_{cd} Q_{ef} + \frac{1}{4} \varepsilon^{\alpha\beta} Q_{ab} Q^a_\alpha Q^b_\beta \label{tensors}
\eea
and similarly for $P^{ab}$ and $P_a^\alpha$. 

\subsubsection*{The central charges}

The 28 complex central charges $Z_{ij}(q,p)$ decompose accordingly into two real components  
\bea \mathpzc{Z}_0 &=& \frac{1}{\sqrt{\det[\upsilon^\dagger \upsilon]}} \Scal{ q_0 + \trace[  \mathpzc{a}\,  \mathpzc{Q}] + \trace[ \mathpzc{a}\times \mathpzc{a}\,  \mathpzc{P} ] - \det[\mathpzc{a}] p^0 } \CR
\mathpzc{Z}^0 &=& \sqrt{\det[\upsilon^\dagger \upsilon]} \, p^0 
\eea
and two Jordan algebra elements 
\bea \mathpzc{Z}_{\bf 27} &=&   \rho(\varsigma)  \frac{1}{\sqrt{\det[\upsilon^\dagger \upsilon]}} : \upsilon \bigl[ \mathpzc{Q} + 2 \mathpzc{a} \times \mathpzc{P} - \mathpzc{a} \times \mathpzc{a}\, p^0 \bigr] \upsilon^\dagger : \CR
 \mathpzc{Z}_{\overline{\bf 27}} &=&    \rho(\varsigma) \sqrt{\det[\upsilon^\dagger \upsilon]}\, : \upsilon^{\dagger \, -1} \bigl[ \mathpzc{P} - \mathpzc{a} \, p^0 \bigr] \upsilon^{-1} : 
\eea 
where $\upsilon$ is formally defined as
\be  \upsilon = \left(\begin{array}{ccc} 
\hspace{0.0mm} e^{-\phi_1}   \hspace{0.0mm}& \hspace{0.0mm} 0  \hspace{0.0mm}  & \hspace{0.0mm} 0  \hspace{0.0mm} \\
\hspace{0.0mm} 0  \hspace{0.0mm}& \hspace{0.0mm} e^{-\phi_2}  \hspace{0.0mm}  & \hspace{0.0mm} 0  \hspace{0.0mm} \\
 \hspace{0.0mm}  0     \hspace{0.0mm}& \hspace{0.0mm} 0 \hspace{0.0mm}  & \hspace{0.0mm} e^{-\phi_3}   \hspace{0.0mm} 
\end{array} \right) \exp \left(\begin{array}{ccc} 
\hspace{0.0mm} 0  \hspace{0.0mm}& \hspace{0.0mm}- \xi   \hspace{0.0mm}  & \hspace{0.0mm}- \zeta^*     \hspace{0.0mm} \\
\hspace{0.0mm} 0   \hspace{0.0mm}& \hspace{0.0mm} 0\hspace{0.0mm}  & \hspace{0.0mm}  -\eta  \hspace{0.0mm} \\
 \hspace{0.0mm} 0  \hspace{0.0mm} & \hspace{0.0mm} 0  \hspace{0.0mm}  & \hspace{0.0mm} 0   \hspace{0.0mm} 
\end{array} \right) \ee
with the product appearing in the exponential understood to be ordered such that each element acts on the charges in a first place, according to (\ref{NormalProd}). As for $\N=2$, it will be more convenient to define the central charges in a complex basis. These combinations will break the symmetry group to the one of the maximal $\N=2$ truncation, \ie to $SU(2) \times_{\mathds{Z}_2} Spin^*(12)\subset E_{7(7)}$ and $S(U(2) \times U(6)) \subset SU(8)$. $ \mathpzc{Z}_0$ and $\mathpzc{Z}^0 $ are only invariant with respect to the subgroup $Sp(4)\subset SU(8)$, and the traces of $ \mathpzc{Z}_{\bf 27}$ and $ \mathpzc{Z}_{\overline{\bf 27}}$
\bea \trace \mathpzc{Z}_{\bf 27} & =&   \frac{1}{\sqrt{\det[\upsilon^\dagger \upsilon]}} \trace \bigl[  ( \upsilon^\dagger \upsilon)  \scal{ \mathpzc{Q} + 2 \mathpzc{a} \times \mathpzc{P} - \mathpzc{a} \times \mathpzc{a}\, p^0} \bigr]  \CR
\trace  \mathpzc{Z}_{\overline{\bf 27}} &=&  \frac{1}{\sqrt{\det[\upsilon^\dagger \upsilon]}} \trace \bigl[  (\upsilon^\dagger \upsilon) \times (\upsilon^\dagger \upsilon)  \scal{ \mathpzc{P} - \mathpzc{a} \, p^0 }\bigr] 
\eea 
are only invariant with respect to the subgroup $Sp(1) \times Sp(3) \subset S(U(2) \times U(6))$. But the `central charge' 
\bea Z(p,q) &=& \frac{1}{4} \scal{   \mathpzc{Z}_0 + i \trace \mathpzc{Z}_{\bf 27} - \trace  \mathpzc{Z}_{\overline{\bf 27}} + i \mathpzc{Z}^0 } \CR
&=& \frac{1}{\sqrt{ 2i \det[ \mathpzc{t} - \bar{\mathpzc{t}}]}} \Scal{ q_0 + \trace \bigl[ \mathpzc{t} \mathpzc{Q}\bigr] + \trace \bigl[  \mathpzc{t} \times \mathpzc{t} \, \mathpzc{P} \Bigr] - \det[ \mathpzc{t}] p^0 } \label{CentralCharge} \eea
is itself invariant with respect to $S(U(2) \times U(6))$. Of course, if one considers the maximal $\N=2$ truncation of the theory by fixing all split octonions to be quaternions, then this `central charge' defines the $\N=2$ central charge. The unconventional factor of $\sqrt{2}$ is due to the fact that we will not consider integral electromagnetic charges, but rather charges valued in $\sqrt{2} \mathds{Z}$, in order to simplify expressions. One can similarly recombine the other components of the central charges $Z_{ij}$ into the Jordan algebra element 
\bea \mathpzc{D} Z(p,q)  &=&    - \frac{1}{2} \scal{  \mathpzc{Z}_{\overline{\bf 27}} + i   \mathpzc{Z}_{\bf 27} } + \frac{1}{4}  \scal{   \mathpzc{Z}_0 + i \trace   \mathpzc{Z}_{\bf 27}  + \trace  \mathpzc{Z}_{\overline{\bf 27}} -i \mathpzc{Z}^0)  }\mathds{1} \CR
&=&  \frac{ \rho(\varsigma)}{4 \sqrt{ \det[ \upsilon^\dagger \upsilon]}}\Bigl( q_0 \mathds{1} -2  i : \upsilon \mathpzc{Q} \upsilon^\dagger : + i \trace[ \upsilon^\dagger \upsilon  \mathpzc{Q}\, ] \mathds{1} + \trace[ \mathpzc{a} \, \mathpzc{Q}\, ] \mathds{1}   \Bigr . \CR 
 && \hspace{15mm} \Bigl .-2  \det[ \upsilon^\dagger \upsilon] : v^{\dagger -1} \mathpzc{P} v^{-1} : + \trace[ (\upsilon^\dagger \upsilon)\times (\upsilon^\dagger \upsilon) \mathpzc{P}] \mathds{1} +  \trace[ \mathpzc{a} \times \mathpzc{a}\,  \mathpzc{P}] \mathds{1} \Bigr . \CR 
 && \hspace{15mm} \Bigl . -  4 i : \upsilon ( \mathpzc{a} \times \mathpzc{P}) \upsilon^\dagger :   + 2 i \trace[ (\upsilon^\dagger \upsilon) \times \mathpzc{a} \, \mathpzc{P}]\mathds{1}  \Bigr . \CR && \hspace{10mm} - \Bigl(  \scal{  i\det[ \upsilon^\dagger \upsilon]  + \trace [ (\upsilon^\dagger \upsilon) \times (\upsilon^\dagger \upsilon)  \mathpzc{a}] + i \trace[ \upsilon^\dagger \upsilon  ( \mathpzc{a} \times \mathpzc{a})] +   \det[\mathpzc{a}]} \mathds{1}  \Bigr .   \Bigr . \CR 
 && \hspace{30mm} \Bigl . \Bigl .  -2 \det[ \upsilon^\dagger \upsilon]  : v^{\dagger -1} \mathpzc{a} v^{-1} :   -2  i : \upsilon ( \mathpzc{a} \times \mathpzc{a}) \upsilon^\dagger  :   \Bigr)  p^0 \Bigr) \eea
which reduces to the K\"{a}hler derivative of the central charge in tangent frame for quaternions. In general it includes 27 complex components, and decomposes into its quaternionic and its $\ell$ component which transform in the ${\bf 15} \oplus {\bf 2} \otimes \bar{\bf 6}$ of $S(U(2)\times U(6))$. In particular, its trace appears to be equal to 
\be \trace \mathpzc{D} Z(p,q)  = - \sqrt{ 2i \det[ \mathpzc{t} - \bar{\mathpzc{t}}]} \trace\Bigl[  ( \mathpzc{t} - \bar{\mathpzc{t}} ) \frac{ \partial \, }{\partial \mathpzc{t} }\Bigr] \frac{ Z}{\sqrt{ 2i \det[ \mathpzc{t} - \bar{\mathpzc{t}}]} } \ee
It breaks $S(U(2)\times U(6))$ to its $Sp(1)\times Sp(3)$ subgroup, and it will appear explicitly in the definition of the ADM mass of the solutions we will describe in this paper.

\subsection{Extraction of the solutions}
A solution is generated from an Ansatz in the symmetric gauge. For a given nilpotent orbit one choses a specific representative ${\bf h}$  of the semi-simple orbit which characterises the nilpotent element ${\bf e}$ through \cite{Collingwood}
\be [ {\bf h} ,{\bf e} ] = 2 {\bf e} \ee
For even orbits, this element also characterises the corresponding nilpotent subalgebra as the subalgebra of strictly positive grade, generated by the elements  
 \be [ {\bf h}  , {\bf e}^\ord{p} ] = 2 p \, {\bf e}^\ord{p}  \ee
where $0< p \le n$ for a finite $n$. We then decompose a general vector $L$ in this nilpotent subalgebra as    
\be {\bf L} = \sum_{p=1}^n {\bf L}^\ord{p} \ee
such that \be [ {\bf h}  , {\bf L} ]  = 2 \sum_{p=1}^n  p \, {\bf L}^\ord{p}\ee A solution is obtained from such vector ${\bf L}$ defined as a function over the punctured $\mathds{R}^3$, which punctures correspond to the black hole horizons. The functions defining ${\bf L}$ satisfy a set of differential equations as explained in details in \cite{BossardRuef}. 

In order to render this section slightly more explicit, let us discuss the particular case of the non-BPS composite system in the $E_{6(6)}$ truncation. In that case, the relevant generator ${\bf h}$ (which can be expressed in a Cartan basis as \DSpIV0020 \cite{E6Djo}) decomposes the $\sp(8,\mathds{R})$ algebra into 
\be \sp(8,\mathds{R}) \cong {\bf 6}^\ord{-2} \oplus ( {\bf 2} \otimes {\bf 3})^\ord{-1} \oplus \scal{ \gl_1 \oplus \sl_2 \oplus \sl_3 }^\ord{0} \oplus ( {\bf 2} \otimes \overline{\bf 3})^\ord{1} \oplus \overline{\bf 6}^\ord{2} \ee
where the nilpotent generators satisfy the algebra 
\be [ {\bf E}_a^\alpha , {\bf E}_b^\beta ] = \varepsilon^{\alpha\beta} {\bf E}_{ab} \ee
The ${\bf 42}$ representation of $Sp(8,\mathds{R})$ decomposes accordingly into 
\be {\bf 42} \cong {\bf 2}^\ord{-3}  \oplus  \overline{ \bf 3}^\ord{-2} \oplus ( {\bf 2} \otimes \overline{\bf 6})^\ord{-1} \oplus {\bf 8}^\ord{0} \oplus ( {\bf 2} \otimes{\bf 6})^\ord{1} \oplus {\bf 3}^\ord{2} \oplus {\bf 2}^\ord{3} \ee
where the nilpotent generators satisfy
\be [ {\bf e}_\alpha^{ab} , {\bf e}_\beta^{cd} ] = \frac{1}{2}  \varepsilon_{\alpha\beta}   \varepsilon^{ace} \varepsilon^{bdf}   {\bf E}_{ef} \; , \qquad [ {\bf E}_{ab} , {\bf e}_\alpha^{cd} ] =\frac{1}{2} (  \delta_{a}^{c} \delta_b^d + \delta_a^d \delta_b^c )  {\bf e}_\alpha \ee
and the grade 2 generators ${\bf e}^a$ commute with the others, but the generators ${\bf E}_a^\alpha$. The complete $\e_{6(6)}$ algebra decomposes as
\begin{multline}  \e_{6(6)} \cong\\ {\bf 2}^\ord{-3} \oplus  ({\bf 3} \otimes {\bf 3})^\ord{-2} \oplus ( {\bf 2} \otimes \overline{\bf 3}\otimes \overline{\bf 3} )^\ord{-1} \oplus \scal{ \gl_1 \oplus \sl_2 \oplus \sl_3\oplus \sl_3  }^\ord{0}  \oplus ( {\bf 2} \otimes{\bf 3} \otimes {\bf 3})^\ord{1} \oplus (\overline{\bf 3}\otimes \overline{\bf 3})^\ord{2} \oplus {\bf 2}^\ord{3} \end{multline}
and the fundamental representation as
\be {\bf 27} \cong (\overline{\bf 3}\otimes{\bf 1})^\ord{-2} \oplus ({\bf 2} \otimes {\bf 1} \otimes {\bf 3})^\ord{-1} \oplus ( {\bf 3} \otimes \overline{\bf 3})^\ord{0} \oplus ( {\bf 2} \otimes \overline{\bf 3} \otimes {\bf 1})^\ord{1} \oplus ( {\bf 1} \otimes {\bf 3})^\ord{2} \ee
It follows from the grading that the generators ${\bf e}_\alpha^{ab}$ are nilpotent of order 5 (\ie $(X_{ab}^\alpha {\bf e}_\alpha^{ab})^5 = 0 \; \forall X_{ab}^\alpha$), the generators ${\bf e}^a$ nilpotent  of order 3, and the generators ${\bf e}_\alpha$ nilpotent  of order 2. Therefore ${\bf L}^5=0$.

It will turn out to be useful to break the $SL(2)$ symmetry rather drastically by taking an Ansatz of the form (breaking also the apparent $SL(3)$ to $SO(3)$)
\begin{multline}  {\bf L} = ( \delta_{ab} + L_{ab} ) {\bf e}^{ab}_1 + K_{ab} {\bf e}^{ab}_2 + Y_a {\bf e}^a \\ + \Scal{ M  - \frac{1}{6} \delta^{ab} K_{ab} +  \frac{1}{6} \varepsilon^{ace} \varepsilon^{bdf} ( \delta_{ab} L_{cd} K_{ef} - L_{ab} L_{cd} K_{ef} ) } {\bf e}_1\\  - \frac{1}{2}  \Scal{ 1 + V - \frac{1}{6}   \varepsilon^{ace} \varepsilon^{bdf} ( 2 \delta_{ab} K_{cd} K_{ef} - L_{ab} K_{cd} K_{ef}) } {\bf e}_2 \label{Specific} \end{multline} 
The algebra is such that the system of equations can be written for ${\bf L}$ as
\be d \star d \, {\bf L} = - \frac{2}{3} [ d {\bf L} , [ {\bf L} , \star d {\bf L}]] \ee
and some algebra permits to obtain that $L_{ab}, K_{ab}$ and $Y_a$ are all harmonic functions, whereas $V$ and $M$ satisfy 
\be d \star d V = -  \varepsilon^{ace} \varepsilon^{bdf}  L_{ab} d K_{cd} \star d K_{ef} \; , \qquad d\star d M =  \varepsilon^{ace} \varepsilon^{bdf}  L_{ab} d L_{cd} \star d K_{ef} \ee

\vskip 2mm

The Ansatz for the solution is then defined for the $E_{6(6)} / \Sp(8,\mathds{R})$ representative as
\bea \label{VVdagmat}
 \V  \upeta \V^t \upeta  &=&  \left(\begin{array}{ccc} 
\hspace{2mm} \times  \hspace{2mm}& \hspace{2mm} e^{-2U} M^{AB} \sigma + \frac{1}{2} e^{-2U} M^{CB}{}  \zeta^{ADE} \zeta_{CDE}     \hspace{2mm}  & \hspace{2mm} \times \hspace{2mm}  \\
\hspace{2mm} \times   \hspace{2mm}& \hspace{2mm}e^{-2U} M^{AB}  \hspace{2mm}  & \hspace{2mm} \times   \hspace{2mm}  \\
\hspace{2mm}  \times     \hspace{2mm}& \hspace{2mm}  e^{-2U} M^{AD}  \zeta_{DBC} \hspace{2mm}  & \hspace{2mm} \times  \hspace{2mm} \end{array} \right) \CR
&=&  \exp(-2{\bf L}) = \mathds{1} - 2  {\bf L} + 2 {\bf L}^2 - \frac{4}{3} {\bf L}^3 + \frac{2}{3} {\bf L}^4
\eea
where $M^{AB} = (v^{t} v)^{AB} $. We first focus on the matrix element $e^{-2U} M^{AB}$. Computing the determinant of this matrix gives the scaling factor 
\be e^{-12U} = \det[ e^{-2U} M ] \ee
One computes in a specific frame that it is given by the cube of 
\be e^{-4U} = V \frac{1}{6}  \varepsilon^{ace} \varepsilon^{bdf}  L_{ab} L_{cd} L_{ef} - \frac{1}{2}  \varepsilon^{ace} \varepsilon^{bdf} L_{ab} L_{cd} Y_e Y_f \ee
for the non-BPS solution. It is then straightforward to write an algorithm which extracts from $M^{AB}$ the fields defined in (\ref{vtExpression}). We obtain in this way their expression in the case of the non-BPS solution 
\be t_{ab} = K_{ab} +\frac{ (  - M + i e^{-2U} )  L_{ab}}{\frac{1}{6}  \varepsilon^{gce} \varepsilon^{hdf}  L_{gh} L_{cd} L_{ef} } + \ell \varepsilon_{abc} \frac{ \frac{1}{2} \varepsilon^{ceg} \varepsilon^{dfh} L_{ef} L_{gh}  }{\frac{1}{6} \varepsilon^{ikm} \varepsilon^{jln}  L_{ij} L_{kl} L_{mn} } Y_d \ee 
where the notation is justified by the property  that the real part of $\mathpzc{t}$ is a symmetric matrix, whereas its imaginary part is antisymmetric. The two scalars $\varsigma_\pm$ are simply zero everywhere. Note that the specific Ansatz (\ref{Specific}) has been chosen in order to get such simple expressions. In general it is necessary to implement non-linear reparametrizations preserving somehow the grading in order to get a solution which is manifestly symmetric with respect to the largest available symmetry group. 

Once the components of $v^t$ have been extracted, the matrix $M$ is trivial to invert, and one can extract the electromagnetic scalars from $ e^{-2U} M^{AD}  \zeta_{DBC}$.

It remains then to extract the three-dimensional vector fields. This can be done straightforwardly by computing the Noether current  
\be 
 d W \equiv \star \V P \V^{-1} = - \sum_{k =0}^{n-1} \frac{(-2)^k}{(k+1)!} \ad_{\bf L}^{\; \; k}  \star d {\bf L} \quad , 
\ee 
In the case of the composite non-BPS solutions 
\bea \star d W &=& - d {\bf L} + [ {\bf L} , d {\bf L} ] - \frac{2}{3} [ {\bf L }, [{\bf L}, d {\bf L}]]\CR
&=& - d L_{ab} {\bf e}_1^{ab} - d K_{ab} \scal{ {\bf e}^{ab}_2 + \delta^{ac} \delta^{bd} {\bf e}_{cd}  - \delta^{ab} (  \delta^{cd} {\bf e}_{cd} - \frac{1}{2} {\bf e}_1 )} - d Y_a {\bf e}^a \CR 
&& \quad  + \frac{1}{2}  \varepsilon^{ace} \varepsilon^{bdf}  \scal{ L_{ab} d K_{cd} - K_{ab} d L_{cd} } \scal{ {\bf e}_{ef} + \delta_{ef} {\bf e}_1} \CR
&& \qquad + \frac{1}{2} \Scal{ d V + \frac{1}{2}  \varepsilon^{ace} \varepsilon^{bdf}  \scal{ L_{ab} d (K_{cd} K_{ef})  - K_{ab} K_{cd} d L_{ef} }}  {\bf e}_2 \CR
&& \quad \qquad - \Scal{ d M - \frac{1}{2}  \varepsilon^{ace} \varepsilon^{bdf} L_{ab} L_{cd} d K_{ef}} {\bf e}_1 \eea
One obtains 
\be \label{eqomega}
 d \omega = \trace {\bf E} \, dW  = \star  e^{-4U} \biggl( d \sigma - \frac{1}{2} \zeta^{ABC} d \zeta_{ABC} \biggr)  \; , 
\ee
which in this case gives
\be \star d \omega = d M - \frac{1}{2}  \varepsilon^{ace} \varepsilon^{bdf} L_{ab} L_{cd} d K_{ef} \ee
and 
\bea \label{eqemvectors}
 d w_{ABC} \!&=& - \frac{1}{4} \trace {\bf E}_{ABC} d W  \CR
&=& e^{-2U} \! M^{-1}_{AD} M^{-1}_{BE} M^{-1}_{CF} \star  d \zeta^{DEF} \! - \! e^{-4U} \! \star \! \biggl( d \sigma \! - \! \frac{1}{2} \zeta^{DEF} d \zeta_{DEF} \biggr)  \zeta_{ABC} \CR
&=& e^{-2U}  M^{-1}_{AD} M^{-1}_{BE} M^{-1}_{CF} \star  d \zeta^{DEF}  - \zeta_{ABC} d \omega \; , 
\eea
which gives 
\bea \star d w_0 &=& d V + \frac{1}{2}  \varepsilon^{ace} \varepsilon^{bdf}  \scal{ L_{ab} d (K_{cd} K_{ef})  - K_{ab} K_{cd} d L_{ef} }\CR
 \star d w^{ab} &=&\star d \omega \, \delta^{ab} +  \varepsilon^{ace} \varepsilon^{bdf}  \scal{ L_{ab} d K_{cd} - K_{ab} d L_{cd} } + \ell \varepsilon^{abc} d Y_c \CR
 \star d v_{ab} &=& d L_{ab} \CR
 \star d v^0 &=& - \star d \omega \eea
In order to restore the manifest covariance, we moreover carry out gauge transformations linear in time in order to remove the $\star d\omega$ components as well as the corresponding non-covariant constant terms in 
\be \zeta^0 = 1- e^{4U} \frac{1}{6}  \varepsilon^{ace} \varepsilon^{bdf} L_{ab} L_{cd} L_{ef} \ee
such that 
\bea  \zeta^0 ( dt + \omega) + v^0 - dt &=& \scal{  1- e^{4U} \frac{1}{6}  \varepsilon^{ace} \varepsilon^{bdf} L_{ab} L_{cd} L_{ef} } ( d t + \omega ) - ( dt +  \omega ) \CR 
&=& - e^{4U} \frac{1}{6}  \varepsilon^{ace} \varepsilon^{bdf} L_{ab} L_{cd} L_{ef} ( d t + \omega )  \eea
and similarly for  $\zeta^{ab} (dt + \omega ) + w^{ab} + d ( \delta^{ab} t ) $.   

\vskip 2mm

Once the fields of the truncated theory have all been extracted, one can simply reorganise them in the Jordan algebra scalar field $\mathpzc{t}$ and equivalently for the vector fields. It appears that all the nilpotent orbits of $\e_{6(6)}$ admit at least two grade zero components  associated to the two extra generators of the Cartan subalgebra of $\e_{6(6)}$ with respect to the Cartan subalgebra of $\sp(8,\mathds{R})$. This suggests that one could chose the coordinates such that the two scalars $\varsigma_\pm = 0 $ in the Ansatz. Equivalently for $\N=8$ supergravity, the maximal nilpotent orbit of $E_{8(8)}$ that admits a non-trivial intersection with the coset component has a grade zero component which is defined by the non-compact generators of the algebra $\so(4,4) \subset \e_{8(8)}$. If the two scalars $\varsigma_\pm = 0 $ in the truncation, it seems reasonable to believe that the 16 scalars $\varsigma^I$ that parametrize the quaternionic space $SO(4,4)/(SO(4)\times SO(4))$ can also all be set to zero. However the situation is not always so simple, and preliminary computations show that the scalar fields $\varsigma_\pm $ get non-trivial values in the maximal solvable systems. It seems nevertheless that the scalar fields parametrizing the submanifold $G_{2(2)} / SO(4) $ are still constant, such that one could obtain the $\N=8$ scalar fields in $SO(4,4)/(SO(4)\times SO(4))$ by promoting the two non-trivial scalars $e^{\ell \varsigma_\pm}$ to uni-modular split octonions. Anyway, the systems we will study in this paper have more symmetries, and in these cases the two scalars  $\varsigma_\pm$ are indeed null. Moreover, one can then check that the scalar fields $\varsigma^I$ are not sourced by the other fields, through a cancellation of the contributions of the other scalar fields and the electromagnetic fields. We will call the scalar fields that are constant in the solutions the `flat directions', referring to the property that they decouple completely from the equations and are simply set to constant. Note however that they do not define the common stabiliser subgroup of all the electromagnetic charges of the black hole constituents. 

In all the systems there is a linearly realised  $[SU(2)]^4$ symmetry. This symmetry permits to determine the ordering ambiguities appearing in the generalisation of an expression written in terms of real functions to the expression written in terms of quaternions. Each quaternion transforms with respect to one specific $SU(2)$ on its left, and one specific $SU(2)$ on its right. The only problem that can appear, is when quantities like
\be x y^* - y x^* \ee
appear explicitly in the solution (where $x$ and $y$ transforms identically with respect to $[SU(2)]^4$), since such expression vanishes identically for reals. In fact one might have to deal with such ambiguities for the most general systems, and it may then become necessary to  compute the solutions within the $E_{7(7)} / SU_{\scriptscriptstyle \rm c}(4,4) $ non-linear sigma model. However, we will mainly focus in this paper on solutions admitting a much larger linearly realised symmetry $SU(2) \times SL(3,\mathds{H})$, such that quaternions all combine into 3 by 3 Hermitian matrices, or 3-vectors over the quaternions. In this case we will see that there is no such ambiguity problem.

\section{Composite non-BPS system}
The nilpotent orbit in which is parametrized the composite non-BPS solution of the STU model \cite{BossardRuef} can be embedded in a larger $\e_{8(8)}$ nilpotent orbit. The nilpotent algebra admits as an automorphism $SL(2) \times SU(2)\times SL(3,\mathds{H})$. The parametrization of the solution breaks $SL(2)$, which is related to Ehlers symmetry. But the automorphism symmetry of the Jordan algebra is preserved because the solution can be explicitly written in terms of scalars, Jordan algebra elements, and 3 vectors over the quaternions. The nilpotent algebra itself is entirely determined by the $\gl_1$ weights and the Jordan algebra cross product, such that the satisfaction of the equations of motion in the $E_{6(6)}$ truncated theory ensures that they are satisfied in $\N=8$ supergravity. The structure is as such extremely constrained, and one can check at each step that the $SL(3,\mathds{R})$ invariant terms generalise in a unique way to the $SL(3,\mathds{H})$ invariant expressions. 

\subsection{The solution Ansatz}
The solution is defined in terms of harmonic functions organised in two 3 by 3 Hermitian matrices over the quaternions, $\mathcal{K}$ and $\mathcal{L}$, and a 3 vector over the quaternions $Y$. There are moreover two sourced scalar functions $V$ and $M$ which solve the equations 
\bea d \star \scal{ d V + \trace \mathcal{L}  \, d ( \mathcal{K} \times \mathcal{K} ) - \trace  \mathcal{K} \times \mathcal{K} \, d \mathcal{L} } &=& 0 \CR
d \star \scal { dM - \trace \L \times \L \, d \K } &=& 0  \eea 
The system is invariant with respect to the $SL(3,\mathds{H})$ symmetry group of the cross product. 

The four-dimensional metric is determined by its scaling factor
\be e^{-4U} = V \det[\L] - \scal{ Y^\dagger \L\times \L \, Y} - M^2 \ee
and the Kaluza--Klein vector 
\be \star d \omega = d M - \trace \L \times \L \, d \K \ee
The scaling factor can be seen to be invariant with respect to $SL(4,\mathds{H})$, by rewriting its two first terms as the determinant of a 4 by 4 Hermitian matrix over the quaternions. This is a consequence of the property that the system reduces for $\K=0$ to the system describing single centre non-BPS black holes which is invariant with respect to this symmetry. 

\vskip 2mm

For the sake of clarity, let us recall the form of these equations in terms of $SU(2) \times SU^*(6)$ tensors, as discussed in equations (\ref{tensors}),
\bea d \star \Scal{ d V + \frac{1}{16} \varepsilon^{abcdef} \scal{  L_{ab} d ( K_{cd} K_{ef}) - K_{cd} K_{ef} d L_{ab}}} &=& 0 \CR
d\star \Scal{ d M - \frac{1}{16}  \varepsilon^{abcdef}  L_{ab} L_{cd} d K_{ef} } &=& 0 \eea
and
\bea e^{-4U} &=& V\frac{1}{48} \varepsilon^{abcdef} L_{ab} L_{cd} L_{ef}  - \frac{1}{16} \varepsilon_{\alpha\beta} \varepsilon^{abcdef} Y_a^\alpha L_{bc} L_{de} Y_f^\beta - M^2 \CR
\star d \omega &=& d M -\frac{1}{16} \varepsilon^{abcdef} L_{ab} L_{cd} d K_{ef} \eea
Nevertheless, we shall only use the Jordan algebra notations in the following. 

\vskip 2mm

The expression of the scalar $\mathpzc{t}$ is 
 \be \mathpzc{t} = \K + \frac{\L}{\det[\L]} \scal{ - M + i e^{-2U} } + \ell  \invo \frac{ \L \times \L}{\det[\L]} \, Y \ee
 and the 16 scalars $\varsigma^I$ are constant. Here one sees that setting $Y$ to zero, the scalar fields are Hermitian matrices over the quaternions and the solution sits in the $\N=2$ truncation of $\N=8$ supergravity. The product with $Y$ is defined as for an ordinary 3 by 3 matrix
\be  \frac{ \L \times \L}{\det[\L]} =  \L^{-1} \ee
The three dimensional vector components of the vector fields are 
\bea \star d w_0 &=&  d V - \trace \mathcal{L}  \, d ( \mathcal{K} \times \mathcal{K} ) +\trace  \mathcal{K} \times \mathcal{K} \, d \mathcal{L} \CR
\star d \mathpzc{w} &=& 2 \L \times d \K -2  \K \times d \L + \ell \invo d Y \CR
\star d \mathpzc{v} &=& d \L \CR
\star d v^0 &=& 0 \eea
and they combine with the scalar components to give the 28 vectors 
\bea A^0 &=& - e^{4U} \det[ \mathcal{L}] ( dt + \omega) + v^0  \CR
\mathcal{A} &=& - e^{4U} \scal{ \det[\mathcal{L}] \mathcal{K} - M \mathcal{L} + \ell \invo ( \mathcal{L}\times \mathcal{L}) Y }( dt + \omega) + \mathpzc{v} \eea
and their 28 duals 
\bea
\tilde{A}_0 &=& e^{4U} \bigl(  MV + M \trace[( \mathcal{K} \times \mathcal{K}) \mathcal{L}] - V \trace[( \mathcal{L} \times \mathcal{L}) \mathcal{K}]
\bigr . \CR&& \hspace{40mm} \bigl . - \det[ \mathcal{L}] \det[\mathcal{K}] + 2 ( Y^\dagger ( \mathcal{K} \times \mathcal{L}) Y)\bigr) ( dt + \omega ) + w_0  \\
\tilde{\mathcal{A}} &=& e^{4U} \bigl( V \mathcal{L} \times \mathcal{L} - 2 M \mathcal{K} \times \mathcal{L} + \det[\mathcal{L}] \mathcal{K} \times \mathcal{K} \bigr . \CR && \hspace{40mm} \bigl .  - 2 \mathcal{L} \times YY^\dagger+ \ell \invo ( - \mathcal{K} ( \mathcal{L} \times \mathcal{L} ) + M ) Y \bigr) ( dt + \omega) + \mathpzc{w} \nn
\eea
They can be rewritten conveniently in terms of the scalar fields and $M$ as
\bea A^0 &=& - 4 \frac{e^{U} }{\sqrt{ 2 i\det[ \mathpzc{t}-\overline{\mathpzc{t}}]}} ( dt + \omega)  \CR
\mathcal{A} &=& -2 \frac{e^{U} }{\sqrt{ 2 i\det[ \mathpzc{t}-\overline{\mathpzc{t}}]}} (  \mathpzc{t} + \overline{ \mathpzc{t}}) ( dt + \omega) + \mathpzc{v} \CR
\tilde{\mathcal{A}} &=& 4 \frac{e^{U} }{\sqrt{ 2 i\det[ \mathpzc{t}-\overline{\mathpzc{t}}]}} ( \mathpzc{t} \times \overline{\mathpzc{t} } )( dt + \omega) + \mathpzc{w} \CR
\tilde{A}_0 &=&  \Bigl( \frac{e^{U} }{\sqrt{ 2 i\det[ \mathpzc{t}-\overline{\mathpzc{t}}]}} \scal{ \det[ \mathpzc{t}]  +  \det[ \overline{\mathpzc{t}}] -  \trace [ \mathpzc{t} \times \mathpzc{t} \overline{\mathpzc{t}}] - \trace[ \overline{\mathpzc{t}} \times \overline{\mathpzc{t}} \mathpzc{t}]  }  \Bigr . \CR
&& \hspace{40mm}   \Bigl . - \frac{e^{3U}}{2} \sqrt{ 2i \det[\mathpzc{t}-\overline{\mathpzc{t}}] }\, M \Bigr)  ( dt + \omega) + w_0  \label{VecScal} 
\eea

\subsection{Solving the differential equations} 

Let us now solve this system. We define the harmonic functions 
\be \L = \mathpzc{l} + \sum_A \frac{\mathpzc{p}_A}{|x-x_{\scriptscriptstyle A}|} \; , \qquad  \K = \mathpzc{k} + \sum_A \frac{\gamma_A \mathpzc{p}_A}{|x-x_{\scriptscriptstyle A}|} \; , \qquad Y = y + \sum_A \frac{q_A}{|x-x_{\scriptscriptstyle A}|} \ee
Note that we already fixed each pole of $\K$ to be proportional to the corresponding pole of $\L$, because this is required for the Noether charge associated to each centre to be nilpotent of order 3, such that it can correspond to a regular black hole solution \cite{BossardRuef}. Although, the solutions for $V$ and $M$ can be found in principle, they do not admit closed form expressions in general. Therefore we will first compute these functions locally near each pole $ x_{\scriptscriptstyle A}$, and in the asymptotic region. We will discuss the global solution when we will compute the angular momentum.

Near the pole $x_{\scriptscriptstyle A}$, $\L$ admits the expansion 
\begin{multline} \L = \frac{\mathpzc{p}_A}{|x-x_{\scriptscriptstyle A}|} + \Scal{  \mathpzc{l} +\sum_{B\ne A} \frac{\mathpzc{p}_B}{|x_{\scriptscriptstyle B}-x_{\scriptscriptstyle A}|} } + \Scal{\sum_{B\ne A} \frac{\mathpzc{p}_B}{|x_{\scriptscriptstyle B}-x_{\scriptscriptstyle A}|^3} (x_{\scriptscriptstyle B}-x_{\scriptscriptstyle A})_i}  ( x -x_{\scriptscriptstyle A})^i \\ + \frac{1}{2}  \Scal{\sum_{B\ne A} \frac{\mathpzc{p}_B}{|x_{\scriptscriptstyle B}-x_{\scriptscriptstyle A}|^5} \scal{ 3 (x_{\scriptscriptstyle B}-x_{\scriptscriptstyle A})_i (x_{\scriptscriptstyle B}-x_{\scriptscriptstyle A})_j - \delta_{ij} |x_{\scriptscriptstyle B}-x_{\scriptscriptstyle A}|^2 } }  ( x -x_{\scriptscriptstyle A})^i  ( x -x_{\scriptscriptstyle A})^j \\ + \mathcal{O}\scal{(x-x_{\scriptscriptstyle A})^3} \end{multline}
where all higher order corrections are symmetric traceless polynomials in $ x -x_{\scriptscriptstyle A}$. $\K$ admits the similar expansion 
\begin{multline} \K = \gamma_A \frac{\mathpzc{p}_A}{|x-x_{\scriptscriptstyle A}|} + \Scal{  \mathpzc{k} +\sum_{B\ne A} \gamma_B \frac{\mathpzc{p}_B}{|x_{\scriptscriptstyle B}-x_{\scriptscriptstyle A}|} } + \Scal{\sum_{B\ne A} \gamma_B \frac{\mathpzc{p}_B}{|x_{\scriptscriptstyle B}-x_{\scriptscriptstyle A}|^3} (x_{\scriptscriptstyle B}-x_{\scriptscriptstyle A})_i}  ( x -x_{\scriptscriptstyle A})^i \\ + \frac{1}{2}  \Scal{\sum_{B\ne A} \gamma_B \frac{\mathpzc{p}_B}{|x_{\scriptscriptstyle B}-x_{\scriptscriptstyle A}|^5} \scal{ 3 (x_{\scriptscriptstyle B}-x_{\scriptscriptstyle A})_i (x_{\scriptscriptstyle B}-x_{\scriptscriptstyle A})_j - \delta_{ij} |x_{\scriptscriptstyle B}-x_{\scriptscriptstyle A}|^2 } }  ( x -x_{\scriptscriptstyle A})^i  ( x -x_{\scriptscriptstyle A})^j \\ + \mathcal{O}\scal{(x-x_{\scriptscriptstyle A})^3} \end{multline}
Using this expression, we find the particular solution for $M$ near $x_\pA$
\begin{multline} M^\ord{1} = \gamma_A \frac{\det[\mathpzc{p}_A]}{| x -x_{\scriptscriptstyle A}|^3} + \gamma_A \trace \Bigl[ \mathpzc{p}_A \times \mathpzc{p}_A \Scal{ 
 \mathpzc{l}  +\sum_{B\ne A} \frac{\mathpzc{p}_B}{|x_{\scriptscriptstyle B}-x_{\scriptscriptstyle A}|} } \Bigr]  \frac{1}{| x -x_{\scriptscriptstyle A}|^2} \\ + \biggl( \sum_{B\ne A} \gamma_B \trace \bigl[ \mathpzc{p}_A \times \mathpzc{p}_A \, \mathpzc{p}_B \bigr] \, \frac{ (x_{\scriptscriptstyle B}-x_{\scriptscriptstyle A})_i}{|x_{\scriptscriptstyle B}-x_{\scriptscriptstyle A}|^3}\biggr)  \frac{ (x -x_{\scriptscriptstyle A})^i}{|x -x_{\scriptscriptstyle A}|^2}  + \mathcal{O}(1) \end{multline}
 to which we can add a homogenous solution
 \be M^\ord{0} = m + \sum_A \frac{\bar p_A^0}{|x-x_{\scriptscriptstyle A}|} -  \sum_A \frac{\J_A^i ( x-x_{\scriptscriptstyle A})_i }{|x-x_{\scriptscriptstyle A}|^3} \ee 
No higher order poles are allowed because they would necessarily produce singularities in the metric. The absence of Dirac--Misner string singularities requires to constraint the simple poles of $M^\ord{0}$ such that 
\be  \bar p_A^0 =  \gamma_A \trace \biggl[ \mathpzc{p}_A   \Scal{  \mathpzc{l} +\sum_{B\ne A} \frac{\mathpzc{p}_B}{|x_{\scriptscriptstyle B}-x_{\scriptscriptstyle A}|} }  \times  \Scal{  \mathpzc{l} +\sum_{C\ne A} \frac{\mathpzc{p}_C}{|x_{\scriptscriptstyle C}-x_{\scriptscriptstyle A}|} }  \biggr]  \ee
and 
\be \omega =  \varepsilon_{ijk} \frac{ \J_A^i (x-x_{\scriptscriptstyle A})^j d x^k }{|x-x_{\scriptscriptstyle A}|^3} + \mathcal{O}(x-x_{\scriptscriptstyle A})  \ee
One finds therefore the expansion of $M$ near $x_{\scriptscriptstyle A}$
\begin{multline} M = \gamma_A \frac{\det[\mathpzc{p}_A]}{| x -x_{\scriptscriptstyle A}|^3} + \gamma_A \trace \Bigl[ \mathpzc{p}_A \times \mathpzc{p}_A \Scal{ 
 \mathpzc{l}  +\sum_{B\ne A} \frac{\mathpzc{p}_B}{|x_{\scriptscriptstyle B}-x_{\scriptscriptstyle A}|} } \Bigr]  \frac{1}{| x -x_{\scriptscriptstyle A}|^2} \\- \frac{\J_A^i ( x-x_{\scriptscriptstyle A})_i }{|x-x_{\scriptscriptstyle A}|^3}  +  \gamma_A \trace \biggl[ \mathpzc{p}_A   \Scal{  \mathpzc{l} +\sum_{B\ne A} \frac{\mathpzc{p}_B}{|x_{\scriptscriptstyle B}-x_{\scriptscriptstyle A}|} }  \times  \Scal{  \mathpzc{l} +\sum_{C\ne A} \frac{\mathpzc{p}_C}{|x_{\scriptscriptstyle C}-x_{\scriptscriptstyle A}|} }  \biggr]  \frac{1}{|x -x_{\scriptscriptstyle A}|} \\
+ \biggl( \sum_{B\ne A} \gamma_B \trace \bigl[ \mathpzc{p}_A \times \mathpzc{p}_A \, \mathpzc{p}_B \bigr] \, \frac{ (x_{\scriptscriptstyle B}-x_{\scriptscriptstyle A})_i}{|x_{\scriptscriptstyle B}-x_{\scriptscriptstyle A}|^3}\biggr)  \frac{ (x -x_{\scriptscriptstyle A})^i}{|x -x_{\scriptscriptstyle A}|^2} 
+  \mathcal{O}(1) \end{multline}
We will solve similarly $V$ by rewriting its equation 
\be d \star d \scal{ V - \trace \L\, \K \times \K } = - 2 d \, \trace \K \times \K \, \star d \L  \ee
We find in the neighbourhood of $x_\pA$ the particular solution 
\begin{multline} \hspace{-4mm} \scal{  V - \trace \L\, \K \times \K }^\ord{1} =-2 \Biggl(  \gamma_A^{\; 2}  \frac{\det[\mathpzc{p}_A]}{| x -x_{\scriptscriptstyle A}|^3} + \gamma_A \trace \Bigl[ \mathpzc{p}_A \times \mathpzc{p}_A \Scal{ 
 \mathpzc{k}  +\sum_{B\ne A} \gamma_B  \frac{\mathpzc{p}_B}{|x_{\scriptscriptstyle B}-x_{\scriptscriptstyle A}|} } \Bigr]  \frac{1}{| x -x_{\scriptscriptstyle A}|^2} \\ + \gamma_A^{\; 2}  \biggl( \sum_{B\ne A} \trace \bigl[ \mathpzc{p}_A \times \mathpzc{p}_A \, \mathpzc{p}_B \bigr] \, \frac{ (x_{\scriptscriptstyle B}-x_{\scriptscriptstyle A})_i}{|x_{\scriptscriptstyle B}-x_{\scriptscriptstyle A}|^3}\biggr)  \frac{ (x -x_{\scriptscriptstyle A})^i}{|x -x_{\scriptscriptstyle A}|^2}  + \mathcal{O}(1) \Biggr) \end{multline}
to which we can add the homogenous solution 
\be  \scal{  V - \trace \L\, \K \times \K }^\ord{0} = h + \sum_A \frac{\bar q_{0 A} }{|x-x_{\scriptscriptstyle A}|} -  \sum_A \frac{\beta_A^i ( x-x_{\scriptscriptstyle A})_i }{|x-x_{\scriptscriptstyle A}|^3} \ee 
The definition of the associated charge $q_{0 A}$ through the explicit form of the vector $w_0$ requires that 
\be \bar q_{0 A} = q_{0 A} - 2  \trace \biggl[ \mathpzc{p}_A   \Scal{  \mathpzc{k} +\sum_{B\ne A} \gamma_B \frac{\mathpzc{p}_B}{|x_{\scriptscriptstyle B}-x_{\scriptscriptstyle A}|} }  \times  \Scal{  \mathpzc{k} +\sum_{C\ne A} \gamma_C \frac{\mathpzc{p}_C}{|x_{\scriptscriptstyle C}-x_{\scriptscriptstyle A}|} }  \biggr]  \ee
such that 
\be d w_0 =  \star d \frac{ q_{0 A}}{ |x-x_{\scriptscriptstyle A}|} + d \, \Scal{   \varepsilon_{ijk} \frac{ \beta_A^i (x-x_{\scriptscriptstyle A})^j d x^k }{|x-x_{\scriptscriptstyle A}|^3} + \mathcal{O}(x-x_{\scriptscriptstyle A}) }  \ee
The expansion of $V$ near $x_{\scriptscriptstyle A}$ gives finally 
\begin{multline} V = \gamma_A^{\; 2 } \frac{\det[\mathpzc{p}_A]}{| x -x_{\scriptscriptstyle A}|^3} + \gamma_A^{\; 2} \trace \Bigl[ \mathpzc{p}_A \times \mathpzc{p}_A \Scal{ 
 \mathpzc{l}  +\sum_{B\ne A} \frac{\mathpzc{p}_B}{|x_{\scriptscriptstyle B}-x_{\scriptscriptstyle A}|} } \Bigr]  \frac{1}{| x -x_{\scriptscriptstyle A}|^2} \\+ \frac{\beta_A^i ( x-x_{\scriptscriptstyle A})_i }{|x-x_{\scriptscriptstyle A}|^3}  + \frac{q_{0 A}}{|x-x_{\scriptscriptstyle A}|} \\ +  \trace \biggl[2 \gamma_A  \mathpzc{p}_A   \Scal{  \mathpzc{l} +\sum_{B\ne A} \frac{\mathpzc{p}_B}{|x_{\scriptscriptstyle B}-x_{\scriptscriptstyle A}|} }  \times  \Scal{  \mathpzc{k} +\sum_{C\ne A} \gamma_C \frac{\mathpzc{p}_C}{|x_{\scriptscriptstyle C}-x_{\scriptscriptstyle A}|} }  \biggr . \hspace{15mm}  \\ \biggl .\hspace{20mm}  -  \mathpzc{p}_A   \Scal{  \mathpzc{k} +\sum_{B\ne A} \gamma_B \frac{\mathpzc{p}_B}{|x_{\scriptscriptstyle B}-x_{\scriptscriptstyle A}|} } \times   \Scal{  \mathpzc{k} +\sum_{C\ne A} \gamma_C \frac{\mathpzc{p}_C}{|x_{\scriptscriptstyle C}-x_{\scriptscriptstyle A}|} } \biggr]  \frac{1}{|x -x_{\scriptscriptstyle A}|} \\
+ \biggl( \sum_{B\ne A} \gamma_A  ( 2 \gamma_B - \gamma_A)  \trace \bigl[ \mathpzc{p}_A \times \mathpzc{p}_A \, \mathpzc{p}_B \bigr] \, \frac{ (x_{\scriptscriptstyle B}-x_{\scriptscriptstyle A})_i}{|x_{\scriptscriptstyle B}-x_{\scriptscriptstyle A}|^3}\biggr)  \frac{ (x -x_{\scriptscriptstyle A})^i}{|x -x_{\scriptscriptstyle A}|^2} 
+\mathcal{O}(1) \end{multline}
\subsection{Near horizon geometry}
Let us now study the form of the scaling factor $e^{-4U}$ near $x_{\scriptscriptstyle A}$. We get 
\be V \det[\L] - M^2 = \frac{\det[\mathpzc{p}_A]}{|x-x_{\scriptscriptstyle A}|^6}  ( \beta_A + 2 \gamma_A \J_A)^i ( x-x_{\scriptscriptstyle A})_i  + \mathcal{O}(|x-x_{\scriptscriptstyle A}|^{-4})\ee
Therefore regularity requires 
\be \beta^i_A = -2 \gamma_A \J_A^i \ee
Using this equation one then gets 
\begin{multline}   V \det[\L] - M^2 = - \frac{\det[\mathpzc{p}_A]}{|x-x_{\scriptscriptstyle A}|^4}  \biggl( -q_{0A} \biggr . \\ \biggl . + \trace \Bigl[ \mathpzc{p}_A \Scal{ \gamma_A \mathpzc{l} - \mathpzc{k} + \sum_{B\ne A} ( \gamma_A - \gamma_B ) \frac{ \mathpzc{p}_B}{|x_{\scriptscriptstyle A}-x_{\scriptscriptstyle B}|} } \times  \Scal{ \gamma_A \mathpzc{l} - \mathpzc{k} + \sum_{C\ne A} ( \gamma_A - \gamma_C ) \frac{ \mathpzc{p}_C}{|x_{\scriptscriptstyle A}-x_{\scriptscriptstyle C}|} }  \Bigr] \biggr) \\
- \frac{ \scal{ \J_A^i ( x - x_{\scriptscriptstyle A})_i}^2}{|x-x_{\scriptscriptstyle A}|^6} + \mathcal{O}(|x-x_{\scriptscriptstyle A}|^{-3}) 
\end{multline} 
To interpret this limit we must compute the charges $\mathpzc{q}_A$ associated to the vector field $\mathpzc{w}$. Near $x_{\scriptscriptstyle A}$ one computes that 
\be \L \times d \K - \K \times d \L  = \Scal{ \gamma_A \mathpzc{l} - \mathpzc{k} + \sum_{B\ne A} ( \gamma_A - \gamma_B ) \frac{ \mathpzc{p}_B}{|x_{\scriptscriptstyle B}-x_{\scriptscriptstyle A}|} }  \times  d \frac{ \mathpzc{p}_A}{|x - x_{\scriptscriptstyle A}|} + \mathcal{O}(1) \ee
and therefore 
\be \mathpzc{q}_A = 2 \mathpzc{p}_A\times \Scal{ \gamma_A \mathpzc{l} - \mathpzc{k} + \sum_{B\ne A} ( \gamma_A - \gamma_B ) \frac{ \mathpzc{p}_B}{|x_{\scriptscriptstyle B}-x_{\scriptscriptstyle A}|} } \label{ChargeContraintes} \ee
We then use the Jordan algebra identity \cite{Jordan}
\be \det[ \mathpzc{p}_A]  \mathpzc{q}_A = 4 \mathpzc{p}_A \times \scal{ \mathpzc{q}_A \times ( \mathpzc{p}_A \times \mathpzc{p}_A ) } - \mathpzc{p}_A \times \mathpzc{p}_A \trace [ \mathpzc{p}_A \mathpzc{q}_A] \ee
to show that 
\begin{multline}   V \det[\L] - M^2 = - \frac{-q_{0A}  \det[\mathpzc{p}_A] +  \trace\bigl[ (  \mathpzc{q}_A \times \mathpzc{q}_A)  ( \mathpzc{p}_A\times\mathpzc{p}_A) \bigr]   - \frac{1}{4} \scal{ \trace[\mathpzc{p}_A \mathpzc{q}_A] }^2   }{|x-x_{\scriptscriptstyle A}|^4}  \\
- \frac{ \scal{ \J_A^i ( x - x_{\scriptscriptstyle A})_i}^2}{|x-x_{\scriptscriptstyle A}|^6} + \mathcal{O}(|x-x_{\scriptscriptstyle A}|^{-3}) 
\end{multline} 
We recognise the first term as a component of the quartic invariant. Using coordinates such that $\J_A^i \partial_i = \J_A \partial_z$ we can rewrite the scaling factor as 
\bea   e^{-4U}  &=& - \frac{-q_{0A}  \det[\mathpzc{p}_A] +  \trace\bigl[ (  \mathpzc{q}_A \times \mathpzc{q}_A)  ( \mathpzc{p}_A\times\mathpzc{p}_A) \bigr]   + \scal{  q_A^\dagger \, ( \mathpzc{p}_A \times \mathpzc{p}_A) \, q_A }  - \frac{1}{4} \scal{ \trace[\mathpzc{p}_A \mathpzc{q}_A] }^2   }{|x-x_{\scriptscriptstyle A}|^4}  \CR
&& \hspace{30mm} 
- \frac{ \scal{ \J_A^i ( x - x_{\scriptscriptstyle A})_i}^2}{|x-x_{\scriptscriptstyle A}|^6} + \mathcal{O}(|x-x_{\scriptscriptstyle A}|^{-3}) \CR
&=& - \frac{I_4[q_A,p_A]+ \J_A^{\; 2} \cos^2(\theta_{\pA}) }{|x-x_{\scriptscriptstyle A}|^4} + \mathcal{O}(|x-x_{\scriptscriptstyle A}|^{-3})
\eea
as expected for a non-BPS extremal black hole \cite{Rasheed:1995zv,Larsen:1999pp,Ferrara:2006em}. The horizon is therefore squashed, and the scalar fields do not take constant values, but rather 
\be \mathpzc{t}_\pA =  \frac{ - \frac{ \partial I_4}{\partial \mathpzc{q}_\pA} + \ell \invo  \frac{ \partial I_4 }{\partial q^\dagger_\pA} + \mathpzc{p}_A \Scal{  \J_\pA \cos( \theta_\pA) + i \sqrt{ -I_4  -  \J_\pA^{\; 2} \cos^2(\theta_\pA) }}}{ - \frac{ \partial I_4 }{\partial q_{\pA 0} } } \label{AttractorNBPS} \ee  
This corroborates the analysis in \cite{DimitruSen}. We will see in the next section that the scalar fields satisfy a generalised attractor equation which is the $U(1)$ Ehlers transform of the standard one, for which the angle depends on the angular momentum $\J_\pA$ and the coordinate $\theta_\pA$. 
\subsection{The asymptotic region} 

Let us now consider the solution in the asymptotic region. In that case one has 
\be \L = \mathpzc{l} + \frac{ \sum_A \mathpzc{p}_A}{|x|} + \frac{  \sum_A \mathpzc{p}_A x^i_{\scriptscriptstyle A} x_i }{|x|^3} + \mathcal{O}(|x|^{-3}) \; , \quad \K = \mathpzc{k} + \frac{ \sum_A \gamma_A \mathpzc{p}_A}{|x|} + \frac{  \sum_A \gamma_A \mathpzc{p}_A x^i_{\scriptscriptstyle A} x_i }{|x|^3} + \mathcal{O}(|x|^{-3}) \ee
The boundary conditions for the functions $V$ and $M$ were fixed at the poles, and without solving the equation completely one only fixes these functions up to arbitrary harmonic functions. Nevertheless, the conservation of charges determines the first pole to be the sum of the poles associated to each black holes and we can determine that 
\bea M &=& m  + \frac{ \trace \bigl[ \mathpzc{l} \times \mathpzc{l} \sum_A \gamma_A \mathpzc{p}_A\bigr] }{|x|} + \mathcal{O}(|x|^{-2}) \CR
 V &=& h  + \frac{ \sum_A q_{0 A} + \trace \bigl[ 2 \mathpzc{l} \times \mathpzc{k} \sum_A \gamma_A \mathpzc{p}_A-\mathpzc{k} \times \mathpzc{k} \sum_A \mathpzc{p}_A \bigr]   }{|x|} + \mathcal{O}(|x|^{-2}) \eea
One can check that this is indeed the case for the global solution we will discuss latter. 
\subsubsection*{The ADM mass}
Using these formulas one obtains that 
\begin{multline}  e^{-4U} = h \det[ \mathpzc{l} ] - \scal{ y^\dagger ( \mathpzc{l}\times \mathpzc{l}) y} - m^2 + \biggl(  h \trace \Bigl[ \mathpzc{l} \times \mathpzc{l} \sum_A \mathpzc{p}_A \Bigr] \biggr . \\ + \det[\mathpzc{l}] \Scal{ \sum_A q_{0 A} + \trace\Bigl[ 2  \mathpzc{l} \times \mathpzc{k} \sum_A \gamma_A \mathpzc{p}_A
- \mathpzc{k} \times \mathpzc{k} \sum_A \mathpzc{p}_A\Bigr] }\\  - \Scal{ y^\dagger ( \mathpzc{l}\times \mathpzc{l}) \sum_A q_A } - \Scal{ \Scal{ \sum_A q_A}^\dagger ( \mathpzc{l} \times \mathpzc{l}) y} - 2 \Scal{ y^\dagger \Scal{ \mathpzc{l} \times \sum_A \mathpzc{p}_A} y}\\ \biggl .  - 2 m \trace \Bigl[ \mathpzc{l} \times \mathpzc{l} \sum_A \gamma_A \mathpzc{p}_A \Bigr] \biggr)\frac{1}{|x|} + \mathcal{O}(|x|^{-2})   \end{multline} 
The conventional boundary condition on the scaling factor of the metric determines 
\be h = \frac{1+   \scal{ y^\dagger ( \mathpzc{l}\times \mathpzc{l}) y} + m^2 }{\det[\mathpzc{l}]} \ee
 To interpret this formula more precisely we will also need the asymptotic moduli 
 \be \mathpzc{t}_\asym = \mathpzc{k} + \frac{ \mathpzc{l}}{\det[\mathpzc{l}]} \scal{ - m + i} + \ell \invo \frac{ \mathpzc{l}\times \mathpzc{l}}{\det[\mathpzc{l}]} y \ee
 and the total electromagnetic charges 
 \bea q_0 &=& \sum_A q_{0 A} \qquad \mathpzc{p} = \sum_A \mathpzc{p}_A \qquad q = \sum_A q_A \CR
 \mathpzc{q} &=& \sum_A \mathpzc{q}_A = 2 \sum_A \scal{ \gamma_A \mathpzc{l} - \mathpzc{k} } \times \mathpzc{p}_A \eea
 Let us first combine the terms 
 \bea &&  \frac{m^2}{\det[\mathpzc{l}]}  \trace \Bigl[ \mathpzc{l} \times \mathpzc{l} \sum_A \mathpzc{p}_A \Bigr] + \det[\mathpzc{l}] \Scal{ \sum_A q_{0 A} + \trace\Bigl[ 2  \mathpzc{l} \times \mathpzc{k} \sum_A \gamma_A \mathpzc{p}_A
- \mathpzc{k} \times \mathpzc{k} \sum_A \mathpzc{p}_A\Bigr] }\CR
&& \hspace{80mm}  - 2 m \trace \Bigl[ \mathpzc{l} \times \mathpzc{l} \sum_A \gamma_A \mathpzc{p}_A \Bigr] \CR
&=& \det[\mathpzc{l}] \trace \biggl[ \Scal{ \mathpzc{k} - \frac{ \mathpzc{l} }{\det[\mathpzc{l}]} m} \times  \Scal{ \mathpzc{k} - \frac{ \mathpzc{l} }{\det[\mathpzc{l}]} m} \sum_A \mathpzc{p}_A \biggr . \CR&& \hspace{50mm} \biggl . + 2  \Scal{ \mathpzc{k} - \frac{ \mathpzc{l} }{\det[\mathpzc{l}]} m}  \Scal{ \mathpzc{l} \times \sum_A \gamma_A \mathpzc{p}_A - \mathpzc{k} \times \sum_A \mathpzc{p}_A } \biggr]  \CR
&=& \det[\mathpzc{l}] \trace \biggl[ \Scal{ \mathpzc{k} - \frac{ \mathpzc{l} }{\det[\mathpzc{l}]} m} \times  \Scal{ \mathpzc{k} - \frac{ \mathpzc{l} }{\det[\mathpzc{l}]} m}  \mathpzc{p}  +   \Scal{ \mathpzc{k} - \frac{ \mathpzc{l} }{\det[\mathpzc{l}]} m}  \mathpzc{q} \biggr] 
\eea
We can now have a look at the terms in $y$
\be \frac{ \scal{ y^\dagger (  \mathpzc{l} \times \mathpzc{l}) y }}{\det[\mathpzc{l}]} \trace \bigl[ \mathpzc{l} \times \mathpzc{l}\,  \mathpzc{p} \bigr] - \scal{ y^\dagger ( \mathpzc{l}\times \mathpzc{l}) q} - \scal{ q^\dagger ( \mathpzc{l} \times \mathpzc{l}) y} - 2 \scal{ y^\dagger ( \mathpzc{l} \times \mathpzc{p}) y} \label{yyterms} \ee
 For the Jordan algebra associated to the associative composition algebras, we have the matrix product identity 
 \be \mathpzc{p} ( \mathpzc{l} \times \mathpzc{l} ) + 2 \mathpzc{l} ( \mathpzc{l} \times \mathpzc{p} ) = \trace [ \mathpzc{l} \times \mathpzc{l} \, \mathpzc{p}] \, \mathds{1}  \ee 
 which is generally valid for any exceptional Jordan algebra for the commutative Jordan product. One can then simplify (\ref{yyterms}) as
 \be \det[\mathpzc{l}] \biggl( \Scal{ \Scal{ \frac{ \mathpzc{l}\times \mathpzc{l}}{\det[\mathpzc{l}]} y}^\dagger \mathpzc{p} \frac{ \mathpzc{l}\times \mathpzc{l}}{\det[\mathpzc{l}]} y} -  \Scal{ \Scal{ \frac{ \mathpzc{l}\times \mathpzc{l}}{\det[\mathpzc{l}]} y}^\dagger q } - \Scal{ q^\dagger \frac{ \mathpzc{l}\times \mathpzc{l}}{\det[\mathpzc{l}]} y} \biggr) \ee
Using the octonionic expression  for the asymptotic moduli and the electric charges 
\be \mathpzc{a}_\asym = \mathpzc{k} - \frac{ \mathpzc{l}}{\det[\mathpzc{l}]}  m + \ell \invo \frac{ \mathpzc{l}\times \mathpzc{l}}{\det[\mathpzc{l}]} y \qquad \mathpzc{Q} = \mathpzc{q} + \ell \invo q \ee
 we can combine these expressions in 
 \be \trace \Bigl[ \Scal{ \mathpzc{k} - \frac{ \mathpzc{l} }{\det[\mathpzc{l}]} m} \times  \Scal{ \mathpzc{k} - \frac{ \mathpzc{l} }{\det[\mathpzc{l}]} m}  \mathpzc{p} \Bigr] + \Scal{ \Scal{ \frac{ \mathpzc{l}\times \mathpzc{l}}{\det[\mathpzc{l}]} y}^\dagger \mathpzc{p} \frac{ \mathpzc{l}\times \mathpzc{l}}{\det[\mathpzc{l}]} y}  = \trace \bigl[ \mathpzc{a}_\asym \times \mathpzc{a}_\asym \, \mathpzc{p} \bigr] \ee
and 
\be \trace \Bigl[  \Scal{ \mathpzc{k} - \frac{ \mathpzc{l} }{\det[\mathpzc{l}]} m}  \mathpzc{q} -  \Scal{ \Scal{ \frac{ \mathpzc{l}\times \mathpzc{l}}{\det[\mathpzc{l}]} y}^\dagger q } - \Scal{ q^\dagger \frac{ \mathpzc{l}\times \mathpzc{l}}{\det[\mathpzc{l}]} y}\Bigr] = \trace \bigl[ \mathpzc{a}_\asym \mathpzc{Q} \bigr] \ee
This way we can rewrite the scaling factor as 
\be e^{-4U} = 1 + \det[\mathpzc{l}]\Scal{ q_0  + \trace \bigl[ \mathpzc{a}_\asym \mathpzc{Q} \bigr] + \trace \Bigl[ \Scal{ \frac{\mathpzc{l}}{\det[\mathpzc{l}]} \times  \frac{\mathpzc{l}}{\det[\mathpzc{l}]}  + \mathpzc{a}_\asym \times \mathpzc{a}_\asym }\, \mathpzc{p} \Bigr] }\frac{1}{|x|} + \mathcal{O}(|x|^{-2}) \ee
 We conclude that the ADM mass is determined in terms of the asymptotic moduli and the total electromagnetic charges as 
 \be M_{\scriptscriptstyle \rm{ADM}} = \frac{\det[\mathpzc{l}]}{4} \Scal{ q_0  + \trace \bigl[ \mathpzc{a}_\asym \mathpzc{Q} \bigr] +\trace \Bigl[ \Scal{ \frac{\mathpzc{l}}{\det[\mathpzc{l}]} \times  \frac{\mathpzc{l}}{\det[\mathpzc{l}]}  + \mathpzc{a}_\asym \times \mathpzc{a}_\asym }\, \mathpzc{p} \Bigr]  }\ee
 This can be written in terms of the $\mathpzc{t}_\asym$ themselves as 
 \be M_{\scriptscriptstyle \rm{ADM}} = \frac{1}{\sqrt{ 2i \det[ \mathpzc{t}_\asym - \bar{\mathpzc{t}}_\asym]}} \Scal{ q_0  + \frac{1}{2} \trace \bigl[ ( \mathpzc{t}_\asym + \bar{\mathpzc{t}}_\asym)  \mathpzc{Q} \bigr] +\trace \bigl[  \mathpzc{t}_\asym \times \bar{\mathpzc{t}}_\asym \, \mathpzc{p} \Bigr]  }\ee
 Although this solution does not sit within an $\N=2$ truncation, it is characterised in this duality frame by the `central charge' (\ref{CentralCharge}) such that 
\be M_{\scriptscriptstyle \rm{ADM}} =  \frac{1}{2}  \Scal{ - Z(q,p) + \trace [  \mathpzc{D} Z(p,q)  ]} \label{ADMnonBPS}  \ee

\subsubsection*{The angular momentum} 
This perturbative expansion does not permit to compute the angular momentum. In order to obtain the global solutions, one needs to solve several Laplace equations with sources. The ones involving only two centres were already solved in \cite{Bena:2009en}, and we will refer to this paper for further details. However, the equation involving three centres
\be \Delta F_{A,BC} =  \frac{1}{|x-x_{\pA}|} \Delta   \biggl( \frac{1}{|x-x_{\pB}||x-x_{\pC}|} \biggr) \ee
 was only solved in the axisymmetric case for which $x_{\pA},x_{\pB},x_{\pC}$ were assumed to be aligned. In this case 
 \be F^{\rm \scriptscriptstyle Axsym}_{A,BC} = \frac{1}{(x_\pB-x_\pA)\hspace{-1mm}\cdot \hspace{-1mm}(x_\pC - x_\pA)} \biggl( \frac{|x-x_\pA|}{|x-x_{\pB}||x-x_{\pC}|} - \frac{|x_\pB-x_\pA|}{|x_\pB-x_{\pC}| |x-x_{\pB}|}-\frac{|x_\pC-x_\pA|}{|x_\pB-x_{\pC}||x-x_{\pC}|}\biggr) \ee
 We will not be able to solve this Laplace equation in general, but one can define
\be  F_{A,BC} = - \int \frac{d^3 y}{2\pi} \frac{ (y-x_\pB)\hspace{-1mm}\cdot \hspace{-1mm}(y - x_\pC)}{|y-x| |y-x_{\pA}| |y-x_{\pB}|^3 |y-x_{\pC}|^3} \ee
which is the unique solution to the Laplace equation that admits no poles and is regular in the asymptotic region. We study this function in the Appendix \ref{Laplace}, where we prove that the integral converges, and we calculate its Taylor expansion in the asymptotic region.

 The exact solution for the function $M$ is then 
 \begin{multline} M = m + \sum_A \gamma_A \biggl( \frac{ \det[\mathpzc{p}_\pA] }{|x-x_\pA|^3} + \frac{\trace \mathpzc{l} \, \mathpzc{p}_A \times \mathpzc{p}_A }{|x-x_\pA|^2} + \frac{\trace \mathpzc{p}_A \mathpzc{l} \times  \mathpzc{l} }{|x-x_\pA|} \biggr ) 
 - \sum_A \frac{ \J_{\pA i } ( x - x_\pA)^i}{ |x-x_\pA|^3}\\ + \sum_{A \ne B} \gamma_A \trace  \mathpzc{l} \, \mathpzc{p}_A \times \mathpzc{p}_B \biggl( \frac{ 1}{|x-x_\pA||x-x_\pB|} + \frac{1}{|x_\pA - x_\pB||x-x_\pA|} - \frac{1}{|x_\pA - x_\pB||x-x_\pB|} \biggr) \\
 + \frac{1}{2} \sum_{A\ne B} \trace  \mathpzc{p}_B \, \mathpzc{p}_A \times \mathpzc{p}_A \biggl( \frac{ \gamma_A + \gamma_B}{|x-x_\pA|^2|x-x_\pB|} + \frac{\gamma_A-\gamma_B}{|x_\pA-x_\pB|^2} \Bigl( \frac{ |x-x_\pB|}{|x-x_\pA|^2} - \frac{1}{|x-x_\pB|}\Bigr) \biggr) \\
 + \sum_{A\ne B \ne C} \gamma_C \trace  \mathpzc{p}_A \, \mathpzc{p}_B \times \mathpzc{p}_C \Bigl( F_{A,BC} + \frac{1}{|x_\pA-x_\pC||x_\pB-x_\pC||x-x_\pC|} \Bigr) 
 \end{multline} 
 We will not compute the Kaluza--Klein vector $\omega$ globally, but the asymptotic behaviour of $F_{A,BC}$ at $x \rightarrow \infty$ is enough to compute the asymptotic behaviour of the vector 
 \be \star d \omega_{AB,C} = d \Scal{ F_{(A,B)C} + \frac{1}{|x_\pA-x_\pC||x_\pB-x_\pC||x-x_\pC|} } - \frac{1}{|x-x_\pA||x-x_\pB|} d \frac{1}{|x-x_\pC|} \ee
 as 
\bea \omega_{AB,C}  &=&  \frac{1}{2} \biggl( \frac{ x^i_\pA - x^i_\pB}{|x_\pB- x_\pA||x_\pA - x_\pC|} + \frac{ x^i_\pB - x^i_\pA}{|x_\pA - x_\pB||x_\pB - x_\pC|}  + \frac{ x^i_\pA + x^i_\pB-2 x^i_\pC }{|x_\pA-x_\pC||x_\pB-x_\pC|} \biggr) \frac{\varepsilon_{ijk} x^j dx^k}{|x|^3}  \CR
&& - \frac{ \Bigl( |x_\pA - x_\pB|^2 ( x^i_\pB - x^i_\pC ) - ( x_\pA - x_\pB ) \hspace{-1mm}\cdot \hspace{-1mm} ( x_\pB - x_\pC ) ( x^i_\pA - x^i_\pB )\Bigr) \varepsilon_{ijk} x^j dx^k }{|x|^3  |x_\pA-x_\pB||x_\pA-x_\pC||x_\pB-x_\pC|\scal{   |x_\pA-x_\pB| + |x_\pA-x_{\pC}| + |x_\pB-x_\pC| }}+ \mathcal{O}(x^{-2})\CR 
\eea
The solution is 
\begin{multline} \omega =  \sum_A \varepsilon_{ijk} \frac{ \J_{\pA}^i ( x - x_\pA)^j dx^k}{ |x-x_\pA|^3} + \sum_{A\ne B\ne C} \gamma_C \trace  \mathpzc{p}_A \, \mathpzc{p}_B \times \mathpzc{p}_C \; \omega_{AB,C} 
\\+ \sum_{A\ne B} \frac{  ( \gamma_A-\gamma_B)   \trace  \mathpzc{l} \, \mathpzc{p}_A \times \mathpzc{p}_B \, \varepsilon_{ijk} ( x_\pA^i - x_\pB^i) ( x^j - x_\pB^j ) dx^k }{|x_\pA - x_\pB||x-x_\pA||x-x_\pB| \scal{ |x-x_\pA| + |x-x_\pB| + |x_\pA - x_\pB|}} \\
+ \sum_{A \ne B} ( \gamma_A-\gamma_B) \frac{ \trace  \mathpzc{p}_A \, \mathpzc{p}_A \times \mathpzc{p}_B \, \varepsilon_{ijk} ( x_\pB^i - x_\pA^i) ( x^j - x_\pA^j ) dx^k }{|x_\pA - x_\pB|^2|x-x_\pA|^2|x-x_\pB|} \end{multline} 
 Using the asymptotic expression of $\omega_{AB,C}$ one obtains the angular momentum 
 \bea  J^i &=&  \sum_A \J_\pA^i  +  \sum_{A>B} \trace \biggl(  \Scal{ \gamma_A \mathpzc{l}  + \sum_{C\ne A} ( \gamma_C - \gamma_A) \frac{ \mathpzc{p}_C}{|x_\pA-x_\pC|}}  \times \mathpzc{p}_A \, \mathpzc{p}_B  \biggr . \\ && \hspace{40mm}  \biggl .  - \Scal{ \gamma_B \mathpzc{l}  + \sum_{C\ne B}  ( \gamma_C - \gamma_B) \frac{ \mathpzc{p}_C}{|x_\pB-x_\pC|}}  \times \mathpzc{p}_B \, \mathpzc{p}_A    \biggr) \frac{ x_\pA^i - x_\pB^i}{|x_\pA-x_\pB|}\CR && 
 \hspace{-10mm}  + \sum_{A\ne B\ne C} \gamma_C \trace \mathpzc{p}_A \mathpzc{p}_B \times \mathpzc{p}_C  \frac{ |x_\pA - x_\pB|^2 ( x^i_\pC - x^i_\pB ) - ( x_\pA - x_\pB ) \hspace{-1mm}\cdot \hspace{-1mm} ( x_\pC - x_\pB ) ( x^i_\pA - x^i_\pB ) }{  |x_\pA-x_\pB||x_\pA-x_\pC||x_\pB-x_\pC|\scal{   |x_\pA-x_\pB| + |x_\pA-x_{\pC}| + |x_\pB-x_\pC| }}  \CR
 &=&   \sum_A \J_\pA^i  + \frac{1}{2}  \sum_{A>B} \Scal{ q_{0A} p_B^0 + \trace [ \mathpzc{Q}\, _A \mathpzc{P}_B] - \trace [\mathpzc{P}_A  \mathpzc{Q}\, _B ]- p^0_{A} q_{0 B}  }  \frac{ x_\pA^i - x_\pB^i}{|x_\pA-x_\pB|} \hspace{35mm}  \CR
 && \hspace{-10mm}  + \sum_{A\ne B\ne C} \gamma_C \trace \mathpzc{p}_A \mathpzc{p}_B \times \mathpzc{p}_C  \frac{ |x_\pA - x_\pB|^2 ( x^i_\pC - x^i_\pB ) - ( x_\pA - x_\pB ) \hspace{-1mm}\cdot \hspace{-1mm} ( x_\pC - x_\pB ) ( x^i_\pA - x^i_\pB ) }{  |x_\pA-x_\pB||x_\pA-x_\pC||x_\pB-x_\pC|\scal{   |x_\pA-x_\pB| + |x_\pA-x_{\pC}| + |x_\pB-x_\pC| }} \nn \label{NonBPSJ} 
 \eea
 The first term is simply the sum of the intrinsic angular momenta of the black holes, and the second is the standard  angular momentum resulting from the non-commuting dyon charges. The third one is rather non-standard, and is generated by interactions between three non-aligned black holes. One can easily compute that the sum over the permutations of $ABC$ of the function 
 \be   |x_\pA - x_\pB|^2 ( x^i_\pC - x^i_\pB ) - ( x_\pA - x_\pB ) \hspace{-1mm}\cdot \hspace{-1mm} ( x_\pC - x_\pB ) ( x^i_\pA - x^i_\pB ) \ee
vanishes. Therefore the associated contribution to the angular momentum clearly vanishes if $\gamma_A=\gamma_B=\gamma_C$. The function of the charges  $ ( \gamma_C + c) \trace \mathpzc{p}_A \mathpzc{p}_B \times \mathpzc{p}_C$, does not obviously look like it is $E_{7(7)}$ invariant  for any given constant $c$. Nonetheless we will show in the next section that it is.

 \section{Duality invariance}
 The composite non-BPS nilpotent orbit is characterised by a semi-simple element ${\bf h}$ of $\so^*(16)$ (which can be expressed in a Cartan basis  as \DSOXVI00000200)  defining the graded decomposition 
 \be \so^*(16) \cong {\bf 15}^\ord{-2} \oplus ( {\bf 2}\otimes {\bf 2} \otimes {\bf 6})^\ord{-1} \oplus \scal{ \gl_1 \oplus \sl_2 \oplus \su(2) \oplus \su^*(6)}^\ord{0} \oplus  ( {\bf 2}\otimes {\bf 2} \otimes \overline{\bf 6})^\ord{1}\oplus  \overline{\bf 15}^\ord{2} \ee
 such that 
 \be {\bf 128} \cong {\bf 2}^\ord{-3} \oplus ({\bf 2} \otimes \overline{ \bf 6})^\ord{-2} \oplus ( {\bf 2} \otimes \overline{\bf 15})^\ord{-1} \oplus ( {\bf 2}\otimes {\bf 20})^\ord{0} \oplus ( {\bf 2} \otimes{\bf 15})^\ord{1} \oplus ({\bf 2}\otimes {\bf 6})^\ord{2} \oplus {\bf 2}^\ord{3} \label{128Decom} \ee
 A generic element of the grade one component is not left invariant by any generator of $\so^*(16)$ of negative grade. To compute the isotropy subalgebra in the grade zero component, it is convenient to decompose 
 \be \su^*(6) \cong ({\bf 2} \otimes \overline{\bf 4})^\ord{-3} \oplus ( \gl_1 \oplus \su(2) \oplus \su^*(4) )^\ord{0} \oplus  ({\bf 2} \otimes {\bf 4})^\ord{3} \ee
 such that 
 \be {\bf 15} \cong {\bf 1}^\ord{-4} \oplus ( {\bf 2}\otimes {\bf 4})^\ord{-1} \oplus {\bf 6}^\ord{2} \ee
 A generic element of the $ ( {\bf 2} \otimes{\bf 15})^\ord{1} $ is then chosen to be a doublet of non-null orthogonal vectors of $SO(1,5) \cong SU^*(4) / \mathds{Z}_2$ with a non-trivial component in the grade $-4$ singlet.  If one vector is time-like, the isotropy subgroup is 
 \be \Scal{ \prod_{\Lambda=0}^3 SU(2)_\Lambda}  \ltimes \Scal{ \bigoplus_{\Xi > \Lambda} {\bf 2}_\Xi \otimes {\bf 2}_\Lambda \oplus \mathds{R}} \subset \Spin^*(16) \ee 
 which defines a Lagrangian submanifold of dimension 83 inside the $E_{8(8)}$ nilpotent orbit (associated to the weighted Dynkin diagram \DSOXVI00000200) of stabiliser 
 \be Spin(4,4) \ltimes \mathds{R}^{6\times 8 + 6} \subset E_{8(8)}\ee
 If the two vectors are space-like, the isotropy subgroup is of the form
 \be \scal{ SU(2)_1 \times SU(2)_2 \times SL(2,\mathds{C})  } \ltimes \Scal{ {\bf 2}_\mathds{C} \otimes ( {\bf 2}_1 \oplus {\bf 2}_2) \oplus {\bf 4}  \oplus \mathds{R}} \subset \Spin^*(16) \ee 
 which defines a Lagrangian submanifold of dimension 83 inside the $E_{8(8)}$ nilpotent orbit (associated to the weighted Dynkin diagram 
 \DSOXVI02000020) of stabiliser
  \be Spin(5,3) \ltimes \mathds{R}^{6\times 8 + 6} \subset E_{8(8)}\ee
For single-centre extremal black holes, the regularity of the solutions requires the two $\so(16,\mathds{C})$ weighted Dynkin diagrams,  determining respectively in which $\Spin(16,\mathds{C})$ orbit and in which $E_{8(8)}$ orbit the Noether charge sits in, to be identical \cite{BossardW}. Only the first representative (with one time-like vector) satisfies to this criterion. In the neighbourhood of a given centre, the semi-simple component of the stabiliser subgroup of the momentum $P\in {\bf 128}$ must necessarily be compact in order to interpolate smoothly  the stabiliser subgroup $Sp(4) \ltimes \mathds{R}^{27}$ associated to a regular solution. Therefore this criterion must still be satisfied in the near horizon regions of the solution. But in principle the signature of the isotropy subgroup of the momentum $P\in {\bf 128}$ is not necessarily constant allover space,\footnote{The composite BPS solutions which total electromagnetic charges have a negative quartic invariant, admit for instance a mometum $P$ of stabiliser $(SU(2) \times SU(6)) \ltimes \mathds{C}^{2\times 6} \oplus \mathds{R} $ in the near horizon regions, and of stabiliser $(SU(2) \times SU^*(6)) \ltimes \mathds{C}^{2\times 6} \oplus \mathds{R}$ in the asymptotic region \cite{BossardRuef}.} and we could have regular solutions for which the total Noether charge defines a representative of the other nilpotent orbit.

A configuration of charges which gives rise to a regular black hole must lie in a smaller orbit. After some computation, one finds that the only configurations corresponding to regular extremal black holes in this system are necessarily corresponding to non-BPS black holes. To illustrate this, let us consider a particularly simple configuration, with one antisymmetric tensor in the grade 1 component $P^+_{ab}$ of weight 1 with respect to a given Cartan generator of $\sl_2$, and a singlet $Q_0^-$ in the grade 3 component and of weight $-1$ with respect to $\sl_2$. Then one computes that the stabiliser of such charge configuration only allows for two components of grade $1$ and $-1$ satisfying 
\be E^{-\alpha b} P^+_{ab} + F^{+\alpha}_{\, a} Q^-_0 = 0 \label{SignCont} \ee
the generators of $\su(2)$, and the generators of $\su^*(6)$ that leave $P^+_{ab}$ invariant, as well as the combination of the weight 2 $\sl_2$ generator $E^{++}$ and the grade 2 component $E^{ab}$ satisfying 
\be E^{++} Q^-_0 + E^{ab} P^+_{ab} = 0 \ee
and the grade 1 component of $\sl_2$ weight 1 $E^{+ \alpha a}$. In order for the grade 0 component of the stabiliser to be compact, we see that $P^+_{ab}$ must have eigen values of the same sign, such that it admits $Sp(3)$ as an isotropy subgroup of $SU^*(6)$. In order for the combination of $E^{-\alpha b} $ and $F^{+\alpha}_{\, a} $ to define a compact generator, we see that we must have 
\be Q^-_0 \varepsilon^{abcdef} P^+_{ab} P^+_{cd} P^+_{ef} > 0 \ee
which is equivalent as to have a negative quartic invariant. The components $E^{++}$, $E^{ab}$ and $E^{+ \alpha a}$, then combine into nilpotent generators transforming in the ${\bf 27}$ of $Sp(4)$, and this reproduces the correct isotropy subgroup of a regular non-BPS black hole \cite{Bossard:2009at}
\be \Sp(4) \ltimes \mathds{R}^{27} \subset \Spin^*(16) \ee

The elements of the ${\bf 2}\otimes {\bf 20}$ grade 0 component in (\ref{128Decom}) correspond by definition to fields which are not sourced by the system of equations. With respect to $Sp(1) \times Sp(3)$ they decompose into ${\bf 2} \otimes {\bf 14}$ non-compact generators and ${\bf 2}\otimes {\bf 6}$ compact ones. The non-compact generators are associated to the scalar fields, and the compact generators to the electromagnetic fields. The `flat directions'  therefore define the quaternionic symmetric space 
\be F_{4(4)} / ( Sp(1) \times_{\mathds{Z}_2} Sp(3) ) \subset E_{7(7)} / \SU \label{FlatF44} \ee
 Indeed, the imaginary part of the scalar field $\mathpzc{t}$ is quaternionic in the Ansatz (does not include components in $\ell$). The corresponding non-compact generators in $\e_{6(6)}$ are realised as off-diagonal anti-Hermtian 3 by 3 matrices over the octonions which are linear in $\ell$. The group of unitary matrices over the split octonions acting on traceless Hermitian 3 by 3 matrices extended by the action of the authomorphism group $G_{2(2)}$ defines the fundamental representation of $F_{4(4)}$. Therefore we see that the scalar fields which are trivial in the solution indeed parametrize the quaternionic manifold (\ref{FlatF44}).

\subsection{Generalised fake superpotential} 
 
 The isotropy subgroup of $SU(8)$ of the element ${\bf h}$ defining the nilpotent subalgebra is $Sp(1) \times Sp(3)$ (because the $U(1)$ factor lies in the Ehlers group). As it was discussed in \cite{BossardRuef}, this implies that the set of inequivalent embeddings for the solvable system described by the 44 functions $\mathcal{L}, \mathcal{K}, Y , V , M$ is parametrized by the 39 angles parametrizing 
 \be SU(8)/ ( Sp(1) \times Sp(3)) \ee
 which indeed add up to the 83 dimensions of the orbit. These angles parametrize two orthogonal antisymmetric tensors of $SU(8)$ satisfying 
 \bea && \omega_{ij} \omega^{ij} = 2 \qquad \omega_{[ij} \omega_{kl]} = 0 \qquad \omega_{ik} \omega^{jk} + \Omega_{ik} \Omega^{jk} = \delta_i^k \qquad \omega^{ik} \Omega_{jk} = 0 \CR && \frac{1}{96} \varepsilon^{ijklmnpq} \omega_{ij} \Omega_{kl} \Omega_{mn} \Omega_{pq} = 1 \label{ConstraintOmega} \eea
 where $\omega^{ij}$ and $\Omega^{ij}$ are the complex conjugates of $\omega_{ij}$ and $\Omega_{ij}$. In the solutions described in the last section, $\omega_{ij}$ defines the `central charge'
 \be \omega^{ij} Z_{ij} = Z \ee
 and $\Omega_{ij}$ the Jordan algbera identity, such that 
 \be \Omega_{ij} Z^{ij} =  \trace [ \mathpzc{D} Z ] \ee
The mass formula can be rewritten as
 \be M_{\scriptscriptstyle \rm ADM} = \frac{1}{2} \scal{ \Omega_{ij} Z^{ij} - \omega^{ij} Z_{ij} } \ee
 Note that within our Ansatz the potential $ \Omega_{ij} Z^{ij} - \omega^{ij} Z_{ij} $ is automatically real, whereas its imaginary part was interpreted as the NUT charge in \cite{BossardRuef}. This discrepancy is due to the fact that we already assumed that the NUT charge vanished when we computed the electromagnetic charges, whereas one would need in general to consider the complete spatial components of the field strengths. 
 
Let us define the generalised fake superpotential 
\be W \equiv   \frac{1}{2} \scal{ \Omega^{ij} Z_{ij} - \omega_{ij} Z^{ij} } \ee
which within our specific Ansatz is defined as
\be W  = \frac{1}{2} \scal{ \trace [ \overline{\mathpzc{D}Z}] - \overline{Z}} \ee
 
 The real part of  $ \mathpzc{D} Z$ is a Hermitian matrix over the quaternions (\ie the components in $\ell$ all vanish), and rewriting this property in terms of the tensors $\omega_{ij},\, \Omega_{ij}$ we conclude that in general 
 \be \omega_{ik} \Omega_{jl} \scal{ Z^{kl} + \omega^{kp} \Omega^{lq} Z_{pq} } = 0 \label{ConstraintReal}  \ee
 These $12$ constraints (in the ${\bf 2}\otimes {\bf 6}$ of $Sp(1) \times Sp(3)$) reduce the number of angles associated to a duality frame for given asymptotic central charges to $27$. One of these angles is moreover determined such that
  \be  \frac{1}{2} \scal{ \Omega_{ij} Z^{ij} - \omega^{ij} Z_{ij} }  \in \mathds{R}_+^* \ee
 In this way the $56$ remaining parameters defining the specific nilpotent element associated to the asymptotic momenta $\V^{-1} d \V|_{\bf 128}$ are the 28 complex central charges. 
 
 \subsubsection*{Relation to the standard fake superpotential}
This potential $W$ somehow generalises the `flat direction dependent'  fake superpotential described in \cite{Ceresole:2009vp}. To see this, let us  consider for a moment the exceptional $\N=2$ supergravity with moduli parametrizing the exceptional special K\"{a}hler space $E_{7(-25)} / ( U(1) \times E_{6(-78)})$ \cite{GunaydinMagic,GunaydinJordan}. In this case our Ansatz applies as well, with all $Y$ set to zero and $\mathcal{L}$ and $\mathcal{K}$ understood to be three by three Hermitian matrices over the octonions. In this case the constraint $(\ref{ConstraintReal})$ disappears and 
\be W \equiv \frac{1}{2} \scal{ \trace[ \Omega  \overline{\mathpzc{D} Z }] - \det[ \Omega] \bar Z } \label{TrueFake} \ee
where  $\Omega$ is a complex Hermitian matrix over the octonions \footnote{The field of complex being  kept distinct from the field of octonions, such that the Hermitian property is defined for the octonions only.}  satisfying 
\be \Omega \times \Omega  =  \det[ \Omega] \, \overline{\Omega}  \; , \quad \trace [ \Omega \overline{\Omega}] = 3 \label{ConstraintOmega2} \ee
$\Omega$ is parametrized by $U(1) \times E_{6(-78)} / F_{4(-52)}$. And so with the additional positivity constraint $W>0$, the remaining free parameters lie in $E_{6(-78)} / F_{4(-52)}$, which defines a compact version of the space of flat directions $E_{6(-26)} / F_{4(-52)}$ of non-BPS charges in the theory \cite{UdualityN2}. Therefore $W(\Omega)$ can be interpreted as a `flat direction dependent' fake superpotential as in \cite{Ceresole:2009vp}. In order to be the case, one should recover the single-centre non-BPS fake superpotential when the auxiliary parameters extremize the potential. However the constraint that $W>0$ is relatively complicated to enforce, and in order to simplify the problem we will consider instead the potential 
\be  W_\alpha \equiv \frac{{\rm Re}[ e^{2i\alpha} W ] }{\cos(2\alpha)} \ee
as a function of $\Omega \in U(1) \times E_{6(-78)} / F_{4(-52)}$, which is equal to $W$ when $W$ is real. It turns out that this potential coincides with $W$ at its extremum, if and only if $\alpha$ is the phase appearing in the definition of the non-standard diagonal form of the central charge and its K\"{a}hler derivative  \cite{BossardW}, and in this way, there is no ambiguity in the definition of $W_\alpha$. Using the property that the constraint (\ref{ConstraintOmega2}) is invariant with respect to the variation of $\Omega$ in $ U(1) \times E_{6(-78)} / F_{4(-52)}$, one relates the variation of $\overline{\Omega}$ to its complex conjugate through 
\be \delta \overline{\Omega} = \frac{1}{\det[ \Omega]} \scal{ 2 \Omega \times \delta \Omega - \overline{\Omega}  \trace[ \Omega \times \Omega \, \delta \Omega] } \ee
and one computes that $\Omega$ extermizes $W_\alpha$ if and only if
\be e^{-2i\alpha} \det[ \overline{\Omega}] \scal{ 2 \Omega \times \mathpzc{D}Z + \overline{\Omega} ( Z - \det[\Omega] \trace [ \overline{\Omega} \mathpzc{D}Z])} + e^{2i\alpha} \scal{ \overline{\mathpzc{D}Z} - \det[ \Omega] \overline{\Omega}  \bar{Z}} = 0 \label{SingleCentre} \ee
Because this equation is $U(1) \times E_{6(-78)}$ invariant, one can solve it in any basis. There always exists an element $g \in E_{6(-78)}$ and a phase $e^{i \beta} \in U(1)$  such that $e^{3i\beta} Z$ and $e^{i\beta} g(\mathpzc{D}Z)  $ are in the non-standard diagonal form defined in \cite{BossardW} 
\bea e^{3i\beta} Z &=& e^{\frac{i\pi}{4}} ( e^{i\alpha} + i e^{-i\alpha}\sin(2\alpha))  \varrho   + e^{\frac{i\pi}{4}-i\alpha } ( \xi_1 + \xi_2 + \xi_3 )  \CR
e^{i\beta} g(\mathpzc{D}Z) &=&  e^{-\frac{i\pi}{4}} ( e^{-i\alpha} - i e^{i\alpha}\sin(2\alpha))  \varrho \,  \mathds{1} - e^{-\frac{i\pi}{4} + i \alpha} \left(\begin{array}{ccc} 
\hspace{0.0mm} \xi_1  \hspace{0.0mm}& \hspace{0.0mm} 0    \hspace{0.0mm}  & \hspace{0.0mm} 0      \hspace{0.0mm} \\
\hspace{0.0mm} 0   \hspace{0.0mm}& \hspace{0.0mm} \xi_2 \hspace{0.0mm}  & \hspace{0.0mm}  0   \hspace{0.0mm} \\
 \hspace{0.0mm} 0  \hspace{0.0mm} & \hspace{0.0mm} 0  \hspace{0.0mm}  & \hspace{0.0mm} \xi_3    \hspace{0.0mm} 
\end{array} \right) \label{nonSdiag} 
\eea
In this basis one computes that  (\ref{SingleCentre}) implies (up to a sign)
\be e^{i\beta} g (\Omega) = e^{-i\alpha - \frac{i\pi}{4}} \, \mathds{1} \ee
and using then (\ref{TrueFake}) that $W = 2 \varrho$. The 27 constraints (\ref{SingleCentre}) altogether with the positivity condition $W>0$ therefore determine $\Omega$ in terms of $Z$ and $\mathpzc{D}Z$, such that $W(\Omega)$ defines the single-centre non-BPS fake superpotential. 

Understanding the link between the potential $W_\alpha(\Omega,Z)$ away from its extremum value $\Omega_*(Z)$ is important in order to be able to generalise the triangular inequality for the central charge \cite{Denef} to show that the energy of a composite non-BPS bound state is inferior to the sum of the energies of its non-BPS constituents. If the extrema of $W_\alpha(\Omega,Z)$ were necessarily maximums, it would follows directly that this is indeed the case since 
\be W_\alpha(\Omega,Z_1+Z_2) = W_\alpha(\Omega,Z_1)  + W_\alpha(\Omega,Z_2) <   W_\alpha(\Omega_*(Z_1),Z_1)  + W_\alpha(\Omega_*(Z_2),Z_2) \ee
but the Hessian $\frac{ \partial^2 W_\alpha}{\partial \Omega \partial \Omega }\big|_{\Omega_*(Z)} $ can admit positive eigenvalues, and this identity is not trivially true. Nevertheless, there ought to be a similar property relying on the compatibility of the two constituents that would imply that this inequality holds.

 \subsubsection*{Extension to the interior of the solution}
 
We will now generalise these equations to describe the flow of the various fields allover space. For this we define the spatial components of the vector field strengths as the horizontal components with respect to the connexion $dt + \omega$ (with $I=1,\, 8$ of $SU(8)$)
\be  F_{IJ} = d w_{IJ} + \zeta_{IJ} d \omega \ee
and define the functions of the scalar fields defined as the central charges acting on these vectors. Within our specific Ansatz (\ref{VecScal}),
\be \begin{split} \zeta^0 &= - 4 e^{U} \frac{ \delta {\rm Re}[ W]}{\delta q_0} \\
\upzeta &= - 4e^{U} \frac{ \delta {\rm Re}[ W]}{\delta \mathpzc{Q}} 
\end{split}\qquad\begin{split}
\tilde{\zeta}_0 &= 4 e^{U}\frac{ \delta {\rm Re}[ W]}{\delta p^0}-  \frac{e^{3U}}{2} \sqrt{ 2i\det[ \mathpzc{t}-\overline{\mathpzc{t}}]} \, M \\
\tilde{\upzeta} &=  4 e^{U} \frac{ \delta {\rm Re}[ W]}{\delta  \mathpzc{P}} 
\end{split} \ee
Using a complex basis for the 56 charges in terms of 28 complex $Q^{IJ}$ and their complex conjugate $Q_{IJ}$, one will have in general
\be \zeta_{IJ} = - e^{U} i \frac{ \delta {\rm Re}[ W] }{\delta Q^{IJ}} -\frac{1}{2} e^{3U} \mathcal{R}_{IJ}  M \ee
with 
\be \mathcal{R}_{IJ} =  u^{ij}{}_{IJ} \frac{1}{2} \scal{ \omega_{ij} + \Omega_{ij} }  - v_{ijIJ} \frac{1}{2} \scal{ \omega^{ij} + \Omega^{ij}} \ee
where we used the standard notation of \cite{NicolaideWit} for the scalar fields such that 
\be Z_{ij} = u_{ij}{}^{IJ} Q_{IJ} + v_{ijIJ} Q^{IJ} \ee
In the single centre case, $\mathcal{R}_{IJ}$ can be identified up to a renormalisation with the small vector appearing in \cite{Galli:2010mg} in the framework of $\N=2$ supergravity. Note that because the duality group acts linearly on the field strength, the direction in $\mathds{R}^{56}$ defined by the tensor $\mathcal{R}_{IJ}$ will always be a constant, as it is in the specific duality frame of the last section. Moreover, (\ref{ConstraintOmega}) implies that $\mathcal{R}_{IJ}$ satisfies the constraints of a 1/2 BPS charge, and its stabiliser subgroup of $E_{7(7)}$ is  \cite{FerraraOrbit}
\be  E_{6(6)} \ltimes  \mathds{R}^{27} \subset E_{7(7)} \ee
The simplicity of the solution we described in the last section comes from the property that we parametrized the moduli in $E_{7(7)} / \SU$ in terms of the specific parabolic subgroup $( \mathds{R}_+^* \times E_{6(6)}) \ltimes  \mathds{R}^{27} $ which coincides with this stabiliser, such that the action of the scalars on $ \mathcal{R}_{IJ} $ amounted to a rescaling by the dilaton and 
\be \frac{1}{2} \scal{ \omega_{ij} + \Omega_{ij} } = u_{ij}{}^{IJ} \mathcal{R}_{IJ} + v_{ijIJ} \mathcal{R}^{IJ} \label{OmegaR} \ee 
was constant. In this way the defining tensors $\omega_{ij}$ and $\Omega_{ij}$ are determined in terms of the scalar fields and the specific direction in $\mathds{R}^{56}$ determined by the small charge $\mathcal{R}_{IJ}$ of unit mass, noting that its norm is fixed by 
\be \frac{1}{2}  Z(\mathcal{R})_{ij} Z(\mathcal{R})^{ij} = 1 \label{NormR} \ee 
It is clear from (\ref{OmegaR}) that in any other duality frame,  $\omega_{ij}$ and $\Omega_{ij}$ will flow allover space accordingly. 

\vskip 2mm

Using some identities valid when $ \upsilon^\dagger \upsilon$ is a Hermitian matrix over the quaternions (no $\ell$ component)\footnote{To prove the second identity, one uses associativity to cancel the $\upsilon$ factors on the left-hand-side, and one computes for the right-hand-side, using $\mathpzc{b} \equiv \upsilon^\dagger \upsilon$ for short
\bea \frac{ \mathpzc{b} \times \mathpzc{b}}{\det[\mathpzc{b}] }( \ell \invo (\mathpzc{b} \times \mathpzc{b}) d Y ) \frac{ \mathpzc{b} \times \mathpzc{b}}{\det[\mathpzc{b}] } &=& - 2 \scal{ \mathpzc{b} \times ( \ell \invo (\mathpzc{b} \times \mathpzc{b}) d Y )} \frac{ \mathpzc{b} (\mathpzc{b} \times \mathpzc{b})} {\det[\mathpzc{b}]^2 }
= - 2 \frac{\mathpzc{b} \times ( \ell \invo ( \mathpzc{b} \times \mathpzc{b}) d Y )}{\det[\mathpzc{b}]} \CR
&=& \frac{ \ell \invo \mathpzc{b} (\mathpzc{b} \times \mathpzc{b} )d Y }{\det[\mathpzc{b}]} 
= \ell \invo d Y 
  \eea}
\bea \det[ \upsilon^\dagger \upsilon ] : \upsilon^{\dagger -1} d \mathcal{K} \upsilon^{-1} :  & = & \trace [ (\upsilon^\dagger \upsilon)\times (\upsilon^\dagger \upsilon) d \mathcal{K} ]\,  \mathds{1} -  2 : \upsilon \scal{  (\upsilon^\dagger \upsilon) \times d \mathcal{K} } \upsilon^\dagger : \CR
: \upsilon \scal{ \ell \invo dY} \upsilon^\dagger : &=& : \upsilon^{\dagger -1} \scal{ \ell \invo ( \upsilon \upsilon^\dagger ) \times  ( \upsilon \upsilon^\dagger ) dY } \upsilon^{-1} \eea
one computes
\be e^U Z(\star F) = \frac{1}{2} \Scal{ d U - \frac{i}{2} e^{2U} \star d \omega } - \frac{i}{4} \trace [ \upsilon^{\dagger -1} d \mathpzc{t} \upsilon^{-1} ] \ee
and using the definition $\mathpzc{t} = \mathpzc{t}_\un + \ell \mathpzc{t}_\deux$, that 
\be e^U \mathpzc{D}Z(\star F) = - \frac{1}{2} \Scal{ d U - \frac{i}{2} e^{2U} \star d \omega +  \frac{i}{2} \trace [ \upsilon^{\dagger -1} d \mathpzc{t} \upsilon ] } \, \mathds{1} + \frac{i}{2} \rho(\varsigma)  : \upsilon^{\dagger -1}   d ( \mathpzc{t}_\un - \ell \mathpzc{t}_\deux)  \upsilon^{-1}   : \ee
Therefore it follows that
 \be dU  - \frac{i}{2} e^{2U} \star d \omega  = \frac{1}{2} e^U \scal{ Z( \star F) - \trace [ \mathpzc{D}Z( \star F)]}  \ee
which generalises (\ref{ADMnonBPS}). Similarly, the scalar fields momenta are defined as 
\be D \mathpzc{t}_\un -\ell D  \mathpzc{t}_\deux = - 2 i e^U \mathpzc{D}Z(\star F) + i e^U \scal{ Z( \star F) + \trace [ \mathpzc{D}Z( \star F)]}  \, \mathds{1} \label{PartiScal} \ee
Note that this equation shows that each split octonion component `$\mathpzc{x}_i + \ell \mathpzc{y}_i $' of $-2i e^U \mathpzc{D}Z(\star F)$ is equal to the `$\ell$ conjugate' of the corresponding split octonion component `$\mathpzc{x}_i - \ell \mathpzc{y}_i $' of $D \mathpzc{t}_\un +\ell D  \mathpzc{t}_\deux$. This implies that their respective contribution to the equations of motion of the scalar $\varsigma^I$ cancel precisely such that the latter are indeed constant. 

The property that $\mbox{Re}[\mathpzc{D}Z(\star F)]$ is defined as a 3 by 3 Hermitian matrix over the quaternions (\ie that its $\ell$ component vanishes) is also valid allover space. We conclude therefore that 
  \be \omega_{ik} \Omega_{jl} \scal{ Z^{kl}(F)  + \omega^{kp} \Omega^{lq} Z_{pq}(F) } = 0 \label{FirstOrder} \ee
also generalises allover space, such that in general one would have
 \be dU  - \frac{i}{2} e^{2U} \star d \omega = - \frac{1}{2} e^U \scal{ \Omega_{ij} Z^{ij}(\star F) - \omega^{ij} Z_{ij}(\star F) } \label{FirstOrderMass}  \ee
To compute the expression of the scalar fields momenta, we will use the property that the  $\e_{(8)} \ominus \so^*(16)$ momenta $P$ is a nilpotent element characterised by a generator ${\bf h}$ of $\so^*(16)$, which can be expressed in a Cartan basis  as \DSOXVI00000200. Defining the chiral $\Spin^*(16)$ spinor $P$ in terms of fermionic oscillators as in \cite{Bossard:2009at} 
\begin{multline}  |P\rangle \equiv    \biggl(  dU + \frac{i}{2} e^{2U} \star d \omega + e^U Z_{ij}(\star F) a^i a^j + \frac{1}{12}   \scal{ u_{ij}{}^{IJ} d v_{klIJ} - v_{ijIJ} d u_{kl}{}^{IJ} } a^i a^j a^k a^l  \biggr . 
\\ \biggl . + \frac{1}{6!} \varepsilon_{ijklpqrs} e^U Z^{ij}(\star F) a^k a^l a^p a^q a^r a^s + \frac{1}{8!} \varepsilon_{ijklpqrs}\Scal{ dU - \frac{i}{2}  e^{2U} \star d \omega}  a^i a^j  a^k a^l  a^p a^q  \biggr)  |0\rangle \end{multline}
 one obtains that ${\bf h}$ can be chosen as
 \be {\bf h} = - \frac{1}{2} \Omega_{ij} a^i a^j + \frac{1}{2} \Omega^{ij} a_i a_j \ee
 such that $|P\rangle = |P^\ord{1} \rangle+  |P^\ord{2}\rangle+  |P^\ord{3}\rangle$ decomposes according to (\ref{128Decom}) and 
 \be {\bf h}\,  |P\rangle =  |P^\ord{1}\rangle +  2 |P^\ord{2}\rangle+  3 |P^\ord{3}\rangle \ee
 Or, alternatively
 \be ( {\bf h} - 1 ) ( {\bf h}-2 ) ( {\bf h} - 3) |P \rangle = 0 \ee
 Solving this equation one gets back (\ref{FirstOrder},\ref{FirstOrderMass}), and moreover that the scalar fields momenta satisfy
 \begin{multline}   u_{ij}{}^{IJ} d v_{klIJ} - v_{ijIJ} d u_{kl}{}^{IJ} = 6 e^{U} \Bigl(  \omega_{[ij} \Omega_{k|p} \Omega_{l]q} Z^{pq}(\star F) - \Omega_{[ij} Z_{kl]}(\star F) \Bigr . \\ \hspace{36mm}  - \frac{1}{4} \omega_{[ij} \Omega_{kl]} \scal{ \Omega_{pq} Z^{pq}(\star F) - \omega^{pq} Z_{pq}(\star F)}  \\
\Bigl .  + \frac{1}{8} \Omega_{[ij} \Omega_{kl]}  \scal{ \Omega^{pq} Z_{pq}(\star F) - \omega_{pq} Z^{pq}(\star F)}  \Bigr) \end{multline}
This expression is complex self-dual thanks to (\ref{FirstOrder}).\footnote{One computes for instance that
\begin{multline} \frac{1}{24} \varepsilon_{ijklmnpq} \scal{ \omega^{mn} \Omega^{pr} \Omega^{qs} Z_{rs} - \Omega^{mn} Z^{pq}} = \omega_{[ij} \Omega_{k|p} \Omega_{l]q} Z^{pq}  - \Omega_{[ij} Z_{kl]} - \frac{1}{2} \omega_{[ij}   \Omega_{kl]} \scal{ \Omega_{pq} Z^{pq}- \omega^{pq} Z_{pq}} \\ + \frac{1}{4} \Omega_{[ij} \Omega_{kl]}  \scal{ \Omega^{pq} Z_{pq} - \omega_{pq} Z^{pq}} + 2 \Omega_{[ij} \omega_{k|p} \Omega_{l]q} \scal{ Z^{pq} + \omega^{pm} \Omega^{qn} Z_{mn}} \end{multline}}
 The component in the $({\bf 2}\otimes{\bf 14}_\trois)_{\mathds{R}}$ of $Sp(1)\times Sp(3)$ vanishes, exhibiting `flat directions' parametrizing the quaternionic space $F_{4(4)} / ( Sp(1) \times_{\mathds{Z}_2} Sp(3))$.  Within our Ansatz $\omega_{ij}$ and $\Omega_{ij}$ are constant allover the flow, as opposed to the phase $\alpha$ which appears in the BPS system as the trivialisation of the `modified K\"{a}hler connexion' \cite{Denef,Bates}. However, it is clear from equation (\ref{OmegaR}) that this is an artefact of the specifically simple duality frame we chose to define the solution.

This equation directly implies that there is always one trivial vector field in the system. Using the imaginary part of (\ref{FirstOrder}), one obtains 
\bea - i  e^U d\omega &=& \frac{1}{2} \scal{ \omega^{ij} + \Omega^{ij} } Z_{ij}(F) - \frac{1}{2}  \scal{ \omega_{ij} + \Omega_{ij} } Z^{ij}(F)\CR
&=& \mathcal{R}^{IJ} F_{IJ} - \mathcal{R}_{IJ} F^{IJ} \eea
and therefore the vector field in the direction Darboux conjugate to $\mathcal{R}_{IJ}$ is vertical, \ie in $\zeta ( dt + \omega)$  only.

 \subsection{Rotating attractor}

 Let us consider the expression of the momentum $P$ in the near horizon region. One computes that 
 \be \Scal{  dU - \frac{i}{2}  e^{2U} \star d \omega}\Big|_{x_\pA}   = \Scal{ 1 - \frac{ i \J_\pA \cos(\theta_\pA)}{ \sqrt{ - I_4 - \J^2_\pA \cos^2 (\theta_\pA) }}} \Scal{ d \mbox{ln}(r) - \frac{i}{2} \frac{ \J_\pA \sin(\theta_\pA) d \theta_\pA }{\sqrt{ - I_4 - \J^2_\pA \cos^2 (\theta_\pA) }}} \ee
 and similarly that 
 \be  \Scal{ \rho(\varsigma)  : \upsilon^{\dagger -1}   d ( \mathpzc{t}_\un - \ell \mathpzc{t}_\deux)  \upsilon^{-1}   : }\Big|_{x_\pA}  = - \Scal{ 1 - \frac{ i \J_\pA \cos(\theta_\pA)}{ \sqrt{ - I_4 - \J^2_\pA \cos^2 (\theta_\pA) }}}  \, \frac{ \J_\pA \sin(\theta_\pA) d \theta_\pA }{\sqrt{ - I_4 - \J^2_\pA \cos^2 (\theta_\pA) }} \mathds{1} \ee
 Using these expressions one gets 
 \bea \Scal{ e^U Z(\star F) }\Big|_{x_\pA}  &=& \frac{1}{2} \Scal{ 1 - \frac{ i \J_\pA \cos(\theta_\pA)}{ \sqrt{ - I_4 - \J^2_\pA \cos^2 (\theta_\pA) }}} \Scal{ d \mbox{ln}(r) + i  \frac{ \J_\pA \sin(\theta_\pA) d \theta_\pA }{\sqrt{ - I_4 - \J^2_\pA \cos^2 (\theta_\pA) }}}\CR
 \Scal{ e^U \mathpzc{D}Z(\star F)}\Big|_{x_\pA}  &=& \frac{1}{2} \Scal{ 1 - \frac{ i \J_\pA \cos(\theta_\pA)}{ \sqrt{ - I_4 - \J^2_\pA \cos^2 (\theta_\pA) }}} \Scal{ -d \mbox{ln}(r) + i  \frac{ \J_\pA \sin(\theta_\pA) d \theta_\pA }{\sqrt{ - I_4 - \J^2_\pA \cos^2 (\theta_\pA) }}} \mathds{1} \CR
  \eea
 Let us introduce the following notation for simplicity 
 \be\begin{split}  {\rm w} &\equiv \Scal{  dU +  \frac{i}{2}  e^{2U} \star d \omega}\Big|_{x_\pA} \\
 \Sigma &\equiv \frac{i}{2} \Scal{ \rho(\varsigma)  : \upsilon^{\dagger -1}   d ( \mathpzc{t}_\un - \ell \mathpzc{t}_\deux)  \upsilon^{-1}   : }\Big|_{x_\pA}  
 \end{split}\qquad \begin{split} 
 Z &\equiv \Scal{ e^U Z(\star F) }\Big|_{x_\pA} \\
 \mathpzc{D}Z &\equiv \Scal{  e^U \mathpzc{D}Z(\star F)}\Big|_{x_\pA} 
 \end{split}\ee
We define the generator ${\bf h}$ (which can be expressed in a Cartan basis  as \DSOXVI00000020) of $\so^*(16)$ 
\bea {\bf h } \; {\rm w} &=& e^{-2i\varpi} Z - \trace [ \overline{  \mathpzc{D}Z}] \CR
{\bf h} \; Z &=&  e^{2i\varpi} {\rm w} - \trace [ \Sigma]   \CR
{\bf h} \;  \overline{  \mathpzc{D}Z} &=&  e^{-2i \varpi} \Sigma -   {\rm w}\, \mathds{1}   + \overline{\Sigma} - \mathds{1} \trace [ \overline{\Sigma}]  \CR
{\bf h}\;  \Sigma  &=&  e^{2i \varpi} \overline{  \mathpzc{D}Z} -   Z\, \mathds{1}   +   \mathpzc{D}Z - \mathds{1} \trace [{  \mathpzc{D}Z}]  \eea
which according to \cite{BossardW} characterises the nilpotent orbit associated to non-BPS extremal black holes. It decomposes the coset component of $\e_{8(8)}$ into representations of $SU^*(8)$ as
\be {\bf 128} \cong {\bf 1}^\ord{-2} \oplus \overline{\bf 28}^\ord{-1} \oplus {\bf 70}^\ord{0} \oplus {\bf 28}^\ord{1} \oplus {\bf 1}^\ord{2} \ee 
Choosing the phase as 
 \be e^{i\varpi} = \frac{ \sqrt{ - I_4 - \J^2_\pA \cos^2 (\theta_\pA) } - i \J_\pA \cos(\theta_\pA) }{\sqrt{ - I_4}} \ee
 one computes that 
 \be {\bf h} \; P_r|_{x_\pA} = 2 P_r|_{x_\pA}\; , \hspace{10mm} {\bf h} \; P_\theta|_{x_\pA} = 4 P_\theta|_{x_\pA}\; \ee
 It appears therefore that $P_r|_{x_\pA}$ lies in the nilpotent orbit associated to a non-BPS extremal black hole, and $P_\theta|_{x_\pA}$ lies in the minimal nilpotent orbit. The outgoing momentum $P_r|_{x_\pA}$ exhibits that the scalar fields take values determined by the local central charges. The pull backs of the field strengths on the horizon define the charges of the black hole, such that 
 \be F_{IJ}|_{x_\pA} = Q_{IJ} \sin(\theta_\pA) d\theta_\pA \wedge d\varphi_\pA  \ee
It follows that the scalar fields (\ref{AttractorNBPS}) at $\theta_\pA$ satisfy the attractor  equation 
\be  e^{2i \varpi} \overline{  \mathpzc{D}Z} -   Z\, \mathds{1}   +   \mathpzc{D}Z - \mathds{1} \trace [{  \mathpzc{D}Z}]= 0 \ee
 which reproduces the standard spherically symmetric non-BPS attractor equation, only at $\varpi=0$. In general it is associated to a symplectic matrix $\Upomega_{ij}$ of $SU(8) / Sp(4)$ such that 
 \be \Upomega_{[ij} \Upomega_{kl]}  +  \frac{e^{2i\varpi}}{24} \varepsilon_{ijklmnpq} \Upomega^{mn} \Upomega^{pq} = 0 \ee
 and the attractor equation reads 
 \be \Upomega_{[ij} Z_{kl]} + \frac{1}{24} \varepsilon_{ijklmnpq} \Upomega^{mn} Z^{pq} = 0 \ee
 The equation transforms covariantly with respect to the $U(1)$ Ehlers symmetry, such that this extra phase is directly related to the phase of ${\rm w}$.   In our case 
 \be \Upomega_{ij} = e^{2i\varpi} \omega_{ij} - \Omega_{ij} \ee

\subsection{Non-BPS interactions}

To begin let us argue that the asymptotic tensors $\omega_{ij}$ and $\Omega_{ij}$ associated to a solution can be determined in terms of the asymptotic moduli and the charges if the solution includes enough many black holes.  In this subsection the moduli and associated central charges (as function of the scalars) will always be understood to be evaluated in the asymptotic region. 

The set of charges $Q_A^{IJ}$ carried by the black holes will span a vector space $\mathds{R}^{43} \subset \mathds{R}^{56}$ which includes a Lagrangian subspace associated to the invariant symplectic form. We choose a Darboux basis associated to this Lagrangian, such that one can then define the orthogonal complement $\mathds{R}^{13}\subset \mathds{R}^{56}$. In our duality frame, these charges are simply the ones admitting only as non-zero components $p^0$ and a 3-vector of quaternions $p$. One can single out the $p^0$ charge in an invariant way by using the property that the second derivative of the quartic invariant restricted to the adjoint representation evaluated for such charges is only zero if the 3-vector of quaternions $p$ is null 
\be \frac{\partial^2 I_4}{\partial Q \partial Q}\Big|_{\bf 133} \propto p p^\dagger \ee 
The condition $\frac{\partial^2 I_4}{\partial Q \partial Q}\Big|_{\bf 133}(\mathcal{R})=0$ is required for the vector $ \mathcal{R}_{IJ}$ to be small \cite{Ferrara:1997ci}.
The singled out direction determines therefore the vector $\mathcal{R}_{IJ}$, and then the combination $\omega_{ij} + \Omega_{ij}$ through (\ref{OmegaR}, \ref{NormR}). The obtained symplectic form  $\omega_{ij} + \Omega_{ij}$ decomposes in a single way into a sum of orthonormal rank 2 antisymmetric tensors as
\be \omega_{ij} + \Omega_{ij} = \sum_{\Lambda = 0 }^3 \omega_{ij}^\ord{\Lambda} \ee
Then using equation (\ref{FirstOrder}) on all the charges will determine $\omega$ as
  \be \omega_{ij} = \omega_{ij}^\ord{\Lambda} \, \qquad |\quad  \forall\,  Q_A \qquad \sum_{\Xi\ne \Lambda} \omega^\ord{\Lambda}_{ik} \omega^\ord{\Xi}_{jl} \scal{ Z^{kl}(Q_A)  + \omega_\ord{\Lambda}^{kp} \omega_\ord{\Xi}^{lq} Z_{pq}(Q_A) } = 0  \ee
This way one determines $\omega_{ij}$ and $\Omega_{ij}$ in terms of the charges $Q_A^{IJ}$ and the asymptotic moduli. We will therefore consider them as functions of the charges and the asymptotic moduli which transform as the central charges $Z_{ij}$ with respect to $E_{7(7)}$ transformations. 

\vskip 2mm
 
 It will turn out to be useful to define the real tensor in the  ${\bf 1}\oplus {\bf 14}$  of $Sp(3)$ 
 \be W_{ij} \equiv \frac{1}{2} \Omega_{ik} \Omega_{jl} \scal{ Z^{kl} + \Omega^{kp} \Omega^{lq} Z_{pq} } - \frac{1}{2} \Omega_{ij} {\rm Re}[ \omega^{kl} Z_{kl}] \ee
 This tensor is described within our Ansatz by the Jordan algebra element 
 \bea \mathpzc{W}(q,p) &\equiv&  {\rm Re}[ \mathpzc{D}Z(q,p) - Z(q,p)  \mathds{1}]\CR
&=&  \frac{1}{2\det[\mathpzc{l}]} \Scal{  \trace[( \mathpzc{l} \times \mathpzc{l} )( \mathpzc{P} - \mathpzc{a} p^0 )] \, \mathds{1} -   \rho(\varsigma) :\upsilon^{\dagger -1} ( \mathpzc{P} - \mathpzc{a} p^0) \,  \upsilon^{-1} :}
\CR
&=&  \frac{1}{2\det[\mathpzc{l}]} \Scal{  \trace[( \mathpzc{l} \times \mathpzc{l} ) \mathpzc{p} ] \, \mathds{1} -   \rho(\varsigma) \upsilon^{\dagger -1} \mathpzc{p}  \,  \upsilon^{-1} }
  \eea
 which is a Hermitian matrix over the quaternions because of (\ref{ConstraintReal}). 

It is now time to have a second look at equation (\ref{ChargeContraintes}). First of all let us note that assuming that $\det[\mathpzc{p}_A] \ne 0 $, we can inverse this relation to
\be \gamma_A \mathpzc{l}  + \sum_{B\ne A} ( \gamma_A - \gamma_B ) \frac{ \mathpzc{p}_B}{|x_{\scriptscriptstyle B}-x_{\scriptscriptstyle A}|} = \frac{ 2 ( \mathpzc{p}_A \times \mathpzc{p}_A) \times \mathpzc{q}_A - \frac{1}{2} \trace[   \mathpzc{p}_A \mathpzc{q}_A]  \,  \mathpzc{p}_A}{\det[  \mathpzc{p}_A] } +  \mathpzc{k} \ee
Then we will contract this equation with $\mathpzc{l}\times\mathpzc{l}$, and recognise each component as being writable in terms of the tensor function $ \mathpzc{W}$ of the moduli. Indeed, one can define the following invariants by using the property that $\trace [ \mathpzc{W}^n ]$ is reproduced in general by the trace of $W_{ik} \Omega^{jk} $ to the power $n$
\bea \frac{\trace[ ( \mathpzc{l} \times \mathpzc{l} )\, \mathpzc{p}_A]}{\det[  \mathpzc{l}]} &=& \trace[ \mathpzc{W}(q_A,p_A) ] \\
 \frac{\trace[ \mathpzc{l} ( \mathpzc{p}_A \times \mathpzc{p}_B )]}{\det[  \mathpzc{l}]} &=&  \trace[ \mathpzc{W}(q_A,p_A) ] \trace[ \mathpzc{W}(q_B,p_B) ] - 2 \trace[\mathpzc{W}(q_A,p_A)\mathpzc{W}(q_B,p_B)] \CR
 \frac{\det[\mathpzc{p}_A]}{\det[  \mathpzc{l}]} &=& 8  \det[ \mathpzc{W}(q_A,p_A) ] + 4   \trace[ \mathpzc{W}(q_A,p_A) ]\trace[ \mathpzc{W}(q_A,p_A)^2 ] -  \trace[ \mathpzc{W}(q_A,p_A) ]^3 \nn
 \eea
 For convenience we will rename these invariants as 
  \bea W_A &\equiv& \frac{\trace[ ( \mathpzc{l} \times \mathpzc{l} )\, \mathpzc{p}_A]}{\det[  \mathpzc{l}]} \CR
 W_{AB} &\equiv&  \frac{\trace[ \mathpzc{l} ( \mathpzc{p}_A \times \mathpzc{p}_B )]}{\det[  \mathpzc{l}]} \CR
 W_{ABC} & \equiv&  \frac{\trace[\mathpzc{p}_A\times\mathpzc{p}_B\, \mathpzc{p}_C]}{\det[  \mathpzc{l}]} \eea
Using the quartic $E_{7(7)}$ invariant $I_4(q,p)$, we can also define the invariant quantity
 \bea J_A &\equiv& - \frac{ \trace\Bigl[ \mathpzc{W}\scal{\frac{\partial I_4}{\partial p_A} , -\frac{\partial I_4}{\partial q_A} }\Bigr] }{W_{AAA}}\\
 & =& \frac{1}{\det[  \mathpzc{p}_A]}  \trace\Bigl[ ( \mathpzc{l} \times \mathpzc{l} )\, \Scal{ 2 ( \mathpzc{p}_A \times \mathpzc{p}_A) \times \mathpzc{q}_A - \frac{1}{2} \trace[   \mathpzc{p}_A \mathpzc{q}_A]  \,  \mathpzc{p}_A + \det[\mathpzc{p}_A] \Scal{ \mathpzc{k} -\frac{ m \mathpzc{l}}{ \det[\mathpzc{l}]} }} \Bigr] \nn \eea
 Finally, we can write the equations 
 \be 3  \Scal{ \gamma_A - \frac{m}{\det[\mathpzc{l}]}} + \sum_{B\ne A} ( \gamma_A - \gamma_B) \frac{W_B }{|x_\pB - x_\pA|} =  \frac{ J_A }{\det[\mathpzc{l}]} \ee
 which permit to determine all $\gamma_A$ coefficients. It turns out that $\gamma_A$ are not covariant quantities, but $\det[\mathpzc{l}] \gamma_A-m$ are. One can define the $N$ by $N$ matrix 
 \be M_{AB} = \left( \begin{array}{cccc} 3 + \sum_{C\ne 1}  \frac{W_C}{|x_\pC - x_1|} & - \frac{W_2}{|x_2 - x_1|}& -\frac{W_3}{|x_3 - x_1|} & \cdots \\
 -\frac{W_1}{|x_1 - x_2|}&3 + \sum_{C\ne 2}  \frac{W_C}{|x_\pC - x_2|}& - \frac{W_3}{|x_3 - x_2|}&\cdots \vspace{2mm} \\
 \vdots & \vdots & \vdots & \end{array}\right) \ee
 and solve 
 \be \det[\mathpzc{l}] \gamma_A-m = \sum_B M^{-1}_{AB} J_B \label{gammaAsol}\ee

 We will now consider the symplectic products of (\ref{ChargeContraintes}) with $\mathpzc{p}_B$, which give
 \begin{multline}  ( \gamma_A - \gamma_B ) \Scal{ \trace [ \mathpzc{l} (  \mathpzc{p}_A \times \mathpzc{p}_B )] + \frac{\trace [ \mathpzc{p}_A (  \mathpzc{p}_B \times \mathpzc{p}_B )] +  \trace [ \mathpzc{p}_B (  \mathpzc{p}_A \times \mathpzc{p}_A )]  }{|x_\pA - x_\pB|}}\\ + \sum_{C\ne A,B} \biggl( ( \gamma_A - \gamma_C) \frac{ \trace[   \mathpzc{p}_A (  \mathpzc{p}_B \times \mathpzc{p}_C )]}{|x_\pA-x_\pC|} -  ( \gamma_B - \gamma_C) \frac{ \trace[   \mathpzc{p}_A (  \mathpzc{p}_B \times \mathpzc{p}_C )]}{|x_\pB-x_\pC|} \biggr) = \frac{1}{2} \trace[ \mathpzc{q}_A \mathpzc{p}_B - \mathpzc{q}_B \mathpzc{p}_A ]  \end{multline} 
 and can be rewritten in general
  \begin{multline}  ( \gamma_A - \gamma_B ) \Scal{W_{AB} + \frac{W_{ABB} + W_{BAA}   }{|x_\pA - x_\pB|}}\\ + \sum_{C\ne A,B} \biggl( ( \gamma_A - \gamma_C) \frac{W_{ABC} }{|x_\pA-x_\pC|} -  ( \gamma_B - \gamma_C) \frac{W_{ABC} }{|x_\pB-x_\pC|} \biggr) =\frac{ \langle Q_\pA,Q_\pB\rangle }{\det[\mathpzc{l}]}  \label{SymplecticConstraints} \end{multline} 
  where we used for short 
  \be \langle Q_\pA ,Q_\pB \rangle =  \frac{1}{2} \scal{ q_{0 A} p^0_B - q_{0B} p^0_A + \trace[ \mathpzc{Q}_{\; A} \mathpzc{P}_B - \mathpzc{Q}_{\; B} \mathpzc{P}_A ] } \ee
Considering $\gamma_A$ to be determined as (\ref{gammaAsol}), one has $26N$ remaining equations in (\ref{ChargeContraintes}) for a solution with $N$ centres. Assuming all $\mathpzc{p}_A$ to be linearly independent, equations (\ref{SymplecticConstraints})  provide $\inf[ N(N-1)/2 , 26N]$  covariant equations which constrain the centres in terms of the angles $\omega_{ij}, \Omega_{ij}$. 
 
 Note that, as opposed to the BPS case, one obtains $\inf[ N(N-1)/2 , 26N] \ge 3N-6 + \delta_{N,2}$ equations, and not only $N-1$, such that the system does not even admit trivially a solution for $N\ge 10$. In general these equations will give $\inf[ N(N-1)/2 , 26N] $ polynomial equations of degree $N-1$ for the $N$ centres. This is because the leading terms of order $N$ in the inverse radius all cancel. In the two centres case one gets 
 \be |x_1-x_2| = \frac{ \langle Q_1,Q_2 \rangle ( W_1 + W_2 ) + ( J_2 - J_1 ) ( W_{211} + W_{122} )}{( J_1-J_2) W_{12} - 3 \langle Q_1,Q_2 \rangle } \ee
 In the three centres case one can find the distance separations as real roots of quartic polynomials. Therefore they admit closed form formula in principle, although these expressions are not really illuminating and we will not display them.

 In general (for more than three centres which are not on the same axe, or more than four which are not on the same plan), equation (\ref{ChargeContraintes}) determines $15 N $ charges in terms of $4N-6$ parameters, and therefore it is clear that the moduli space of such composite black hole solutions is only defined for a much smaller submanifold of allowed charges. In the axisymmetric situation, $15N$ charges are determined in terms of $2N-1$ parameters (respectively $3N-3$ for the plan). It is therefore clear that considering the largest possible landscape of allowed charges requires to consider non-axisymmetric configurations, as opposed to the BPS case for which one could always rotate a configuration of charged black holes to an axisymmetric one.  
  
 This property is confirmed by the appearance of a new contribution to the total angular momentum, when at least three centres are not aligned. In such a case, the charge factor reads 
 \bea && \gamma_C \trace [ \mathpzc{p}_A\times   \mathpzc{p}_B \mathpzc{p}_C] - \frac{m}{\det[\mathpzc{l}]} \trace [ \mathpzc{p}_A\times   \mathpzc{p}_B \mathpzc{p}_C]  \CR
  &=&   \biggl( -3  \trace[  \mathpzc{W}(q_A,p_A) ] \trace[  \mathpzc{W}(q_B,p_B) ]\trace[  \mathpzc{W}(q_C,p_C) ] \biggr . \CR && \hspace{5mm}+4  \trace[ \mathpzc{W}(q_A,p_A) ]\trace[ \mathpzc{W}(q_B,p_B)\mathpzc{W}(q_C,p_C) ]  \CR &&\hspace{7mm} +4  \trace[ \mathpzc{W}(q_B,p_B) ]\trace[ \mathpzc{W}(q_C,p_C)\mathpzc{W}(q_A,p_A) ] \CR &&\hspace{9mm}  +4 \trace[ \mathpzc{W}(q_C,p_C) ]\trace[ \mathpzc{W}(q_A,p_A)\mathpzc{W}(q_B,p_B) ] \CR && \hspace{17mm}  + 8 \trace[ \mathpzc{W}(q_A,p_A)\mathpzc{W}(q_B,p_B)\mathpzc{W}(q_C,p_C) ]\biggr)  \sum_D M^{-1}_{CD} J(q_D,p_D) \eea
and because the second term does not contribute by symmetry, we obtain that the contribution to the angular momentum is a duality invariant function of the charges and the asymptotic moduli.

In principle one would expect the expression of the angular momentum to not only be a duality invariant, but to moreover be independent of the asymptotic moduli such that it could take an integer valued expression for arbitrary moduli. However, given the complexity of the complete expression of (\ref{NonBPSJ}) in terms of the charges and the asymptotic moduli, and the property that the charges are themselves constrained, it is very hard to prove that this is indeed the case. The first step would be to actually understand clearly the constraints on the charges for the existence of a solution in terms of duality invariant equations. On the other hand, note that whenever one gets such contribution to the angular momentum, the different contributions to the angular momentum are then not aligned by definition, and there is no reason for the sum of the standard contributions to give rise to an integer valued total angular momentum anyway.  

Let us note nevertheless the following observation. We define 
\be \mathpzc{d}_A \equiv \gamma_A \mathpzc{l}  + \sum_{B\ne A} ( \gamma_A - \gamma_B ) \frac{ \mathpzc{p}_B}{|x_{\scriptscriptstyle B}-x_{\scriptscriptstyle A}|}  \ee
such that (\ref{ChargeContraintes}) reads simply $\mathpzc{q}_A = 2 \mathpzc{p}_A \times \mathpzc{d}_A$. For generic charges $\mathpzc{p}_A$, we know that the interactions disappear if and only if all $\gamma_A$ are equal. Then one has  $\mathpzc{d}_A = \mathpzc{d}_B $ since 
\be \mathpzc{d}_A - \mathpzc{d}_B  =(  \gamma_A - \gamma_B ) \Scal{ \mathpzc{l} + \frac{  \mathpzc{p}_A + \mathpzc{p}_B}{ |x_{\scriptscriptstyle B}-x_{\scriptscriptstyle A}|}} + \sum_{C\ne A,B} \Scal{ ( \gamma_A - \gamma_C ) \frac{  \mathpzc{p}_C}{ |x_{\scriptscriptstyle C}-x_{\scriptscriptstyle A}|} -  ( \gamma_B - \gamma_C ) \frac{  \mathpzc{p}_C}{ |x_{\scriptscriptstyle C}-x_{\scriptscriptstyle B}|}}  \ee
One checks straightforwardly that this implies that the charges are mutually local 
\be \langle Q_\pA , Q_\pB \rangle = \trace[ \mathpzc{p}_A \times \mathpzc{p}_B ( \mathpzc{d}_A - \mathpzc{d}_B) ] \ee
which is the standard property for non-interacting black holes. Moreover, one can check that the sextic invariant proposed in \cite{Andrianopoli:2011gy,Ferrara:2011di} to describe interacting solutions also vanishes 
\begin{multline}  \Bigl\langle \frac{ \partial I_4}{\partial Q_\pA}  , \frac{ \partial I_4 }{\partial Q_\pB} \Bigr\rangle =  8 \det[ \mathpzc{p}_A] \det[ \mathpzc{p}_B]  \scal{ \det[ \mathpzc{d}_A] -  \det[ \mathpzc{d}_B] } + 2 \det[ \mathpzc{p}_B] \scal{ q_A^\dagger ( \mathpzc{p}_A \times ( \mathpzc{d}_A - \mathpzc{d}_B)) q_A} \\ + 2 \det[ \mathpzc{p}_A] \scal{ q_B^\dagger ( \mathpzc{p}_B \times ( \mathpzc{d}_A - \mathpzc{d}_B)) q_B} 
- \scal{ q_{0B} \det[ \mathpzc{p}_A] + q_{0A}  \det[ \mathpzc{p}_B]}  \trace [  \mathpzc{p}_A \times \mathpzc{p}_B ( \mathpzc{d}_A-\mathpzc{d}_B)]   \\- 2 \scal{ q_B^\dagger  ( \mathpzc{p}_B \times \mathpzc{p}_B) ( \mathpzc{d}_A-\mathpzc{d}_B)  ( \mathpzc{p}_A \times \mathpzc{p}_A) q_A } - 2 \scal{ q_A^\dagger  ( \mathpzc{p}_A \times \mathpzc{p}_A) ( \mathpzc{d}_A-\mathpzc{d}_B)  ( \mathpzc{p}_B \times \mathpzc{p}_B) q_B } \end{multline} 
when $\mathpzc{d}_A = \mathpzc{d}_B $. This suggests that this invariant might play a role in the description of these solutions.

\section{Almost BPS system}
 Let us now discuss the generalisation of the almost BPS system of equations derived in \cite{Goldstein:2008fq} to $ \N=8$ supergravity. 
 
\subsection{The solution Ansatz}
 
 The system includes one real harmonic function $V$, a Jordan algebra valued (\ie a 3 by 3 Hermitian matrix over the quaternions) harmonic function $\mathcal{K}$, and a 3-vector over the quaternions of harmonic functions $Y$. It also includes a Jordan algebra valued sourced function $\L$ satisfying 
 \be d \star \scal{ d \L + \K \times \K \,  dV - V d ( \K \times \K ) } = 0 \ee
 and a real function $M$ sourced according to
 \be d \star \scal{ d M - V \trace \L\, d \K + \scal{ Y^\dagger ( d \K) Y}} = 0 \ee
 It will be convenient to define 
 \be \lozenge \equiv V \det[\L] - \scal{ Y^\dagger \L\times \L Y} \ee
 and its derivatives 
 \be \frac{\partial \lozenge}{\partial \L} = V \L\times \L - 2 \L \times ( YY^\dagger) \ee
 and 
 \be  \frac{\partial^2 \lozenge}{\partial \L \times \partial \L} =  V \L  - (Y Y^\dagger) \ee
 where $(YY^\dagger)$ is the Hermitian matrix over the quaternions built from the 3-vector $Y$. 
 
 \vskip 2mm
 
 The metric is defined by the scaling factor 
 \be e^{-4U} = \lozenge - M^2\ee
 and the Kaluza--Klein vector satisfying 
 \be \star d \omega = d M - \trace  \frac{\partial^2 \lozenge}{\partial \L \times \partial \L}  d \K  \ee
 The moduli are defined by 
 \be \mathpzc{t} = \K + \frac{-M+ ie^{-2U}}{V} \frac{\partial \mbox{ln}[\lozenge]}{\partial \L} + \ell \invo \frac{Y}{V} \ee
 The Kaluza--Klein components of the vector fields are 
\bea \star d w_0 &=&  \trace\bigl[    \L d \K-\K d\L \bigr] +  V d \det[\K]- \det[\K] d V   \CR
\star d \mathpzc{w} &=& d\L+ \K\times \K d V  - V d ( \K\times \K) +  \ell \invo \scal{  (d\K) Y -  \K d Y } \CR
\star d \mathpzc{v} &=& V d \K- \K d V   - \ell \invo dY \CR
\star d v^0 &=& -dV \eea
Of course setting $Y=0$ one gets back the system of $\N=2$ supergravity \cite{Goldstein:2008fq}.
\subsection{Solving the differential equations}
Algebra shows directly that the poles of $V$ and $Y$ are incompatible with the poles of $\mathcal{K}$ for the Noether charge carried by the pole to correspond to a regular orbit. The condition for regularity is the same and we will already assume that the function $\mathcal{K}$ admits pole at the BPS centres only. This is in agreement with the result obtained in \cite{Bena:2009ev} for the STU model. For simplicity we will consider one single BPS centre at $x=0$, such that 
\be \mathcal{K} =\mathpzc{k} + V(0)^{-1}  \frac{ \mathpzc{p}_* }{|x|} \ee
 whereas $V$ and $Y$ will admit poles at the (arbitrary $x_\pA\ne 0$) non-BPS centres 
 \be V = h - \sum_A \frac{p_A^0}{|x-x_{\scriptscriptstyle A}|} \; , \qquad Y = y -  \sum_A \frac{p_A}{|x-x_{\scriptscriptstyle A}|} \ee
 This restriction will somehow be complementary of the case discussed in \cite{Bena:2009en}, where they considered one single non-BPS centre and arbitrary many BPS ones. 

One can then solve 
\be \mathcal{L} = V(0)^{-2} \Scal{ h -  \sum_A \frac{p_A^0\, |x-x_{\scriptscriptstyle A}|}{|x_{\scriptscriptstyle A}|^2}} \frac{  \mathpzc{p}_* \times \mathpzc{p}_* }{|x|^2} + \frac{ \mathpzc{q}_* + 2 \mathpzc{k} \times \mathpzc{p}_*}{|x|} + \sum_A \frac{ \overline{\mathpzc{q}}_A}{|x-x_\pA|}  + \mathpzc{l} \ee
The choice of parametrization is justified by the value of the electromagnetic charges at each centre. One computes the charges of the BPS black hole as 
\bea q_{* 0} &=& - \trace\Bigl[ \mathpzc{k} \, \mathpzc{q}_* + \Scal{   \mathpzc{k} \times \mathpzc{k} - \frac{\mathpzc{l} + \sum_A \frac{ \overline{\mathpzc{q}}_A}{|x_\pA|}  }{V(0)} }  \mathpzc{p}_* \Bigr]  - V(0)^{-3} \det[\mathpzc{p}_*] \sum_A \frac{ p_A^0}{|x_\pA|^3} \CR
\mathpzc{Q}_* &=& \mathpzc{q}_*  +  \ell \invo \mathpzc{p}_*  \frac{ Y(0) }{V(0)} \CR
\mathpzc{P}_* &=& \mathpzc{p}_* \CR
p^0_* &=& 0
\eea
and the charges of the non-BPS black holes as
\bea  q_{0 A} &=& - \trace [ \mathcal{K}(x_\pA) \overline{\mathpzc{q}}_A] + \det[ \mathcal{K}(x_\pA) ] p^0_A \CR
\mathpzc{Q}_{\,\,  A} &=&  \overline{\mathpzc{q}}_A - \mathcal{K}(x_\pA) \times \mathcal{K}(x_\pA)  p^0_A  +   \ell \invo \mathcal{K}(x_\pA) p_A \CR
\mathpzc{P}_A &=&  \mathcal{K}(x_\pA)  p^0_A  + \ell \invo p_A \CR
p^0_A &=& p^0_A \eea

\vskip 2mm

In order to compute the expression of the function $M$, we will need to define the function 
\be K_{AB} \equiv \int \frac{d^3 y}{4\pi} \frac{1}{|x-y||y|^3}  \Scal{ y^i \frac{ \partial \, }{\partial y^i}} \frac{ \scal{ |x_\pA||y-x_\pB| - |x_\pB||y-x_\pB|}^2 }{|x_\pA| |x_\pB||y-x_\pA||y-x_\pB||y|^2} \ee
We will not compute this integral, however one can show that it converges for any value of $x$, such that the function $K_{AB}$ is regular everywhere, and satisfies the differential equation 
\be \Delta K_{AB} = \nabla \frac{ \scal{ |x_\pA||x-x_\pB| - |x_\pB||x-x_\pB|}^2 }{|x_\pA| |x_\pB||x-x_\pA||x-x_\pB||x|^2} \cdot \nabla \frac{1}{|x|} \ee
Using this function one computes the solution for $M$
\begin{multline}  M =m + V(0)^{-1} \frac{ \det[ \mathpzc{p}_*]}{|x|^3} + \frac{1}{2} \trace [ ( \mathpzc{q}_*  + 2 \mathpzc{k} \times \mathpzc{p}_* ) \mathpzc{p}_*] \frac{ h - \sum_A \frac{ p_A^0 }{|x-x_\pA|}  \scal{ 1 - \frac{ x\cdot x_\pA}{|x_\pA|^2} }}{V(0) |x|^2}\\  + \frac{V(0)^{-1}}{2} \sum_A \scal{ \trace[ ( h  \overline{\mathpzc{q}}_A- p_A^0 \mathpzc{l})  \mathpzc{p}_* ]  + ( y^\dagger \mathpzc{p}_* p_A) + ( p_A^\dagger \mathpzc{p}_* y) } \frac{|x-x_\pA| + |x_\pA| - |x| }{|x_\pA| |x||x-x_\pA|}   \\ -V(0)^{-1}  \sum_A \scal{  p_A^0 \trace [ \overline{\mathpzc{q}}_A \mathpzc{p}_*] + ( p_A^\dagger \mathpzc{p}_* p_A) } \frac{1}{|x-x_\pA|^2 |x|}  \Scal{ 1 - \frac{ x\cdot x_\pA}{|x_\pA|^2} }\\
- \frac{ \det[ \mathpzc{p}_*]}{2 V(0)^3 } \sum_A h \p_A^0 \Scal{ \frac{ 3 |x-x_\pA|^2 + |x_\pA|^2 - 4 |x_\pA| |x-x_\pA| }{|x_\pA|^2 |x-x_\pA| |x|^3} + \frac{ 1}{|x||x_\pA|^3 } - \frac{1}{|x-x_\pA||x_\pA|^3} } \\
-V(0)^{-1}  \sum_{A>B}  \scal{ p_A^0 \trace[ \overline{\mathpzc{q}}_B \mathpzc{p}_*] + p_B^0 \trace[ \overline{\mathpzc{q}}_A \mathpzc{p}_*] + ( p_A^\dagger \mathpzc{p}_* p_B) + (  p_B^\dagger \mathpzc{p}_* p_A)} \Scal{ F_{(A,B)0} + \frac{1}{|x_\pA||x_\pB||x|}} \Bigr . \\ \Bigl . + V(0)^{-3} \det[ \mathpzc{p}_*]  \sum_{A>B}   p_A^0 p_B^0 \Scal{  K_{AB} + \frac{ |x_\pA - x_\pB|^2}{ |x_\pA|^3 |x_\pB|^3 |x|} }  \\ - V(0)^{-3}  \det[ \mathpzc{p}_*]\Scal{ \sum_B \frac{ p_B^0}{|x_\pB|}} \Scal{ \sum_A p_A^0 \frac{ x \cdot x_\pA}{|x_\pA|^3 |x|^3}} \\
+ V(0)^{-1} \bigl( h \trace[ \mathpzc{l} \mathpzc{p}_*] - ( y^\dagger \mathpzc{p}_* y) \bigr) \frac{1}{ |x|} - \sum_A \frac{ \J_A^i (x-x_\pA)_i}{|x-x_\pA|^3 }  
 \end{multline} 
where we chose the dipole harmonic function at $x=0$ such as to cancel the divergence in $\frac{x^i}{|x|^5}$ in the scaling factor $e^{-4U}$, and each pole such that $\omega$ is globally defined. $F_{(A,B)0}$ is the function defined in the Appendix \ref{Laplace} for $x_\pC=0$. We refer to \cite{Bena:2009ev,Bena:2009en} for a more detailed discussion in the case of the STU model. 

\subsection{Near horizon geometry}
\subsubsection*{The BPS centre}
In the BPS centre near horizon geometry, one computes that 
\be \omega = \frac{\varepsilon_{ijk} \J_*^i x^j dx^k}{|x|^3} + \mathcal{O}(1) \ee
and
\begin{multline}  e^{-4U} = \frac{ - q_{* 0} \det[ \mathpzc{p}_*] + \trace[ ( \mathpzc{q}_* \times \mathpzc{q}_*)( \mathpzc{p}_* \times \mathpzc{p}_*)] + (q^\dagger_* (  \mathpzc{p}_* \times \mathpzc{p}_*) q_* ) - \frac{1}{4} ( \trace [ \mathpzc{q}_* \mathpzc{p}_* ])^2  }{|x|^4} \\ + 2 \frac{ |\J_*|^2  }{|x|^4}  - 
\Scal{ \frac{ \J_{* i} x^i }{|x|^3}}^2 + \mathcal{O}(x^{-3})
 \end{multline} 
 with 
\be \J_*^i \equiv  - \frac{\det[ \mathpzc{p}_*] }{V(0)^2} \sum_A  \frac{ p_A^0 \, x^i_\pA}{|x_\pA|^3} \ee
an angular momentum induced by the interactions with the non-BPS black holes.  The horizon area is thus 
\be S_*  = 4\pi \sqrt{ I_4(q,p) +  |\J_*|^2 } \ee
The scalar fields turn out to do not sit at the attractor values, in fact they are defined at the horizon as (choosing coordinates such that $\J_*^i \partial_i  = \J_* \partial_z$) 
\be \mathpzc{t}_* = \frac{ - \frac{ \partial I_4}{\partial \mathpzc{q}_*} + \ell \invo  \frac{ \partial I_4 }{\partial q^\dagger_*} + \mathpzc{p}_* \Scal{  \J_* \cos( \theta_*) + i \sqrt{ I_4 + |\J_*|^2 + |\J_*|^2 \sin^2(\theta_*) }}}{ - \frac{ \partial I_4 }{\partial q_{* 0} } } \ee  
Indeed one can check that the BPS attractor equation is not satisfied 
\be \mathpzc{D}Z_*  =\frac{ \J_*^2 + \J_*^2 \sin^2(\theta_*) + i \J_* \sin(\theta_*)  \sqrt{  I_4 +    \J_*^2 + \J_*^2 \sin^2(\theta_*) }}{ 2 ( I_4 +    \J_*^2 + \J_*^2 \sin^2(\theta_*) )^{\frac{3}{4}}} \mathds{1} \ne 0 \ee
although the scalar fields are entirely determined by the electromagnetic charges and the local angular momentum. Note that the solution is not quaternionic at $x=0$ (it admits non-trivial components in $\ell$) and therefore does not sit obviously in the $\N=2$ truncation in which $Z(q,p)$ defines the central charge; all these components can nonetheless be eliminated by a duality transformation that shifts the axions by $- \ell \invo \frac{Y(0)}{V(0)}$, such that it does. 

There is therefore no enhancement of supersymmetry in the near horizon geometry. The horizon is itself a squashed sphere with an angular momentum induced by the interactions with the non-BPS back holes. As a consequence, the horizon area is not determined by the quartic invariant of its charges, but is increased by the norm squared of its induced local angular momentum. This horizon is larger than the BPS horizon in all directions, because its metric 
\be ds^2 = \sqrt{ I_4 + |\J_*|^2 + |\J_*|^2 \sin^2(\theta_*)} \, d\theta_*^2  + \frac{ I_4 + |\J_*|^2 }{ \sqrt{ I_4 + |\J_*|^2 + |\J_*|^2 \sin^2(\theta_*)}} \sin^2(\theta_*) d\varphi^2 \ee
verifies that 
\be \frac{ I_4 + |\J_*|^2 }{ \sqrt{ I_4 + |\J_*|^2 + |\J_*|^2 \sin^2(\theta_*)}} > \sqrt{ I_4  } \ee
The existence of interactions seems therefore to create a squashed horizon that surounds the BPS horizon. 

\subsubsection*{Non-BPS centres}
The near horizon geometry of the non-BPS black holes is much simpler to extract. One directly obtains that 
\be  \omega =  \varepsilon_{ijk}  \frac{ \J^i_A  (x-x_\pA)^j dx^k}{|x-x_\pA|^3} + \mathcal{O}(1) \ee
and
\be e^{-4U} = - \frac{ p_A^0  \det[ \overline{\mathpzc{q}}_A] + (p_A^\dagger ( \overline{\mathpzc{q}}_A \times \overline{\mathpzc{q}}_A) p_A )    }{|x|^4}   - 
\Scal{\frac{\J_{A i} ( x-x_\pA)^i }{|x|^3}}^2  + \mathcal{O}(x^{-3})
 \ee
Using the property that the charges of the non-BPS black holes are obtained by `T-duality' of parameter $\mathcal{K}(x_\pA)$, one concludes that
\be S_\pA = 4 \pi  \sqrt{ - I_4(q_A,p_A) -  |\J_A|^2 } \ee
as expected \cite{Rasheed:1995zv,Larsen:1999pp,Ferrara:2006em}. 
\subsection{The asymptotic region}
\subsubsection*{The ADM mass}
To compute the ADM mass, it will be simpler to consider the asymptotic behaviour of the various functions in terms of the total electromagnetic charges
\be V = h - \frac{ p^0}{r} + \mathcal{O}(r^{-2}) \;, \qquad Y = y - \frac{p}{r} + \mathcal{O}(r^{-2}) \; , \qquad \K = \mathpzc{k} + \frac{  \mathpzc{p} - \mathpzc{k}\,  p^0}{r} + \mathcal{O}(r^{-2}) \ee
and 
\be \L = \mathpzc{l} + \frac{ \mathpzc{q} + 2 \mathpzc{k} \times \mathpzc{p} - \mathpzc{k} \times\mathpzc{k} \,  p^0 }{r} + \mathcal{O}(r^{-2}) \ee
The total $q_0$ and $q$ charges are determined such that 
\bea q_0 + \trace [ \mathpzc{k} \, \mathpzc{q} ] + \trace\Bigl[ \Scal{ \mathpzc{k} \times \mathpzc{k}  - \frac{ \mathpzc{l}}{h} } \mathpzc{p} \Bigr] - \Scal{ \det[ \mathpzc{k} ] - \frac{1}{h} \trace[ \mathpzc{l} \, \mathpzc{k} ] } p^0 &=& 0 \CR
q - \mathpzc{k} \,  p - \frac{1}{h} ( \mathpzc{p} - \mathpzc{k} \, p^0) y &=& 0 \label{VanishingCharges}\eea 
Requiring the absence of total NUT charge implies 
\be M = m + \frac{ \trace [ \mathpzc{l} (\mathpzc{p} - \mathpzc{k} p^0)] - \frac{1}{h} ( y^\dagger ( \mathpzc{p} - \mathpzc{k}\,  p^0)y) }{r} + \mathcal{O}(r^{-2}) \ee
This is indeed the case provided 
\be F_{(A,B)0}(x) + \frac{1}{|x_\pA||x_\pB| \, r} =  \mathcal{O}(r^{-2}) \; , \qquad  K_{AB}(x) + \frac{ |x_\pA - x_\pB|^2}{ |x_\pA|^3 |x_\pB|^3 \, r}  =  \mathcal{O}(r^{-2}) \ee
in the asymptotic region. We prove explicitly the first limit in Appendix \ref{Laplace}, and we will assume that the second is valid. This is ensured by the conservation of the three-dimensional $\e_{8(8)}$ current, since there is no poles outside the black hole horizons. 

\vskip 2mm

By definition 
\be e^{-4U} = 1 + 4 \frac{M_{\rm \scriptscriptstyle ADM} }{r} + \mathcal{O}(r^{-2}) \ee
which implies 
\be h \det[ \mathpzc{l}]  = 1 + m^2 + ( y^\dagger ( \mathpzc{l}\times \mathpzc{l}) y) \ee
Using this equation, the asymptotic scalar fields simplify to 
\be \mathpzc{t}_\asym = \mathpzc{k} + \frac{ -m + i}{1+m^2} \Scal{ \mathpzc{l}\times \mathpzc{l} - \frac{2}{h}  \mathpzc{l}\times y y^\dagger} + \ell \invo \frac{y}{h} \ee
One computes that 
\begin{multline} 
M_{\rm \scriptscriptstyle ADM}  = \frac{1}{4} \Bigl( - \det[ \mathpzc{l}] p^0 + h \trace [ ( \mathpzc{l}\times \mathpzc{l}) ( \mathpzc{q} + 2 \mathpzc{k}\times \mathpzc{p} - \mathpzc{k} \times \mathpzc{k} \, p^0 )] \Bigr .\\   - 2 \scal{ y^\dagger ( \mathpzc{q} + 2 \mathpzc{k}\times \mathpzc{p} - \mathpzc{k} \times \mathpzc{k} \, p^0 )y} + \scal{ y^\dagger ( \mathpzc{l}\times \mathpzc{l}) p + p^\dagger ( \mathpzc{l}\times \mathpzc{l}) y} \\ \Bigl . + 2m \trace [ \mathpzc{l} \mathpzc{k} \, p^0 - \mathpzc{l} \mathpzc{p}] + \frac{2m}{h} \scal{ y^\dagger ( \mathpzc{p} - \mathpzc{k}\, p^0) y} \Bigr) \end{multline} 
After some algebra, one computes that it can be rewritten as
\begin{multline} 
M_{\rm \scriptscriptstyle ADM}  = \frac{h}{4} \biggl( \trace \Bigl[ \Scal{ \mathpzc{l} \times \mathpzc{l} - \frac{2}{h} \mathpzc{l} \times y y^\dagger } \mathpzc{Q} \Bigr] + 2 \trace \Bigl[ \Bigl( \Scal{ \mathpzc{k} + \ell \invo \frac{y}{h}} \times \Scal{ \mathpzc{l} \times \mathpzc{l}   - \frac{2}{h} \mathpzc{l} \times y y^\dagger } \biggr .\Bigr . \\ \Bigl .  - \frac{2 m}{1+m^2} \Scal{ \mathpzc{l} \times \mathpzc{l} - \frac{2}{h} \mathpzc{l} \times y y^\dagger } \times \Scal{ \mathpzc{l} \times \mathpzc{l} - \frac{2}{h} \mathpzc{l} \times y y^\dagger } \Bigr) \mathpzc{P} \Bigr] \\ 
- \Bigl( \frac{1}{1+m^2} \det\Bigl[ \mathpzc{l} \times \mathpzc{l} - \frac{2}{h} \mathpzc{l} \times y y^\dagger \Bigr] + \trace \Bigl[ \Scal{ \mathpzc{k} + \ell \invo \frac{y}{h}} \times \Scal{ \mathpzc{k} + \ell \invo \frac{y}{h}} \times \Scal{ \mathpzc{l} \times \mathpzc{l}   - \frac{2}{h} \mathpzc{l} \times y y^\dagger } \Bigr] \Bigr . \\ \biggl .  \Bigl .  - \frac{2m}{1+m^2} \trace \Bigl[  \Scal{ \mathpzc{k} + \ell \invo \frac{y}{h}} \times \Scal{ \mathpzc{l} \times \mathpzc{l}   - \frac{2}{h} \mathpzc{l} \times y y^\dagger } \times \Scal{ \mathpzc{l} \times \mathpzc{l}   - \frac{2}{h} \mathpzc{l} \times y y^\dagger } \Bigr] \Bigr) p^0 \biggr) 
\end{multline} 
where we used in particular that 
\bea  \Scal{ \mathpzc{l} \times \mathpzc{l}   - \frac{2}{h} \mathpzc{l} \times y y^\dagger } \times  \Scal{ \mathpzc{l} \times \mathpzc{l}   - \frac{2}{h} \mathpzc{l} \times y y^\dagger } &=& \Scal{ \det[\mathpzc{l}] - \frac{1}{h}  ( y^\dagger ( \mathpzc{l}\times \mathpzc{l}) y) } \Scal{ \mathpzc{l} - \frac{yy^\dagger}{h}} \CR
\det\Bigl[ \mathpzc{l} \times \mathpzc{l}   - \frac{2}{h} \mathpzc{l} \times y y^\dagger \Bigr] &=& \Scal{ \det[\mathpzc{l}] - \frac{1}{h}  ( y^\dagger ( \mathpzc{l}\times \mathpzc{l}) y) }^2 
\eea
Before to be able to rewrite this expression in terms of the asymptotic central charges, we will first observe that 
\begin{multline}  N = \frac{h\sqrt{1+m^2} }{4} \biggl( q_0 + \trace \Bigl[ \Scal{ \mathpzc{k} + \ell \invo \frac{y}{h}}  \mathpzc{Q} \Bigr] + \trace \Bigl[ \Bigl( \Scal{ \mathpzc{k} + \ell \invo \frac{y}{h}}  \times \Scal{ \mathpzc{k} + \ell \invo \frac{y}{h}} \Bigr . \Bigr . \biggr . \\ \Bigl . \Bigl . 
\hspace{30mm}  - \frac{1}{1+m^2} \Scal{ \mathpzc{l} \times \mathpzc{l}   - \frac{2}{h} \mathpzc{l} \times y y^\dagger }\times \Scal{ \mathpzc{l} \times \mathpzc{l}   - \frac{2}{h} \mathpzc{l} \times y y^\dagger } \Bigr) \mathpzc{P} \Bigr] \\  - \Bigl( \det\Bigl[ \mathpzc{k} + \ell \invo \frac{y}{h}  \Bigr] \hspace{85mm} \Bigr . \\ \biggl . \Bigl .  - \frac{1}{1+m^2} \trace \Bigl[ \Scal{ \mathpzc{l} \times \mathpzc{l}   - \frac{2}{h} \mathpzc{l} \times y y^\dagger }\times \Scal{ \mathpzc{l} \times \mathpzc{l}   - \frac{2}{h} \mathpzc{l} \times y y^\dagger } \times \Scal{ \mathpzc{k} + \ell \invo \frac{y}{h}} \Bigr] \Bigr) p^0 \biggr) \end{multline} 
vanishes identically according to (\ref{VanishingCharges}). Then one can compute that 
\bea M_{\rm \scriptscriptstyle ADM}   &=& M_{\rm \scriptscriptstyle ADM}   - \frac{ 3 e^{i\alpha} + e^{-3i\alpha} }{4} N \CR
&=& - \frac{1}{4} \Scal{ 3 e^{i\alpha} Z(q,p) +  e^{-3i\alpha} \bar Z(q,p) - e^{-i\alpha} \scal{ \trace[ \mathpzc{D} Z(q,p) ] - \trace[ \overline{\mathpzc{D} Z}(q,p) ] }} \CR
&=&  - e^{i\alpha} Z(q,p)  + \frac{i}{2} e^{-i\alpha} {\rm Im}\bigl[ e^{2i\alpha} Z(q,p)  + \trace[ \mathpzc{D} Z(q,p) ] \bigr] \eea
for 
\be e^{i\alpha} = \frac{ m + i }{\sqrt{1+m^2}} \ee
This form indeed generalises the formula derived in \cite{BossardRuef} for the STU model. 

\subsubsection*{Angular momentum}

Using the property that the angular momentum is defined by the asymptotic expression of the Kaluza--Klein vector 
\be \omega =  \frac{ \varepsilon_{ijk} J^i x^j dx^k}{|x|^3} + \mathcal{O}(x^{-2}) \ee
one can compute that it also defines the asymptotic component of the function $M$ 
\be M = m - J_i \frac{x^i}{|x|^3}+ \frac{1}{2 V(0) |x|} \scal{ \trace[( V \mathcal{L}+h \mathpzc{l})  \mathpzc{p}_*] - ( Y^\dagger \mathpzc{p}_* Y ) - ( y^\dagger \mathpzc{p}_* y)  }  + \mathcal{O}(x^{-3}) \ee 
where the last term includes the terms in $\frac{1}{|x|}$ and  $\frac{1}{|x|^2}$. To compute this expression, we would need the asymptotic behaviour of the function $K_{AB}$. We will consider instead a undetermined parameter $j_{AB}^i$ such that 
\be K_{AB} = - \frac{|x_\pA - x_\pB|^2}{|x_\pA|^3 |x_\pB|^3 |x|} - \frac{1}{2} \Scal{ \frac{1}{|x_\pA|} - \frac{1}{|x_\pB|}}^3 \frac{ x_\pA - x_\pB}{|x_\pA- x_\pB|} \cdot \frac{x}{|x|^3 } - j_{AB\, i} \frac{x^i}{|x|^3} + \mathcal{O}(x^{-3}) \ee
The specific choice for the second term (that could have been absorbed in the constants $j_{AB}^i$) will become clear shortly. Using this expression one computes 
\begin{multline}  - J^i = \frac{1}{2 V(0)} \trace[ ( \mathpzc{q}_* + 2 \mathpzc{k} \times \mathpzc{p}_* ) \mathpzc{p}_* ] \sum_A \frac{ p_A^0 x_\pA^i }{|x_\pA|^2}  \\  - \frac{1}{2V(0)} \sum_A \Bigl(  \bigl( \trace[( h \overline{\mathpzc{q}}_A - p_A^0 \mathpzc{l}) \mathpzc{p}_* ]  + ( y^\dagger \mathpzc{p}_* p_A)  + ( p_A^\dagger \mathpzc{p}_* y) \bigr) \frac{x_\pA^i}{|x_\pA|} \Bigr)  \\
+ \frac{1}{V(0)} \sum_A \Scal{ \scal{ p_A^0 \trace[ \overline{\mathpzc{q}}_A \mathpzc{p}_*] + ( p_A^\dagger \mathpzc{p}_* p_A) } \frac{x_\pA^i}{|x_\pA|^2}} \\ + \frac{ \det[ \mathpzc{p}_*]}{2 V(0)^3 } \Scal{ h - 2 \sum_B \frac{p_B^0}{|x_\pB|}}\Scal{ \sum_A \frac{ p_A x_\pA^i }{|x_\pA|^3} }  - \sum_A \J_A^i \\
+ \frac{1}{2 V(0) }  \sum_{A>B}  \biggl( \scal{ p_A^0 \trace[ \overline{\mathpzc{q}}_B \mathpzc{p}_*] + p_B^0 \trace[ \overline{\mathpzc{q}}_A \mathpzc{p}_*] + ( p_A^\dagger \mathpzc{p}_* p_B) + (  p_B^\dagger \mathpzc{p}_* p_A)} \biggr . \\ \biggl . 
 \Bigl(  \frac{ x^i_\pA - x^i_\pB}{|x_\pA - x_\pB|} \Scal{ \frac{1}{|x_\pA|} -  \frac{1}{|x_\pB|}}  + \frac{  x^i_\pA + x^i_\pB}{|x_\pA||x_\pB|} \Bigr. \\  \biggr . \Bigr .  -  \frac{ |x_\pA - x_\pB|^2   x^i_\pB - ( x_\pA - x_\pB ) \hspace{-1mm}\cdot \hspace{-1mm}  x_\pB  ( x^i_\pA - x^i_\pB )  }{ |x_\pA-x_\pB||x_\pA||x_\pB|\scal{   |x_\pA-x_\pB| + |x_\pA| + |x_\pB| }} \Bigr) \biggr) \\
 - \frac{\det[ \mathpzc{p}_*]}{2 V(0)^3} \sum_{A>B} p_A^0 p_B^0 \Scal{ \Scal{  \frac{1}{|x_\pA|} -  \frac{1}{|x_\pB|}}^3 \frac{ x^i_\pA - x^i_\pB}{|x_\pA - x_\pB|} + 2 j_{AB}^i } 
\end{multline} 
After some algebra one obtains that the angular momentum can be rewritten as
\begin{multline}  J^i = \frac{1}{2} \sum_A \Scal{ q_{0A} p_*^0 + \trace [\mathpzc{Q}\, _A \mathpzc{P}_*]- \trace[ \mathpzc{P}_A \mathpzc{Q}_*] - p^0_A q_{0*}} \frac{x_\pA^i}{|x_\pA|} \\ + \frac{1}{2} \sum_{A>B} \Scal{ q_{0A} p_B^0 + \trace [ \mathpzc{Q}\, _A \mathpzc{P}_B] - \trace [\mathpzc{P}_A  \mathpzc{Q}\, _B ]- p^0_{A} q_{0 B}  } \frac{ x_\pA^i - x_\pB^i}{|x_\pA - x_\pB|} \\
+ \sum_A \J_A^i -  \J_*^i  
 +  \frac{\det[ \mathpzc{p}_*]}{ V(0)^3} \sum_{A>B} p_A^0 p_B^0  j_{AB}^i    \\
+  \frac{1}{2 V(0) }  \sum_{A>B}  \biggl( \scal{ p_A^0 \trace[ \overline{\mathpzc{q}}_B \mathpzc{p}_*] + p_B^0 \trace[ \overline{\mathpzc{q}}_A \mathpzc{p}_*] + ( p_A^\dagger \mathpzc{p}_* p_B) + (  p_B^\dagger \mathpzc{p}_* p_A)} \biggr .\\  \biggr .  \frac{ |x_\pB|^2 x_\pA^i - ( x_\pB \hspace{-1mm}\cdot \hspace{-1mm}  x_\pA) x_\pB^i + |x_\pA|^2 x_\pB^i - ( x_\pA \hspace{-1mm}\cdot \hspace{-1mm}  x_\pB) x_\pA^i }{ |x_\pA-x_\pB||x_\pA||x_\pB|\scal{   |x_\pA-x_\pB| + |x_\pA| + |x_\pB| }} \biggr) \end{multline} 
The first two lines correspond to the usual electromagnetic dipole angular momentum. Note that although the charges of the non-BPS black holes would have been trivially commuting in the absence of a BPS centre, the presence of the latter rotates the charges differently depending of their distance from the BPS centre. The third lines contains the angular momenta that can be felt in the near horizon region of each black hole, although note that the angular momentum induced by the non-BPS black holes in the near horizon region of the BPS one has the opposite sign with respect to the contribution to the angular momentum in the asymptotic region. The last line corresponds to a contribution to the angular momentum that would vanish if the centres were all aligned, similar as the one that exists for non-BPS black hole composites. Relying on duality invariance, it is clear that the constants $j_{AB}^i$ should vanish for an axisymmetric solution, and that they should produce a contribution similar to the last line contribution in general. But we will not prove this property in this paper.

\section{Locally BPS system}

Let us now describe a locally BPS system, which is defined by the usual BPS system of differential equations with additional harmonic functions sourcing axions that do not lie in the $\N=2$ truncation. We consider therefore two scalar functions $L_0$ and $K^0$, and two Hermitian matrices over the quaternions $\L$ and $\K$, which altogether would define the harmonic functions of the maximal $\N=2$ truncation. In addition we will consider a 3-vector of quaternions of harmonic functions $Y$, such that only $K^0, \K$ and $Y$ are harmonic functions, whereas $L_0$ and $\L$ satisfy 
\bea d \star \scal{ d L_0 + Y^\dagger ( d \K) Y - Y^\dagger \K dY - ( dY^\dagger) \K Y }  &=& 0 \CR
d \star \scal{ d\L + K^0 d ( Y Y^\dagger ) - Y Y^\dagger d K^0 } &=&  0 \eea
The metric is then defined as in the BPS system by 
\be e^{-4U} = I_4 = K^0 \det[ \L] - L_0 \det[ \K] + \trace [ ( \L \times \L)(\K \times \K) ] - \frac{1}{4} \scal{ K^0 L_0 + \trace [ \K\L]}^2 \ee
and 
\be \star d \omega = \frac{1}{2} \scal{ K^0 d L_0 + \trace[ \K d \L ] - \trace [ \L d \K] - L_0 d K^0 } \ee
which is well defined because 
\be K^0 \Delta L_0 + \trace[ \K \Delta \L ] = K^0 \trace [\K \Delta ( YY^\dagger )] + \trace[ \K \scal{ - K^0 \Delta (YY^\dagger)}] = 0 \ee
The scalar fields are defined as 
\be \mathpzc{t} = \frac{ \frac{ \partial \sqrt{I_4}}{\partial \L} - i \K}{ \frac{ \partial \sqrt{I_4}}{\partial L_0} - i K^0} + \ell \invo Y \ee
and the three dimensional vector components of the vector fields are 
\bea \star d w_0 &=&  L_0 + Y^\dagger ( d \K) Y - Y^\dagger \K dY - ( dY^\dagger) \K Y   \CR
\star d \mathpzc{w} &=& d \L + K^0 d ( Y Y^\dagger ) - Y Y^\dagger d K^0   + \ell \invo \scal{ ( d \K ) Y - \K d Y }  \CR
\star d \mathpzc{v} &=& d \K + \ell \invo \scal{  Y d K^0 - K^0 d Y }   \CR
\star d v^0 &=& d K^0  \eea
However, there is no regular composite black hole solutions which involve the functions $Y$ in a non-trivial manner, because a pole in $Y$ necessarily renders the corresponding axion field  singular.

\section{Maximal nilpotent orbits}

The three solvable systems of differential equations we discussed in this paper were all associated to nilpotent orbits which exist in $D_4$, \ie that they can be represented by elements of $\so(8,\mathds{C})$ appropriately embedded inside $\e_8$. In fact the almost BPS system is associated to the principal orbit of $SO(4,4)$, and the non-BPS composite (as well as the locally BPS) are associated to the subregular orbit of $SO(4,4)$. They correspond to two inequivalent nilpotent orbits in the coset component $\e_8\ominus \so(16,\mathds{C})$. Since they correspond to direct generalisations of the STU models, it was somehow natural that their expression was very similar to $\N=2$ supergravity expressions.

In fact there are orbits of type $D_5$ which generalise these three orbits. The embedding of the principal nilpotent orbit of $SO(5,5)$ defines a system which generalises the almost BPS system, and the embedding of its subregular nilpotent orbit defines two inequivalent systems generalising the non-BPS composite system and the locally BPS system. Then the linearly realised $SU(2) \times SL(3,\mathds{H})$ symmetry is broken to $SU(2) \times SU(2) \times SL(2,\mathds{H})$ and these three systems include 8 more functions which transform in the fundamental of $SL(2,\mathds{H})$. The $SO(5,5) / ( SO(3,2) \times SO(2,3))$ model defines a truncation of the $SO(8,5+n)/( SO(6,2)\times SO(2,3+n))$ one that captures most information about $\N=4$ supergravity black holes solutions. Indeed, the maximal nilpotent orbit of $\so(8,5+n)$ which admits a non-trivial intersection with the coset component is of type $D_5$, and corresponds to the maximal nilpotent orbit of $SO(5,5)$. 

The maximal orbit of $E_{8(8)}$ which admits an intersection with the coset component orthogonal to $\so^*(16)$ is of type $E_6$, which justifies the use of the $E_{6(6)} / Sp_{\scriptscriptstyle \rm c}(8,\mathds{R})$ to obtain black hole solutions in $\N=8$ supergravity. Once again the principal nilpotent orbit of $E_6$ appropriately embedded inside $\e_{8(8)}$ defines a generalisation of the almost BPS system, and the subregular orbit of $E_6$ defines two inequivalent nilpotent orbits associated respectively to a generalisation of the composite non-BPS system and of the locally BPS system. In this case the linearly realised symmetry of these systems is broken to $\prod_{i=0}^4 SU(2)_i$, and the latters include 4 more functions defining a quaternion, that is 56 functions in whole. We will describe in this section the three nilpotent orbits of type $E_6$ in $\e_{8(8)}$.
 
It is useful to consider the decomposition of $\so^*(16)$ 
\bea \so^*(16) &\cong& \bigoplus_{i=0}^3 \so^*(4)_i \oplus \bigoplus_{i>j} ( {\bf 4}_i \otimes {\bf 4}_j) \CR
&\cong&  \bigoplus_{i=0}^3 \sl_2^\ord{i} \oplus  \bigoplus_{i=0}^3 \su(2)_i  \oplus \bigoplus_{i>j} {\bf 2}^\ord{i} \otimes {\bf 2}^\ord{j} \otimes ({\bf 2}_i \otimes {\bf 2}_j)
\eea 
with respect to which the coset component ${\bf 128}$ decomposes as
\be {\bf 128} \cong {\bf 2}^\ord{0} \otimes {\bf 2}^\ord{1} \otimes {\bf 2}^\ord{2} \otimes {\bf 2}^\ord{3} \oplus \bigoplus_{i>j\ne k>l} {\bf 2}^\ord{i} \otimes {\bf 2}^\ord{j} \otimes ({\bf 2}_k \otimes {\bf 2}_l) \oplus ({\bf 2}_0 \otimes {\bf 2}_1 \otimes {\bf 2}_2 \otimes {\bf 2}_3) \ee
The four $\sl_2^\ord{i}$ correspond precisely to the STU truncation, as well as the ${\bf 2}^\ord{0} \otimes {\bf 2}^\ord{1} \otimes {\bf 2}^\ord{2} \otimes {\bf 2}^\ord{3}$ of the coset component. We will call $H_i$ Cartan generators of the $\sl_2^\ord{i}$. The $SU(2)_i$ factors cannot be further decomposed within a real graded decomposition (\ie with respect to a $\gl_1$ and not a $\mathfrak{u}(1)$ which would require complexification). So for any graded decomposition, the $({\bf 2}_0 \otimes {\bf 2}_1 \otimes {\bf 2}_2 \otimes {\bf 2}_3)$ will always be of grade zero, and the corresponding graded decomposition of $\e_{8(8)}$ will accordingly always includes $\so(4,4)$ in its grade zero component. 

This property implies that the nilpotent orbits of $\e_8$ in the coset component orthogonal to $\so(16,\mathds{C})$ which admit a non-trivial intersection with the real ${\bf 128}$ inside $\e_{8(8)}$, are necessarily associated to weighted Dynkin diagram of $\so(16,\mathds{C})$ for which the component associated to compact generators of the Cartan subalgebra of $\so^*(16)$ are all null. They are of type {{\tiny $ {  \vspace{-2mm} \left[ \begin{array}{ccccccccc}  && \mathfrak{0} \hspace{-0.7mm}&&&&&& \vspace{ -1.5mm} \\ \cdot \hspace{-0.5mm}& \mathpzc{a} \hspace{-0.6mm} &\mathpzc{b} \hspace{-0.6mm} &  \mathfrak{0} \hspace{-0.7mm}& \mathpzc{c} \hspace{-0.6mm} & \mathfrak{0}\hspace{-0.7mm}&\mathpzc{d}\hspace{-0.6mm}&\mathfrak{0} \end{array}\right] }$}} with $\mathpzc{a}, \, \mathpzc{b}, \, \mathpzc{c}, \, \mathpzc{d} \in \mathds{N}$ \cite{MoiN4}. 

We refer to \cite{Levi,E8strat} for the complete classification of the nilpotent orbits of $E_{8(8)}$. 
\subsection{Almost BPS solvable algebra}
The maximal nilpotent orbit of $E_{8(8)}$ which does admit a non-trivial intersection with the coset component is associated to the weighted Dynkin diagram of $\e_{8(8)}$ \DEVIII20000222 which defines the graded decomposition
\begin{multline} \e_{8(8)} \cong {\bf 1}^\ord{-11} \oplus {\bf 1}^\ord{-10} \oplus {\bf 1}^\ord{-9} \oplus {\bf 8}_v^\ord{-8} \oplus ({\bf 1}\oplus {\bf 8}_a )^\ord{-7} \oplus  ({\bf 1}\oplus {\bf 8}_c )^\ord{-6} \oplus ({\bf 1}\oplus {\bf 1} \oplus  {\bf 8}_c )^\ord{-5}  \\ \oplus ({\bf 1}\oplus {\bf 8}_a \oplus  {\bf 8}_c )^\ord{-4} \oplus ({\bf 1}\oplus {\bf 8}_v \oplus  {\bf 8}_a )^\ord{-3}  \oplus ({\bf 1}\oplus {\bf 8}_a \oplus  {\bf 8}_v )^\ord{-2}  \oplus ({\bf 1}\oplus  {\bf 1}\oplus {\bf 8}_v \oplus  {\bf 8}_c )^\ord{-1}\\ \oplus \scal{ \gl_1\oplus \gl_1 \oplus \gl_1 \oplus \gl_1 \oplus \so(4,4) }^\ord{0} 
 \oplus ({\bf 1}\oplus  {\bf 1}\oplus {\bf 8}_v \oplus  {\bf 8}_c )^\ord{1} \\
 \oplus ({\bf 1}\oplus {\bf 8}_a \oplus  {\bf 8}_v )^\ord{2}   \oplus ({\bf 1}\oplus {\bf 8}_v \oplus  {\bf 8}_a )^\ord{3}\oplus ({\bf 1}\oplus {\bf 8}_a \oplus  {\bf 8}_c )^\ord{4} \\ \oplus ({\bf 1}\oplus {\bf 1} \oplus  {\bf 8}_c )^\ord{5}    \oplus  ({\bf 1}\oplus {\bf 8}_c )^\ord{6} \oplus ({\bf 1}\oplus {\bf 8}_a )^\ord{7}\oplus {\bf 8}_v^\ord{8} \oplus {\bf 1}^\ord{9}\oplus  {\bf 1}^\ord{10} \oplus {\bf 1}^\ord{11} \end{multline} 
A representative element of the orbit is a generic element of grade 1, for which all four components are non-zero and such that the vector and the spinor have a non-vanishing norm. The four components have all different weights with respect to the four $\gl_1$ such that no combination of them leaves the element invariant. The vector is only preserved by $Spin(3,4) \subset Spin(4,4)$, and the chiral spinor by $G_{2(2)} \subset Spin(3,4)$. The corresponding nilpotent orbit has dimension 216, and is isomorphic to 
\be E_{8(8)} / \scal{ G_{2(2)}  \ltimes \mathds{R}^{2\times 7 + 4}} \ee
Its intersection with the coset component is associated to the $\so^*(16)$ weighted Dynkin diagram \DSOXVI04020240, (\ie $H_1 + 2H_2 + 3 H_3 + 5 H_0$ in the $STU$ basis) which defines the graded decomposition
\begin{multline} \so^*(16) \cong  {\bf 1}^\ord{-10} \oplus ({\bf 2}_0\otimes {\bf 2}_3)^\ord{-8} \oplus ({\bf 2}_0\otimes {\bf 2}_2)^\ord{-7} \oplus  ({\bf 1}\oplus {\bf 2}_0\otimes {\bf 2}_1)^\ord{-6} \oplus ({\bf 2}_2\otimes {\bf 2}_3)^\ord{-5}  \\ \oplus ({\bf 1}\oplus {\bf 2}_1\otimes {\bf 2}_3 \oplus {\bf 2}_0\otimes {\bf 2}_1 )^\ord{-4} \oplus ({\bf 2}_1\otimes {\bf 2}_2 \oplus {\bf 2}_0\otimes {\bf 2}_2)^\ord{-3}  \oplus ({\bf 1}\oplus {\bf 2}_1\otimes {\bf 2}_3\oplus {\bf 2}_0\otimes {\bf 2}_3)^\ord{-2}  \oplus ({\bf 2}_1\otimes {\bf 2}_2\oplus {\bf 2}_2\otimes {\bf 2}_3)^\ord{-1}\\ \oplus \scal{ \gl_1\oplus \gl_1 \oplus \gl_1 \oplus \gl_1 \oplus \bigoplus_{i=0}^3 \su(2)_i  }^\ord{0} 
\oplus ({\bf 2}_1\otimes {\bf 2}_2\oplus {\bf 2}_2\otimes {\bf 2}_3)^\ord{1} \\
 \oplus ({\bf 1}\oplus {\bf 2}_1\otimes {\bf 2}_3\oplus {\bf 2}_0\otimes {\bf 2}_3)^\ord{2}  \oplus ({\bf 2}_1\otimes {\bf 2}_2 \oplus {\bf 2}_0\otimes {\bf 2}_2)^\ord{3}\oplus ({\bf 1}\oplus {\bf 2}_1\otimes {\bf 2}_3 \oplus {\bf 2}_0\otimes {\bf 2}_1 )^\ord{4} \\ \oplus ({\bf 2}_2\otimes {\bf 2}_3)^\ord{5}    \oplus   ({\bf 1}\oplus {\bf 2}_0\otimes {\bf 2}_1)^\ord{6} \oplus ({\bf 2}_0\otimes {\bf 2}_2)^\ord{7}\oplus ({\bf 2}_0\otimes {\bf 2}_3)^\ord{8} \oplus  {\bf 1}^\ord{10} \end{multline} 
for which the ${\bf 128}$ decomposes as
\begin{multline} {\bf 128} \cong  {\bf 1}^\ord{-11} \oplus {\bf 1}^\ord{-9} \oplus ({\bf 2}_1\otimes {\bf 2}_2)^\ord{-8} \oplus ({\bf 1}\oplus {\bf 2}_1\otimes {\bf 2}_3)^\ord{-7} \oplus  ({\bf 2}_2\otimes {\bf 2}_3)^\ord{-6} \oplus ({\bf 1}\oplus{\bf 1}\oplus{\bf 2}_0\otimes {\bf 2}_1)^\ord{-5}  \\ \oplus ( {\bf 2}_0\otimes {\bf 2}_2 \oplus {\bf 2}_2\otimes {\bf 2}_3 )^\ord{-4} \oplus ({\bf 1}\oplus {\bf 2}_0\otimes {\bf 2}_3 \oplus {\bf 2}_1\otimes {\bf 2}_3)^\ord{-3}  \oplus( {\bf 2}_0\otimes {\bf 2}_2\oplus {\bf 2}_1\otimes {\bf 2}_2)^\ord{-2}  \oplus ({\bf 1} \oplus {\bf 1} \oplus {\bf 2}_0\otimes {\bf 2}_3\oplus {\bf 2}_0\otimes {\bf 2}_1)^\ord{-1}\\ \oplus ({\bf 2}_0\otimes{\bf 2}_1\otimes{\bf 2}_2\otimes{\bf 2}_3  )^\ord{0} 
\oplus ({\bf 1} \oplus {\bf 1} \oplus {\bf 2}_0\otimes {\bf 2}_3\oplus {\bf 2}_0\otimes {\bf 2}_1)^\ord{1} \\
 \oplus ( {\bf 2}_0\otimes {\bf 2}_2\oplus {\bf 2}_1\otimes {\bf 2}_2)^\ord{2}  \oplus ({\bf 1} \oplus  {\bf 2}_0\otimes {\bf 2}_3 \oplus {\bf 2}_1\otimes {\bf 2}_3)^\ord{3}\oplus ( {\bf 2}_0\otimes {\bf 2}_2 \oplus {\bf 2}_2\otimes {\bf 2}_3 )^\ord{4} \\ \oplus ({\bf 1} \oplus {\bf 1} \oplus {\bf 2}_0\otimes {\bf 2}_1)^\ord{5}    \oplus   ( {\bf 2}_2\otimes {\bf 2}_3)^\ord{6} \oplus ({\bf 1} \oplus {\bf 2}_1\otimes {\bf 2}_3)^\ord{7}\oplus ({\bf 2}_1\otimes {\bf 2}_2)^\ord{8} \oplus  {\bf 1}^\ord{9}\oplus {\bf 1}^\ord{11}  \end{multline} 
where $ {\bf 2}_i \otimes {\bf 2}_j$ is the real vector representation of $SO(4)_{ij}\cong SU(2)_i \times_{\IZ_2} SU(2)_j$. Again an element of the orbit carries a non-trivial component in each irreducible representation of the grade one component of the ${\bf 128}$. Therefore no combination of the $\gl_1$ generators leaves it invariant, and only the diagonal of the three $SU(2)_i$ (for $i=0, 1,3$) leaves both the ${\bf 2}_1\otimes {\bf 2}_2$ and the ${\bf 2}_2\otimes {\bf 2}_3$ invariant. The corresponding orbit of $\Spin^*(16)$ is of dimension 108, and is isomorphic to
\be \Spin^*(16) / \scal{SU(2) \times SU(2) \ltimes \mathds{R}^{3+3}} \ee
Any element of the positive grade component is also an element of this orbit, as long as the grade one component admits generic elements, for every value of the higher order components.  The components of positive grades define a solvable algebra of dimension 108 which intersection with the coset component is of dimension 56, and therefore which permits to define a system of solvable equations for 56 sourced harmonic functions, \ie one for each electromagnetic component. This system contains the almost BPS system. 

\vskip 2mm

We will now describe this nilpotent algebra $\mathfrak{n}$ in terms of the associated differential graded algebra  $\bigwedge \mathfrak{n}^*$, as we did in section \ref{E7Special}.

Before to consider the maximal nilpotent orbit, let us restrict ourselves in a first step to the maximal nilpotent orbit of $SO(4,4)$ in $\e_{8(8)}$ (discussed in more details in this paper), \ie of weighted Dynkin diagram \DEVIII00000022 \DSOXVI02000040. In that case the relevant solvable algebra inside $\e_{8(8)}$ admits an $E_{6(6)}$ automorphism and can be written in terms of the exceptional Jordan algebra. 

For the solvable algebra 
\be ({\bf 1} \oplus {\bf 27})^\ord{1} \oplus {\bf 27}^\ord{2} \oplus \overline{\bf 27}^\ord{3} \oplus {\bf 1}^\ord{4} \oplus {\bf 1}^\ord{5} \ee
we define $V$, $E_0$ and $M$ associated to the grade 1, 4 and 5 singlets respectively, and ${\bf K}$, ${\bf E}$ and ${\bf L}$ the grade 1, 2 and 3 Hermitian matrices. We chose the notations to be as similar as possible to the ones used in the paper such that one can recognise to which generators are associated the various functions of the associated solvable systems. The differential reads 
\be \delta {\bf E} = V {\bf K} \; , \quad \delta {\bf L} = {\bf E} \times {\bf K} \; , \quad \delta E_0  = \trace  {\bf L} {\bf K}   \; , \quad \delta M = E_0 V +    \trace {\bf E}  {\bf L}  \ee
It is nilpotent because the cross product $\times$ is symmetric. Decomposing these expressions in terms of split real and split octonion numbers according to
\be {\bf K} = \left(\begin{array}{ccc} \hspace{0.5mm} K^1 \hspace{0.5mm} & \hspace{0.5mm} {\bf k}_3  \hspace{0.5mm} & \hspace{0.5mm} {\bf k}_2^* \hspace{0.5mm} \\
 \hspace{0.5mm} {\bf k}_3^*   \hspace{0.5mm} & \hspace{0.5mm} K^2 \hspace{0.5mm} & \hspace{0.5mm} {\bf k}_1  \hspace{0.5mm} \\ \hspace{0.5mm} {\bf k}_2   \hspace{0.5mm} & \hspace{0.5mm} {\bf k}_1^*  \hspace{0.5mm} & \hspace{0.5mm} K^3 \hspace{0.5mm} \end{array}\right) \ee
 and analogously for ${\bf E} = ( E^i , {\bf e}_i )$ and ${\bf L} = ( L_i , {\bf l}^i )$, one gets 
\bea \delta E^i &=& V K^i \; , \quad \delta {\bf e}_i = V {\bf k}_i \; ;  \CR
\delta L_i &=& E^{i+1} K^{i+2} +  E^{i+2} K^{i+1} - {\bf e}_i {\bf k}_i^* + {\bf k}_i {\bf e}_i^*\; ,   \CR
\delta {\bf l}^i &=& - E^i {\bf k}_i + K^i {\bf e}_i + {\bf e}_{i+2}^* {\bf k}_{i+1}^* - {\bf k}_{i+2}^* {\bf e}_{i+1}^*\; , \CR
\delta E_0 &=& \sum_i \scal{ L_i K^i + {\bf l}^i {\bf k}_i^* - {\bf k}_i {\bf l}^{i\, *} } \; , \CR
 \delta M &=& E_0 V +  \sum_i \scal{ E^i L_i + {\bf e}_i {\bf l}^{i\, *} - {\bf l}^i {\bf e}_i^*  } \; . 
 \eea
 where $i=1,2,3$ and repeated indices are not summed. The nilpotency can be checked explicitly by using the property that the real part of a square is symmetric, and that the real part of a cube is cyclic. To decompose the algebra in its $\so^*(16)$ and coset components, one must decompose the split octonions in terms of quaternions
\be {\bf k}_i = k_i + \ell s_i \; , \quad {\bf e}_i = e_i + \ell y_i \; , \quad {\bf l}_i = l_i + \ell x_i \ee
and use the Cayley product (\ref{Cayley}).

The algebra then decomposes in terms of quaternions as  
\bea \delta E^i &=& V K^i \; , \quad \delta e_i = V k_i \; , \quad \delta y_i = V s_i \; ; \CR
\delta L_i &=& E^{i+1} K^{i+2} +  E^{i+2} K^{i+1} - e_i k_i^* + k_i e_i^* + y_i s_i^* - s_i y_i^* \; , \CR
\delta l^i &=& - E^i k_i + K^i e_i + e_{i+2}^* k_{i+1}^* - k_{i+2}^* e_{i+1}^*- s_{i+1} y_{i+2}^* + y_{i+1} s_{i+2}^*\; ,  \CR
\delta { x}^i &=& - E^i s_i + K^i y_i + k_{i+1}^* y_{i+2} + k_{i+2} y_{i+1}  - e_{i+2} s_{i+1} - e_{i+1}^* s_{i+2} \; ; \CR
\delta E_0 &=& \sum_i \scal{ L_i K^i + l^i k_i^* - k_i l^{i\, *} - { x}^i s_i^* + s_i { x}^{i\, *} } \; ; \CR
 \delta M &=& E_0 V +  \sum_i \scal{ E^i L_i + e_i l^{i\, *} - l^i e_i^* + { x}^i y_i^* - y_i { x}^{i\, *} } \; . 
 \eea
which disentangles the elements $V, K^i, k_i, y_i, L_i, l^i , M$ associated to the coset component from the elements $s_i, E^i, e_i, x_i, E_0$ of the $\so^*(16)$ subalgebra. The automorphism group then reduces to $SU(2) \times SL(3,\IH)$.

One obtains the maximal algebra by adding three generators associated to split octonions ${\bf r}_i \equiv r_i + \ell t_i $. $t_i$ will be generators in the coset component ${\bf 128}$ whereas $r_i$ will correspond to generators of $\so^*(16)$. The $E_{6(6)}$ automorphism will then be reduced to $Spin(4,4)$, such that octonions only get multiplied through the triality invariant products which preserve $Spin(4,4)$ in the three trial representations $\rho_i$ \footnote{Here $\rho_1, \rho_2, \rho_3$ state for the same element of $Spin(4,4)$ in the chiral spinor, the antichiral spinor and the vector representation, respectively.}
\be \scal{  \rho_i ( {\scriptstyle (} {\bf x} {\bf y}{\scriptstyle )}^* ) }^*= \rho_{i+1}({\bf x}) \rho_{i+2}({\bf y}) \ee
According to this rule, the grading and the Jacobi identity determines completely the algebra up to a redefinition of ${\bf r}_i$, as
\bea
 \delta E^1 &=& V K^1 \; , \quad \delta {\bf k}_3 = K^1 {\bf r}_3  \; , \quad \delta {\bf r}_2 = - {\bf r}_1^* {\bf r}_3^*\; ; \CR
 \delta K^2 &=& {\bf k}_3 {\bf r }_3^* - {\bf r}_3 {\bf k}_3^* \; , \quad \delta {\bf k}_2 = K^1 {\bf r}_2 - {\bf r}_1^* {\bf k}_3^*  \; , \quad \delta  {\bf e}_3 = V {\bf k}_3  + E^1 {\bf r}_3 \; ; \CR
\delta E^2 &=& V K^2 - {\bf r}_3 {\bf e}_3^* + {\bf e}_3 {\bf r}_3^* \; , \quad \delta {\bf k}_1 = K^2 {\bf r}_1 + {\bf k}_3^* {\bf r}_2^* - {\bf r}_3^* {\bf k}_2^* \; , \quad \delta {\bf e}_2 = V {\bf k}_2 + E^1 {\bf r}_2 - {\bf r}_1^* {\bf e}_3^* \; ;   \CR
 \delta K^3 &=& {\bf k}_1 {\bf r }_1^* - {\bf r}_1 {\bf k}_1^*+ {\bf k}_2 {\bf r }_2^* - {\bf r}_2 {\bf k}_2^* \; , \quad
 \delta L_3= E^{1} K^{2} +  E^{2} K^{1} - {\bf e}_3 {\bf k}_3^* + {\bf k}_3 {\bf e}_3^* \; , \CR
  \delta  {\bf e}_1 &=& V {\bf k}_1  + E^2 {\bf r}_1 + {\bf e}_3^* {\bf r}_2^* - {\bf r}_3^* {\bf e}_2^* \; ;  \CR
  \delta E^3 &=& V K^3 - {\bf r}_1 {\bf e}_1^* + {\bf e}_1 {\bf r}_1^*- {\bf r}_2 {\bf e}_2^* + {\bf e}_2 {\bf r}_2^* \; ,\quad \delta {\bf l}^1 = - E^1 {\bf k}_1 + K^1 {\bf e}_1 + {\bf e}_{3}^* {\bf k}_{2}^* - {\bf k}_{3}^* {\bf e}_{2}^* - L_3 {\bf r}_1 \; ; \nn \\
  \delta L_2 &=& E^{3} K^{1} +  E^{1} K^{3} - {\bf e}_2 {\bf k}_2^* + {\bf k}_2 {\bf e}_2^* - {\bf l}_1 {\bf r}_1^* + {\bf r}_1 {\bf l}_1^*  \; ,  \CR
 \delta {\bf l}^2 &=& - E^2 {\bf k}_2 + K^2 {\bf e}_2 + {\bf e}_{1}^* {\bf k}_{3}^* - {\bf k}_{1}^* {\bf e}_{3}^* - L_3 {\bf r}_2 - {\bf l}_1^* {\bf r}_3^*\; ;  \CR
\delta {\bf l}^3 &=& - E^3 {\bf k}_3 + K^3 {\bf e}_3 + {\bf e}_{2}^* {\bf k}_{1}^* - {\bf k}_{2}^* {\bf e}_{1}^*- L_2 {\rm r}_3 + {\bf r}_2^* {\bf l}^{1\, *} - {\bf l}^{2\, *} {\bf r}_1^* \; ; \CR
\delta L_1 &=& E^{2} K^{3} +  E^{3} K^{2} - {\bf e}_1 {\bf k}_1^* + {\bf k}_1 {\bf e}_1^*  - {\bf l}^2 {\bf r}_2^* + {\bf r}_2 {\bf l}^{2\, *} - {\bf l}^3 {\bf r}_3^* + {\bf r}_3 {\bf l}^{3\, *}  \; ; \CR
\delta E_0 &=& \sum_i \scal{ L_i K^i + {\bf l}^i {\bf k}_i^* - {\bf k}_i {\bf l}^{i\, *} } \; ; \CR
 \delta M &=& E_0 V +  \sum_i \scal{ E^i L_i + {\bf e}_i {\bf l}^{i\, *} - {\bf l}^i {\bf e}_i^*  } \; . 
 \eea

Note that there is a particular truncation of this algebra for which there is an enhanced $SO(5,5)$ symmetry, namely for ${\bf r}_3 = 0 $. The algebra then corresponds to the nilpotent orbit \DEVIII00000022, \DSOXVI02000040. A representative element of the maximal nilpotent orbit in the coset component has  $V,K^1,t_3,t_1$ all different from zero.

\subsection{Locally BPS solvable algebra}
There is another pertinent orbit which also defines a system involving 56 harmonic functions, and which contains the BPS system. It can be obtained from the nilpotent orbits of $\e_{8(8)}$ weighted Dynkin diagram \DEVIII20000202 which defines the graded decomposition
\begin{multline} \e_{8(8)} \cong {\bf 1}^\ord{-8} \oplus {\bf 2}^\ord{-7} \oplus {\bf 8}_v^\ord{-6} \oplus ({\bf 2}\oplus {\bf 8}_a )^\ord{-5} \oplus  ({\bf 1}\oplus {\bf 2} \otimes {\bf 8}_c )^\ord{-4} \oplus ({\bf 2}\oplus {\bf 8}_a  \oplus  {\bf 8}_c )^\ord{-3}  \\  \oplus ({\bf 1}\oplus {\bf 2} \otimes {\bf 8}_a \oplus  {\bf 8}_v )^\ord{-2}  \oplus ({\bf 2} \oplus {\bf 2} \otimes {\bf 8}_v \oplus  {\bf 8}_c )^\ord{-1} \oplus \scal{ \gl_1\oplus \gl_1 \oplus \gl_1 \oplus \sl_2 \oplus \so(4,4) }^\ord{0} \\
 \oplus ({\bf 2} \oplus {\bf 2} \otimes {\bf 8}_v \oplus  {\bf 8}_c )^\ord{1} 
 \oplus({\bf 1}\oplus {\bf 2} \otimes {\bf 8}_a \oplus  {\bf 8}_v )^\ord{2}   \oplus ({\bf 2}\oplus {\bf 8}_a  \oplus  {\bf 8}_c )^\ord{3} \\ \oplus ({\bf 1}\oplus {\bf 2} \otimes {\bf 8}_c )^\ord{4}  \oplus ({\bf 2}\oplus {\bf 8}_a )^\ord{5}    \oplus  {\bf 8}_v^\ord{6} \oplus {\bf 2}^\ord{7}\oplus {\bf 1}^\ord{8}  \end{multline} 
Such a nilpotent element admits a non-trivial component in each irreducible representation of the grade one component, including an element of the ${\bf 2}$, a doublet of two linearly independent non null vectors and a non null chiral spinor. The two orbits associated to this graded decompositions are distinguished by the property that the two vectors are of the same signature or not. If they are, the doublet of vectors is left invariant by an $SU(2,2) \subset Spin(4,4)$, and the chiral spinor by $SU(2,1) \subset SU(2,2)$. Otherwise, they are left invariant by an $SL(4) \subset Spin(4,4)$, and the chiral spinor by $SL(3)  \subset SL(4)$.  These two orbits are of dimension 214, and are isomorphic to 
\be E_{8(8)} / \scal{ SU(2,1) \ltimes \IR^{26}} \, , \qquad  E_{8(8)} / \scal{ SL(3) \ltimes \IR^{26}} \ee
respectively. There is one  intersection of each of these orbits with the coset component. The intersection of the second is  associated to the weighted Dynkin diagram \DSOXVI02020220. Its positive grade components do not include the 8 charges of the $STU$ truncation, and we will only discuss it in the following subsection. The intersection of the former is associated to the weighted Dynkin diagram \DSOXVI04020200 (\ie $H_1 + 2H_2 + 4 H_3$) which defines the graded decomposition 
\begin{multline} \so^*(16)  \cong {\bf 1}^\ord{-8}  \oplus ({\bf 2}_2\otimes {\bf 2}_3)^\ord{-6} \oplus ({\bf 2}_1\otimes {\bf 2}_3)^\ord{-5} \oplus  ({\bf 1}\oplus {\bf 2} \otimes {\bf 2}_0\otimes {\bf 2}_3 )^\ord{-4} \oplus ({\bf 2}_1\otimes {\bf 2}_2 \oplus  {\bf 2}_1\otimes {\bf 2}_3)^\ord{-3}  \\  \oplus ({\bf 1}\oplus {\bf 2} \otimes {\bf 2}_0\otimes {\bf 2}_2 \oplus  {\bf 2}_2\otimes {\bf 2}_3)^\ord{-2}  \oplus ( {\bf 2} \otimes {\bf 2}_0\otimes {\bf 2}_1 \oplus  {\bf 2}_1\otimes {\bf 2}_2 )^\ord{-1} \oplus \scal{ \gl_1\oplus \gl_1 \oplus \gl_1 \oplus \sl_2 \oplus \bigoplus_{i=0}^3 \su(2)_i }^\ord{0} \\
 \oplus ( {\bf 2} \otimes {\bf 2}_0\otimes {\bf 2}_1 \oplus  {\bf 2}_1\otimes {\bf 2}_2 )^\ord{1} 
 \oplus({\bf 1}\oplus {\bf 2} \otimes {\bf 2}_0\otimes {\bf 2}_2 \oplus  {\bf 2}_2\otimes {\bf 2}_3 )^\ord{2}   \oplus ({\bf 2}_1\otimes {\bf 2}_2\oplus  {\bf 2}_1\otimes {\bf 2}_3)^\ord{3} \\ \oplus ({\bf 1}\oplus {\bf 2} \otimes {\bf 2}_0\otimes {\bf 2}_3)^\ord{4}  \oplus ( {\bf 2}_1\otimes {\bf 2}_3 )^\ord{5}    \oplus  ({\bf 2}_2\otimes {\bf 2}_3)^\ord{6} \oplus {\bf 1}^\ord{8}  \end{multline} 
with respect to which the coset component decomposes as
\begin{multline} {\bf 128} \cong {\bf 2}^\ord{-7} \oplus ({\bf 2}_0\otimes {\bf 2}_1)^\ord{-6} \oplus ({\bf 2} \oplus  {\bf 2}_0\otimes {\bf 2}_2)^\ord{-5} \oplus  ( {\bf 2} \otimes {\bf 2}_1\otimes {\bf 2}_2 )^\ord{-4} \oplus ({\bf 2} \oplus {\bf 2}_0\otimes {\bf 2}_3 \oplus  {\bf 2}_0\otimes {\bf 2}_2)^\ord{-3}  \\  \oplus ({\bf 2} \otimes {\bf 2}_1\otimes {\bf 2}_3 \oplus  {\bf 2}_0\otimes {\bf 2}_1)^\ord{-2}  \oplus ( {\bf 2} \oplus  {\bf 2} \otimes {\bf 2}_2\otimes {\bf 2}_3 \oplus  {\bf 2}_0\otimes {\bf 2}_3 )^\ord{-1} \oplus ({\bf 2}_0\otimes{\bf 2}_1\otimes{\bf 2}_2\otimes{\bf 2}_3  )^\ord{0} \\
 \oplus ( {\bf 2} \oplus {\bf 2} \otimes {\bf 2}_2\otimes {\bf 2}_3 \oplus  {\bf 2}_0\otimes {\bf 2}_3 )^\ord{1} 
 \oplus({\bf 2} \otimes {\bf 2}_1\otimes {\bf 2}_3 \oplus  {\bf 2}_0\otimes {\bf 2}_1 )^\ord{2}   \oplus ({\bf 2} \oplus {\bf 2}_0\otimes {\bf 2}_3\oplus  {\bf 2}_0\otimes {\bf 2}_2)^\ord{3} \\ \oplus ({\bf 2} \otimes {\bf 2}_1\otimes {\bf 2}_2)^\ord{4}  \oplus ({\bf 2} \oplus  {\bf 2}_0\otimes {\bf 2}_2 )^\ord{5}    \oplus  ({\bf 2}_0\otimes {\bf 2}_1)^\ord{6} \oplus {\bf 2}^\ord{7}  \end{multline} 
A representative of the nilpotent orbit is then defined as an element of the grade one component with a non-zero ${\bf2}$ element, a doublet of linearly independent vectors of $SO(4)_{23}$ and a vector of $SO(4)_{03}$. The latter is only left invariant by the diagonal $SU(2)_{03}$ subgroup of $SO(4)_{03}$, and the doublet of vectors is only left invariant by a subgroup $SO(2)\times SO(2) \subset SL(2) \times SU(2)_2 \times SU(2)_{03}$. The element in the ${\bf 2}$ is not left invariant by the $SO(2)$ factor, such that the orbit is isomorphic to 
\be \Spin^*(16) / \scal{U(2) \ltimes \IR^9} \ee
and is of dimension 107. The associated system involves 56 harmonic functions and includes the BPS system. The property that one needs a doublet of linearly independent quaternions shows that such element does not exist inside the real $\e_{6(6)}$, and therefore the $E_{6(6)} / Sp_{\scriptscriptstyle \rm c}(8,\mathds{R})$ truncation might not be enough to describe this system of equations. 

\vskip 2mm

Similarly as for the almost BPS system, let us consider in a first place the  $E_{6(6)}$ invariant  $D_4$ type nilpotent algebra associated to the orbit \DEVIII00000020 \DSOXVI02000020. In that case one simply replaces the generator $V$ associated to a $\overline{\rm  D6}$ by a $\bar V$ associated to a D6, which gives 
\be \delta {\bf K} = \bar V {\bf E} \; , \quad \delta {\bf L} = {\bf E} \times {\bf K} \; , \quad \quad \delta M =  \trace {\bf E}  {\bf L}  \; , \quad  \delta E_0  = M \bar V + \trace  {\bf L}  {\bf K}    \ee
or after decomposition 
\bea \delta K^i &=& \bar V E^i \; , \quad \delta k_i = \bar V e_i \; \quad \delta s_i = \bar V y_i \CR
\delta L_i &=& E^{i+1} K^{i+2} +  E^{i+2} K^{i+1} - e_i k_i^* + k_i e_i^* + y_i s_i^* - s_i y_i^* \CR
\delta l^i &=& e_{i+2}^* k_{i+1}^* - k_{i+2}^* e_{i+1}^*- E^i k_i + K^i e_i - s_{i+1} y_{i+2}^* + y_{i+1} s_{i+2}^* \CR
\delta { x}^i &=& K^i y_i + k_{i+1}^* y_{i+2} + k_{i+2} y_{i+1} - E^i s_i - e_{i+2} s_{i+1} - e_{i+1}^* s_{i+2} \CR
 \delta M &=& \sum_i \scal{ E^i L_i + e_i l^{i\, *} - l^i e_i^* + { x}^i y_i^* - y_i { x}^{i\, *} } \CR
 \delta E_0 &=& M \bar V + \sum_i \scal{ L_i K^i + l^i k_i^* - k_i l^{i\, *} - { x}^i s_i^* + s_i { x}^{i\, *} } 
 \eea
The maximal solvable algebra can be obtained by adding again the same three octonions ${\bf r}_i$. And because the corresponding generators commute with the generator associated to $\bar V$ (as can be checked from the grading), one gets the algebra
 \bea
\delta E^2 &=&  - {\bf r}_3 {\bf e}_3^* + {\bf e}_3 {\bf r}_3^* \; , \quad   \delta {\bf k}_3 = \bar V {\bf k}_3 +  K^1 {\bf r}_3  \; , \quad \delta {\bf r}_2 = - {\bf r}_1^* {\bf r}_3^* \; , \quad \delta {\bf e}_2 =  - {\bf r}_1^* {\bf e}_3^* \; ; \CR
 \delta K^2 &=&\bar V E^2 +  {\bf k}_3 {\bf r }_3^* - {\bf r}_3 {\bf k}_3^* \; , \quad  \delta L_3= E^{2} K^{1} - {\bf e}_3 {\bf k}_3^* + {\bf k}_3 {\bf e}_3^*  \; ,\CR
   \delta  {\bf e}_1 &=&E^2 {\bf r}_1 + {\bf e}_3^* {\bf r}_2^* - {\bf r}_3^* {\bf e}_2^*  \; , \quad \delta {\bf k}_2 =\bar V {\bf e}_2+  K^1 {\bf r}_2 - {\bf r}_1^* {\bf k}_3^* \; ;  \CR
   \delta E^3 &=&  - {\bf r}_1 {\bf e}_1^* + {\bf e}_1 {\bf r}_1^*- {\bf r}_2 {\bf e}_2^* + {\bf e}_2 {\bf r}_2^* \; ,\quad \delta {\bf l}^1 = K^1 {\bf e}_1 + {\bf e}_{3}^* {\bf k}_{3}^* - {\bf k}_{3}^* {\bf e}_{3}^* - L_3 {\bf r}_1 \; , \CR  
   \delta {\bf k}_1 &=& \bar V {\bf e}_1  + K^2 {\bf r}_1 + {\bf k}_3^* {\bf r}_2^* - {\bf r}_3^* {\bf k}_2^* \; ; \CR
   \delta K^3 &=&\bar V E^3 +  {\bf k}_1 {\bf r }_1^* - {\bf r}_1 {\bf k}_1^*+ {\bf k}_2 {\bf r }_2^* - {\bf r}_2 {\bf k}_2^* \; , \CR
    \delta L_2 &=& E^{3} K^{1} +  E^{1} K^{3} - {\bf e}_2 {\bf k}_2^* + {\bf k}_2 {\bf e}_2^* - {\bf l}_1 {\bf r}_1^* + {\bf r}_1 {\bf l}_1^*  \; ,  \CR
 \delta {\bf l}^2 &=& - E^2 {\bf k}_2 + K^2 {\bf e}_2 + {\bf e}_{1}^* {\bf k}_{3}^* - {\bf k}_{1}^* {\bf e}_{3}^* - L_3 {\bf r}_2 - {\bf l}_1^* {\bf r}_3^*\; ;  \CR
\delta {\bf l}^3 &=& - E^3 {\bf k}_3 + K^3 {\bf e}_3 + {\bf e}_{2}^* {\bf k}_{1}^* - {\bf k}_{2}^* {\bf e}_{1}^* - L_2 {\rm r}_3 + {\bf r}_2^* {\bf l}^{1\, *} - {\bf l}^{2\, *} {\bf r}_1^* \; ; \CR
\delta L_1 &=& E^{2} K^{3} +  E^{3} K^{2} - {\bf e}_1 {\bf k}_1^* + {\bf k}_1 {\bf e}_1^*  - {\bf l}^2 {\bf r}_2^* + {\bf r}_2 {\bf l}^{2\, *} - {\bf l}^3 {\bf r}_3^* + {\bf r}_3 {\bf l}^{3\, *}  \; , \CR
 \delta M &=&   E^2 L_2 + E^3 L_3 +\sum_i \scal{  {\bf e}_i {\bf l}^{i\, *} - {\bf l}^i {\bf e}_i^*  } \; ; \CR
 \delta E_0 &=& M \bar V +  \sum_i \scal{ L_i K^i + {\bf l}^i {\bf k}_i^* - {\bf k}_i {\bf l}^{i\, *} } \; . 
 \eea
Where we have ordered the terms in function of the grading, and we set $E^1$ to zero because it is of grade zero. Note that a closer look to the algebra permits to recognise the $SL(2)$ automorphism of the maximal BPS system (generated by $E^1$ and its conjugate). 

\subsection{Non BPS solvable algebra}
Now let us come back to the orbit associated to the weighted Dynkin diagram \DSOXVI02020220  (\ie $-\frac{1}{2}H_0 + \frac{3}{2}H_1+ \frac{5}{2}H_2+ \frac{7}{2}H_3$), for which $\so^*(16)$ decomposes as
\begin{multline} \so^*(16)  \cong {\bf 1}^\ord{-7}  \oplus ({\bf 2}_2\otimes {\bf 2}_3)^\ord{-6} \oplus ({\bf 1} \oplus {\bf 2}_1\otimes {\bf 2}_3)^\ord{-5} \oplus  ( {\bf 2}_1\otimes {\bf 2}_2 \oplus {\bf 2}_0 \otimes {\bf 2}_3)^\ord{-4} \\ \oplus ({\bf 1} \oplus  {\bf 2}_0\otimes {\bf 2}_3 \oplus  {\bf 2}_0\otimes {\bf 2}_2)^\ord{-3}    \oplus ( {\bf 2}_0\otimes {\bf 2}_1 \oplus  {\bf 2}_1\otimes {\bf 2}_3\oplus  {\bf 2}_0\otimes {\bf 2}_2)^\ord{-2}  \oplus ( {\bf 1}\oplus  {\bf 2}_0\otimes {\bf 2}_1 \oplus  {\bf 2}_1\otimes {\bf 2}_2 \oplus  {\bf 2}_2\otimes {\bf 2}_3 )^\ord{-1} \\ \oplus \scal{ \gl_1\oplus \gl_1 \oplus \gl_1 \oplus \gl_1 \oplus \bigoplus_{i=0}^3 \su(2)_i }^\ord{0}
 \oplus ( {\bf 1}\oplus  {\bf 2}_0\otimes {\bf 2}_1 \oplus  {\bf 2}_1\otimes {\bf 2}_2 \oplus  {\bf 2}_2\otimes {\bf 2}_3 )^\ord{1} 
\\ \oplus( {\bf 2}_0\otimes {\bf 2}_1 \oplus  {\bf 2}_1\otimes {\bf 2}_3\oplus  {\bf 2}_0\otimes {\bf 2}_2)^\ord{2}   \oplus ({\bf 1} \oplus  {\bf 2}_0\otimes {\bf 2}_3 \oplus  {\bf 2}_0\otimes {\bf 2}_2)^\ord{3} \\ \oplus ( {\bf 2}_1\otimes {\bf 2}_2 \oplus {\bf 2}_0 \otimes {\bf 2}_3)^\ord{4}  \oplus ({\bf 1} \oplus {\bf 2}_1\otimes {\bf 2}_3)^\ord{5}    \oplus  ({\bf 2}_2\otimes {\bf 2}_3)^\ord{6} \oplus {\bf 1}^\ord{7}  \end{multline} 
and respectively does the ${\bf 128}$
\begin{multline} {\bf 128}  \cong {\bf 1}^\ord{-8}  \oplus {\bf 1}^\ord{-7}  \oplus ({\bf 2}_0\otimes {\bf 2}_1)^\ord{-6} \oplus ({\bf 1} \oplus {\bf 2}_0\otimes {\bf 2}_2)^\ord{-5} \oplus  ( {\bf 1} \oplus {\bf 2}_0\otimes {\bf 2}_3 \oplus {\bf 2}_1 \otimes {\bf 2}_2)^\ord{-4} \\ \oplus ({\bf 1} \oplus  {\bf 2}_1\otimes {\bf 2}_2 \oplus  {\bf 2}_1\otimes {\bf 2}_3)^\ord{-3}    \oplus ( {\bf 1} \oplus  {\bf 2}_2\otimes {\bf 2}_3 \oplus  {\bf 2}_0\otimes {\bf 2}_2\oplus  {\bf 2}_1\otimes {\bf 2}_3)^\ord{-2}  \oplus ( {\bf 1}\oplus  {\bf 2}_2\otimes {\bf 2}_3 \oplus  {\bf 2}_0\otimes {\bf 2}_3 \oplus  {\bf 2}_0\otimes {\bf 2}_1 )^\ord{-1} \\ \oplus \scal{{\bf 1} \oplus  {\bf 2}_0\otimes{\bf 2}_1\otimes{\bf 2}_2\otimes{\bf 2}_3\oplus {\bf 1} }^\ord{0}
 \oplus ( {\bf 1}\oplus  {\bf 2}_2\otimes {\bf 2}_3 \oplus  {\bf 2}_0\otimes {\bf 2}_3 \oplus  {\bf 2}_0\otimes {\bf 2}_1 )^\ord{1} 
\\ \oplus( {\bf 1} \oplus  {\bf 2}_2\otimes {\bf 2}_3 \oplus  {\bf 2}_0\otimes {\bf 2}_2\oplus  {\bf 2}_1\otimes {\bf 2}_3)^\ord{2}   \oplus ({\bf 1} \oplus  {\bf 2}_1\otimes {\bf 2}_2 \oplus  {\bf 2}_1\otimes {\bf 2}_3)^\ord{3} \\ \oplus ( {\bf 1} \oplus  {\bf 2}_0\otimes {\bf 2}_3 \oplus {\bf 2}_1 \otimes {\bf 2}_2)^\ord{4}  \oplus ({\bf 1} \oplus {\bf 2}_0\otimes {\bf 2}_2)^\ord{5}    \oplus  ({\bf 2}_0\otimes {\bf 2}_1)^\ord{6} \oplus {\bf 1}^\ord{7} \oplus {\bf 1}^\ord{8} \end{multline} 
An element of the orbit is defined as a generic element of grade one. The four components are not left invariant by any combination of the $\gl_1$ and the three ${\bf 2}_i\otimes {\bf 2}_j$ are only left invariant by the diagonal $SU(2)$ of the four $SU(2)_i$, such that the orbit is of dimension 107 and isomorphic to 
\be \Spin^*(16) / \scal{SU(2) \ltimes \IR^{3\times 3+1}} \ee
The positive grade component defines a solvable algebra which almost contains the one associated to non-BPS solutions in the STU model, but a singlet generator associated to one ${\rm D2}$ is in the grade zero component. However, adding such an element to a general element of grade one does not modify the stabiliser subgroup. Let us write $L^\ord{n}, L^\ord{n}_{ij}$ the coefficients of the elements $({\bf 1}\oplus {\bf 2}_i \otimes {\bf 2}_j)^\ord{n}$ in $\so^*(16)$, and $X, X_{ij}$ the elements of the grade one component of the ${\bf 128}$. We will consider the element of the $ {\bf 2}_i \otimes {\bf 2}_j$ as $2\times 2$ matrix which multiply themselves through the contraction of their indices associated to the same $SU(2)$. Note that any $SO(4)$ vector is invertible as a $2\times 2$ matrix. One computes that the elements of the positive grade component of   $\so^*(16)$ which leave invariant a generic element of the grade one component of the ${\bf 128}$ are defined from 10 parameters as
\bea  L^\ord{6}_{23} \cdot X_{23} &=& 0 \, , \qquad L^\ord{5}_{13} = X_{03}{}^{-1} X_{01} L^\ord{5} \, , \qquad L^\ord{4}_{03} \cdot X_{03} = 0 \, , \quad L^\ord{4}_{12} = X_{03}{}^{-1} X_{23} L^\ord{4}_{03} \, , \CR L_{01}^\ord{2} \cdot X_{01} &=& 0 \, , \quad L_{13}^\ord{2} = X^{-1} X_{03} L_{01}^\ord{2} \, , \quad L_{02}^\ord{2} = X^{-1} X_{01}{}^{-1} X_{23} X_{03} L^\ord{2}_{01} \eea
An element which includes moreover a non-trivial element of grade $0$ associated to a ${\rm D2}$ with coefficient $Y$ is also trivially invariant with respect to the diagonal $SU(2)$, and is also left invariant by the positive grade elements as long as 
\be L_{01}^\ord{1} = X^{-1} Y L_{01}^\ord{2} \, , \qquad  L_{03}^\ord{3} = X^{-1} Y L_{03}^\ord{4} \ee
It follows that one can associate the nilpotent element to the deformed grading associated to 
\be -\frac{1}{2}H_0 + \frac{3}{2}H_1+ \frac{5}{2}H_2+ \frac{7}{2}H_3 + \frac{\varepsilon}{2} \scal{ H_0 + H_1 + H_2 + H_3 } \ee
which gives the following graded decomposition of $\so^*(16)$
\begin{multline} \so^*(16)  \cong {\bf 1}^\ord{-7-\varepsilon}  \oplus ({\bf 2}_2\otimes {\bf 2}_3)^\ord{-6-\varepsilon} \oplus ({\bf 1} \oplus {\bf 2}_1\otimes {\bf 2}_3)^\ord{-5-\varepsilon} \oplus  ( {\bf 2}_1\otimes {\bf 2}_2 )^\ord{-4-\varepsilon} \oplus ( {\bf 2}_0 \otimes {\bf 2}_3)^\ord{-4} \\ \oplus ({\bf 1} \oplus  {\bf 2}_0\otimes {\bf 2}_3 )^\ord{-3-\varepsilon} \oplus  ({\bf 2}_0\otimes {\bf 2}_2)^\ord{-3}  \oplus  ({\bf 2}_0\otimes {\bf 2}_2)^\ord{-2-\varepsilon}  \oplus ( {\bf 2}_0\otimes {\bf 2}_1 \oplus  {\bf 2}_1\otimes {\bf 2}_3)^\ord{-2} \oplus  ({\bf 2}_0\otimes {\bf 2}_1)^\ord{-1-\varepsilon}  \\ \oplus (  {\bf 2}_1\otimes {\bf 2}_2 \oplus  {\bf 2}_2\otimes {\bf 2}_3 )^\ord{-1} \oplus {\bf 1}^\ord{-1+\varepsilon}  \oplus \scal{ \gl_1\oplus \gl_1 \oplus \gl_1 \oplus \gl_1 \oplus \bigoplus_{i=0}^3 \su(2)_i }^\ord{0} \oplus {\bf 1}^\ord{1-\varepsilon} \\ \oplus (  {\bf 2}_1\otimes {\bf 2}_2 \oplus  {\bf 2}_2\otimes {\bf 2}_3 )^\ord{1}  \oplus  ({\bf 2}_0\otimes {\bf 2}_1)^\ord{1+\varepsilon} \oplus ( {\bf 2}_0\otimes {\bf 2}_1 \oplus  {\bf 2}_1\otimes {\bf 2}_3)^\ord{2} \oplus  ({\bf 2}_0\otimes {\bf 2}_2)^\ord{2+\varepsilon}  \oplus  ({\bf 2}_0\otimes {\bf 2}_2)^\ord{3} \\ \oplus ({\bf 1} \oplus  {\bf 2}_0\otimes {\bf 2}_3 )^\ord{3+\varepsilon}\oplus ( {\bf 2}_0 \otimes {\bf 2}_3)^\ord{4}  \oplus  ( {\bf 2}_1\otimes {\bf 2}_2 )^\ord{4+\varepsilon}\oplus ({\bf 1} \oplus {\bf 2}_1\otimes {\bf 2}_3)^\ord{5+\varepsilon}\oplus ({\bf 2}_2\otimes {\bf 2}_3)^\ord{6+\varepsilon} \oplus {\bf 1}^\ord{7+\varepsilon}  \end{multline} 
and of the ${\bf 128}$
\begin{multline} {\bf 128}  \cong {\bf 1}^\ord{-8-\varepsilon}  \oplus {\bf 1}^\ord{-7-2\varepsilon}  \oplus ({\bf 2}_0\otimes {\bf 2}_1)^\ord{-6-\varepsilon} \oplus ( {\bf 2}_0\otimes {\bf 2}_2)^\ord{-5-\varepsilon} \oplus {\bf 1} ^\ord{-5} \oplus ( {\bf 1} \oplus  {\bf 2}_0\otimes {\bf 2}_3)^\ord{-4-\varepsilon} \\ \oplus  ( {\bf 2}_1 \otimes {\bf 2}_2)^\ord{-4}   \oplus ( {\bf 2}_1\otimes {\bf 2}_2 )^\ord{-3-\varepsilon}  \oplus ({\bf 1} \oplus  {\bf 2}_1 \otimes {\bf 2}_3)^\ord{-3} \oplus ({\bf 1}\oplus  {\bf 2}_1 \otimes {\bf 2}_3)^\ord{-2-\varepsilon}    \\  \oplus ( {\bf 2}_2\otimes {\bf 2}_3 \oplus  {\bf 2}_0\otimes {\bf 2}_2)^\ord{-2} \oplus (  {\bf 2}_2\otimes {\bf 2}_3)^\ord{-1-\varepsilon}  \oplus ( {\bf 1} \oplus  {\bf 2}_0\otimes {\bf 2}_3 \oplus  {\bf 2}_0\otimes {\bf 2}_1 )^\ord{-1}  \oplus {\bf 1}^\ord{-\varepsilon}  \\ \oplus \scal{ {\bf 2}_0\otimes{\bf 2}_1\otimes{\bf 2}_2\otimes{\bf 2}_3}^\ord{0} \oplus {\bf 1}^\ord{\varepsilon} 
 \oplus ( {\bf 1}  \oplus  {\bf 2}_0\otimes {\bf 2}_3 \oplus  {\bf 2}_0\otimes {\bf 2}_1 )^\ord{1} \oplus (  {\bf 2}_2\otimes {\bf 2}_3)^\ord{1+\varepsilon} 
\\ \oplus(   {\bf 2}_2\otimes {\bf 2}_3 \oplus  {\bf 2}_0\otimes {\bf 2}_2)^\ord{2}\oplus ({\bf 1}\oplus  {\bf 2}_1\otimes {\bf 2}_3)^\ord{2+\varepsilon}   \oplus ({\bf 1} \oplus  {\bf 2}_1\otimes {\bf 2}_3)^\ord{3} \oplus ( {\bf 2}_1\otimes {\bf 2}_2 )^\ord{3+\varepsilon} \\ \oplus ( {\bf 2}_1 \otimes {\bf 2}_2)^\ord{4} \oplus ( {\bf 1} \oplus  {\bf 2}_0\otimes {\bf 2}_3)^\ord{4+\varepsilon}  \oplus {\bf 1}^\ord{5} \oplus ( {\bf 2}_0\otimes {\bf 2}_2)^\ord{5+\varepsilon}   \oplus  ({\bf 2}_0\otimes {\bf 2}_1)^\ord{6+\varepsilon} \oplus {\bf 1}^\ord{7+2\varepsilon} \oplus {\bf 1}^\ord{8+\varepsilon}  \end{multline} 
Although this is slightly more subtle in this case, one gets in this way a solvable algebra which defines a solvable system of differential equations involving 56 harmonic functions which generalises the composite non-BPS system of the STU model. Of course we could fixe $\varepsilon$ to any fixed value, for instance $\varepsilon=1$ permits to exhibit an additional $SL(2)$ symmetry of the system. The graded decomposition is then defined with respect to $2H_1 + 3H_2 + 4 H_3$, and gives 
\begin{multline} \so^*(16)  \cong {\bf 1}^\ord{-8}  \oplus ({\bf 2}_2\otimes {\bf 2}_3)^\ord{-7} \oplus ({\bf 1} \oplus {\bf 2}_1\otimes {\bf 2}_3)^\ord{-6} \oplus  ( {\bf 2}_1\otimes {\bf 2}_2 )^\ord{-5}  \\ \oplus ({\bf 1} \oplus  {\bf 2}\otimes {\bf 2}_0\otimes {\bf 2}_3 )^\ord{-4} \oplus  ({\bf 2} \otimes {\bf 2}_0\otimes {\bf 2}_2)^\ord{-3}  \oplus ({\bf 2}\otimes  {\bf 2}_0\otimes {\bf 2}_1 \oplus  {\bf 2}_1\otimes {\bf 2}_3)^\ord{-2}  \\ \oplus (  {\bf 2}_1\otimes {\bf 2}_2 \oplus  {\bf 2}_2\otimes {\bf 2}_3 )^\ord{-1}  \oplus \scal{ \gl_1\oplus \gl_1 \oplus \gl_1 \oplus \sl_2 \oplus \bigoplus_{i=0}^3 \su(2)_i }^\ord{0}  \\ \oplus (  {\bf 2}_1\otimes {\bf 2}_2 \oplus  {\bf 2}_2\otimes {\bf 2}_3 )^\ord{1}  \oplus ({\bf 2}\otimes  {\bf 2}_0\otimes {\bf 2}_1 \oplus  {\bf 2}_1\otimes {\bf 2}_3)^\ord{2} \oplus  ({\bf 2} \otimes {\bf 2}_0\otimes {\bf 2}_2)^\ord{3} \\ \oplus ({\bf 1} \oplus  {\bf 2}\otimes {\bf 2}_0\otimes {\bf 2}_3 )^\ord{4}  \oplus  ( {\bf 2}_1\otimes {\bf 2}_2 )^\ord{5}\oplus ({\bf 1} \oplus {\bf 2}_1\otimes {\bf 2}_3)^\ord{6}\oplus ({\bf 2}_2\otimes {\bf 2}_3)^\ord{7} \oplus {\bf 1}^\ord{8}  \end{multline} 
and of the ${\bf 128}$
\begin{multline} {\bf 128}  \cong {\bf 2}^\ord{-9}    \oplus ({\bf 2}_0\otimes {\bf 2}_1)^\ord{-7} \oplus ( {\bf 2}_0\otimes {\bf 2}_2)^\ord{-6} \oplus  ( {\bf 2} \oplus  {\bf 2}_0\otimes {\bf 2}_3)^\ord{-5} \oplus  ( {\bf 2} \otimes {\bf 2}_1 \otimes {\bf 2}_2)^\ord{-4}   \\   \oplus ({\bf 2} \oplus  {\bf 2}\otimes {\bf 2}_1 \otimes {\bf 2}_3)^\ord{-3}   \oplus ({\bf 2}\otimes  {\bf 2}_2\otimes {\bf 2}_3 \oplus  {\bf 2}_0\otimes {\bf 2}_2)^\ord{-2}  \oplus ( {\bf 2} \oplus  {\bf 2}_0\otimes {\bf 2}_3 \oplus  {\bf 2}_0\otimes {\bf 2}_1 )^\ord{-1}   \\ \oplus \scal{ {\bf 2}_0\otimes{\bf 2}_1\otimes{\bf 2}_2\otimes{\bf 2}_3}^\ord{0}  \oplus ( {\bf 2}  \oplus  {\bf 2}_0\otimes {\bf 2}_3 \oplus  {\bf 2}_0\otimes {\bf 2}_1 )^\ord{1} \oplus ({\bf 2}\otimes  {\bf 2}_2\otimes {\bf 2}_3 \oplus  {\bf 2}_0\otimes {\bf 2}_2)^\ord{2}\oplus({\bf 2} \oplus  {\bf 2}\otimes {\bf 2}_1 \otimes {\bf 2}_3)^\ord{3}  \\  \oplus( {\bf 2} \otimes {\bf 2}_1 \otimes {\bf 2}_2)^\ord{4} \oplus ( {\bf 2} \oplus  {\bf 2}_0\otimes {\bf 2}_3)^\ord{5} \oplus ( {\bf 2}_0\otimes {\bf 2}_2)^\ord{6}   \oplus  ({\bf 2}_0\otimes {\bf 2}_1)^\ord{7} \oplus {\bf 2}^\ord{9} \end{multline}

\vskip 2mm

In order to obtain the `maximal' non-BPS nilpotent algebra from the almost BPS one, one can again simply replace the ${\bf K}$ generators associated to D4's by conjugates $\bar {\bf K}$ associated to $\overline{\rm D4}$. Using the already derived commutation relations, it remains only to compute the ones involving ${\bf r}_i$ and $\bar {\bf K}$. But noting that the algebra admits an $SL(2)$ automorphism with respect to which $(\bar {\bf K} , {\bf L})$ transforms as a doublet and ${\bf r}_i$ is invariant, one gets these relations without effort. 
\bea
\delta {\bf r}_2 &=& - {\bf r}_1^* {\bf r}_3^*\;  , \quad  \delta {\bf k}^1 = - K_3 {\bf r}_1 \;  , \quad  \delta {\bf l}^1 = - L_3 {\bf r}_1 \; ; \CR
  \delta K_2 &=&  - {\bf k}_1 {\bf r}_1^* + {\bf r}_1 {\bf k}_1^*  \; ,  \quad   \delta L_2 =  - {\bf l}_1 {\bf r}_1^* + {\bf r}_1 {\bf l}_1^* \; ,  \CR
    \delta {\bf k}^2 &=&  - K_3 {\bf r}_2 - {\bf k}_1^* {\bf r}_3^*  \; ,  \quad  \delta {\bf l}^2 =  - L_3 {\bf r}_2 - {\bf l}_1^* {\bf r}_3^*\; ;   \CR
   \delta E^1 &=& K_2 L_3 + K_3 L_2 - {\bf k}^1 {\bf l}^{1\, *} + {\bf l}^1 {\bf k}^{1\, *}  \; , \CR
 \delta {\bf k}^3 &=& - K_2 {\rm r}_3 + {\bf r}_2^* {\bf k}^{1\, *} - {\bf k}^{2\, *} {\bf r}_1^*  \; , \quad \delta {\bf l}^3 = - L_2 {\rm r}_3 + {\bf r}_2^* {\bf l}^{1\, *} - {\bf l}^{2\, *} {\bf r}_1^* \; ; \CR
 \delta K_1 &=&  - {\bf k}^2 {\bf r}_2^* + {\bf r}_2 {\bf k}^{2\, *} - {\bf k}^3 {\bf r}_3^* + {\bf r}_3 {\bf k}^{3\, *}  \; , \quad  \delta L_1 =  - {\bf l}^2 {\bf r}_2^* + {\bf r}_2 {\bf l}^{2\, *} - {\bf l}^3 {\bf r}_3^* + {\bf r}_3 {\bf l}^{3\, *}  \; , \CR
 \delta  {\bf e}_3 &=& - K_3 {\bf l}^3 + L_3 {\bf k}^3 + {\bf k}^{2\, *}{\bf l}^{1\, *} -  {\bf l}^{2\, *}{\bf k}^{1\, *}   + E^1 {\bf r}_3 \; ; \nn \\ 
\delta E^2 &=&   K_3 L_1 + K_1 L_3 - {\bf k}^2 {\bf l}^{2\, *} + {\bf l}^2 {\bf k}^{2\, *}  - {\bf r}_3 {\bf e}_3^* + {\bf e}_3 {\bf r}_3^* \; , \CR
 \delta {\bf e}_2 &=&  - K_2 {\bf l}^2 + L_2 {\bf k}^2 + {\bf k}^{1\, *}{\bf l}^{3\, *} -  {\bf l}^{1\, *}{\bf k}^{3\, *}   + E^1 {\bf r}_2 - {\bf r}_1^* {\bf e}_3^* \; ;   \CR
   \delta  {\bf e}_1 &=&  - K_1 {\bf l}^1 + L_1 {\bf k}^1 + {\bf k}^{3\, *}{\bf l}^{2\, *} -  {\bf l}^{3\, *}{\bf k}^{2\, *}   + E^2 {\bf r}_1 + {\bf e}_3^* {\bf r}_2^* - {\bf r}_3^* {\bf e}_2^* \; ;  \CR
  \delta E^3 &=& K_1 L_2 + K_2 L_1 - {\bf k}^3 {\bf l}^{3\, *} + {\bf l}^3 {\bf k}^{3\, *} - {\bf r}_1 {\bf e}_1^* + {\bf e}_1 {\bf r}_1^*- {\bf r}_2 {\bf e}_2^* + {\bf e}_2 {\bf r}_2^* \;  ; \CR
   \delta V &=&  \sum_i \scal{ E^i K_i + {\bf e}_i {\bf k}^{i\, *} - {\bf k}^i {\bf e}_i^*  } \; , \quad
 \delta M =  \sum_i \scal{ E^i L_i + {\bf e}_i {\bf l}^{i\, *} - {\bf l}^i {\bf e}_i^*  } \; . 
 \eea
 
The role of the three new quaternionic functions $T_i$ in the system is  clearly to relax the constraint (\ref{FirstOrder}), and to allow the scalar fields $\mathpzc{t}$ to carry a non-trivial $\ell$ component in its imaginary part. The Ansatz is significantly more complicated in this case. Moreover the system of differential equations is also much more involved, and the functions $V$ and $M$ are now sourced by polynomials of order seven in the harmonic functions.  The definition and the analysis of the regularity of these solutions requires further studies.

\section*{Acknowledgments}

We would like to thank  Karine Beauchard, Iosif Bena,  Anna Ceresole, Murat G\"{u}naydin,  Stefanos Katmadas, Jan Manschot, Alessio Marrani, Ilarion Melnikov, Hermann Nicolai, Boris Pioline, Armen Yeranyan and particularly Cl\'ement Ruef for useful discussions. This work was supported by the ERC Advanced Grant  226371, the ITN programme PITN-GA-2009-237920, the IFCPAR CEFIPRA programme 4104-2 and the ANR programme NT09-573739 ``string cosmo''.

\appendix 
\section{Solution of the Laplace equation with three point sources}
\label{Laplace}
We define the solution to the  Laplace equation
\be \Delta F_{A,BC}  =  \frac{ 2 (x-x_\pB)\hspace{-1mm}\cdot \hspace{-1mm}(x - x_\pC)}{|x-x_{\pA}| |x-x_{\pB}|^3 |x-x_{\pC}|^3} \ee
as the integral 
 \be F_{A,BC} = - \int \frac{d^3 y}{2\pi} \frac{ (y-x_\pB)\hspace{-1mm}\cdot \hspace{-1mm}(y - x_\pC)}{|y-x| |y-x_{\pA}| |y-x_{\pB}|^3 |y-x_{\pC}|^3} \ee
This integral is convergent for any value of $x$. Being regular everywhere, this integral is determined to be the solution of the Laplace equation which admits no poles and which vanishes in the asymptotic region $x\rightarrow \infty$.  

As such, it follows that if $x_\pA, x_\pB$ and $x_\pC$ are aligned, the function is given by the axisymmetric solution 
 \be F^{\rm \scriptscriptstyle Axsym}_{A,BC} = \frac{1}{(x_\pB-x_\pA)\hspace{-1mm}\cdot \hspace{-1mm}(x_\pC - x_\pA)} \biggl( \frac{|x-x_\pA|}{|x-x_{\pB}||x-x_{\pC}|} - \frac{|x_\pB-x_\pA|}{|x_\pB-x_{\pC}| |x-x_{\pB}|}-\frac{|x_\pC-x_\pA|}{|x_\pB-x_{\pC}||x-x_{\pC}|}\biggr) \label{AxiABC}  \ee
We note also that by definition the function is invariant with respect to the exchange of $x$ and $x_\pA$, therefore whenever $x$ is aligned with respect to $x_\pB$ and $x_\pC$, it reduces to 
 \be F_{A,BC} = \frac{1}{(x-x_\pB)\hspace{-1mm}\cdot \hspace{-1mm}(x-x_\pC)} \biggl( \frac{|x-x_\pA|}{|x_\pA\hspace{-1mm}-x_{\pB}||x_\pA\hspace{-1mm} -x_{\pC}|} - \frac{|x_\pB-x|}{|x_\pB\hspace{-1mm}-x_{\pC}| |x_\pA\hspace{-1mm}-x_{\pB}|}-\frac{|x_\pC-x|}{|x_\pB\hspace{-1mm}-x_{\pC}||x_\pA\hspace{-1mm}-x_{\pC}|}\biggr) \label{IntegralFormulaFabc} \ee
The symmetry with respect to the interchange $x \leftrightarrow x_\pA$ will play an important role in the following. 

However, if the axisymmetric solution is not too hard to compute, the general solution is rather difficult to obtain. We will study the asymptotic expansion of the general solution in the asymptotic region in this appendix. But let us first show that the integral (\ref{IntegralFormulaFabc}) indeed converges for all values of $x$. To see this, let us study the limit $y\rightarrow x_\pB$ in the most dangerous case, \ie when $x\rightarrow x_\pB$ as well. 

\subsubsection*{Limit $x \rightarrow x_\pB$}
We consider the integral on a very small ball $\mathcal{B}_\pB$ of radius $\epsilon$ surrounding $x_\pB$, such that $x$ lies in the ball. In this case we can expand the integrand in $y-x_\pB$,  (note that $|x-x_\pB|<\epsilon$) 
 \bea && - \int_{{\mathcal{B}_\pB}}  \frac{d^3 y}{2\pi} \frac{ (y-x_\pB)\hspace{-1mm}\cdot \hspace{-1mm}(y - x_\pC)}{|y-x| |y-x_{\pA}| |y-x_{\pB}|^3 |y-x_{\pC}|^3}  + \mathcal{O}(\epsilon) 
\CR
& =& \frac{ ( x-x_\pB) \hspace{-1mm}\cdot \hspace{-1mm}(x_\pC - x_\pB)}{ |x_\pA-x_\pB||x_\pC-x_\pB|^3 |x-x_\pB|}  \int_0^\epsilon dr \int_{-1}^1d\cos\theta \frac{ \cos \theta}{ \sqrt{ r^2 - 2 \cos \theta \, r |x-x_\pB| + |x-x_\pB|^2} } \CR
&=&  \frac{ ( x-x_\pB) \hspace{-1mm}\cdot \hspace{-1mm}(x_\pC - x_\pB)}{ |x_\pA-x_\pB||x_\pC-x_\pB|^3 |x-x_\pB|} \Bigl( \int_0^{|x-x_\pB|}  dr \frac{2r}{3 |x-x_\pB|^2} +  \int_{|x-x_\pB|}^{\epsilon}  dr \frac{2|x-x_\pB|}{3 r^2} \Bigr) \CR
&=&  \frac{ ( x-x_\pB) \hspace{-1mm}\cdot \hspace{-1mm}(x_\pC - x_\pB)}{ |x_\pA-x_\pB||x_\pC-x_\pB|^3 |x-x_\pB|} \Bigl( 1 - \frac{2  |x-x_\pB|}{3 \epsilon} \Bigr) \label{InteBall}
 \eea
Let us now compute the explicit expression of the function $F_{A,BC}$ at $x \rightarrow x_\pB$. In order to do this computation we will first decompose $\mathds{R}^3$ into the interior of the ball $\mathcal{B}_\pB$ surrounding $x$ in the limit $|x-x_\pB| << \epsilon$, and the exterior of this ball. In the interior one gets the same result as in (\ref{InteBall}), with  $\frac{|x-x_\pB|}{\epsilon} \rightarrow  0 $. In the exterior, one can expand the integrand in the harmonics centred at $x=x_\pB$, such that 
\bea  &&- \int_{\mathds{R}^3 \setminus {\mathcal{B}_\pB}}  \frac{d^3 y}{2\pi} \frac{ (y-x_\pB)\hspace{-1mm}\cdot \hspace{-1mm}(y - x_\pC)}{|y-x| |y-x_{\pA}| |y-x_{\pB}|^3 |y-x_{\pC}|^3}    \CR
&=& - \int_{\mathds{R} \setminus {\mathcal{B}_\pB}}  \frac{d^3 y}{2\pi} \frac{ (y-x_\pB)\hspace{-1mm}\cdot \hspace{-1mm}(y - x_\pC)}{|y-x_{\pA}| |y-x_{\pB}|^4 |y-x_{\pC}|^3} + \mathcal{O}(x-x_\pB) \eea 
The singular component in $\epsilon$ vanishes by symmetry, and one can therefore define the limit  at $x \rightarrow x_\pB$ of $F_{A,BC}$ as
\be \lim_{x\rightarrow x_\pB} F_{A,BC} =   \frac{ ( x-x_\pB) \hspace{-1mm}\cdot \hspace{-1mm}(x_\pC - x_\pB)}{ |x_\pA-x_\pB||x_\pC-x_\pB|^3 |x-x_\pB|}  - \int_{\mathds{R}^3}  \frac{d^3 y}{2\pi} \frac{ (y-x_\pB)\hspace{-1mm}\cdot \hspace{-1mm}(y - x_\pC)}{|y-x_{\pA}| |y-x_{\pB}|^4 |y-x_{\pC}|^3} \ee
Up to a harmonic function in $x_\pA$, one obtains 
\begin{multline}  F_{A,BC}  = \frac{  (x-x_\pB)\hspace{-1mm}\cdot \hspace{-1mm}(x_\pC - x_\pB)}{ |x-x_\pB| |x_\pA-x_\pB| |x_\pB - x_\pC|^3}\\ +  \frac{  (x_\pA-x_\pB)\hspace{-1mm}\cdot \hspace{-1mm}(x_\pC - x_\pB)}{ |x_\pA-x_\pB|^2|x_\pA-x_\pC|  |x_\pB - x_\pC|^2} - \frac{1}{|x_\pA - x_\pC| |x_\pB - x_\pC|^2}+\mathcal{O}(x-x_\pB) \label{LimitxB}  \end{multline}
Using the property that this expression agrees with the axisymmetric one (\ref{AxiABC}), one concludes that the possible additional harmonic function in  $x_\pA$ must vanish when $x_\pA$ is on the line $(x_\pB,x_\pC)$. Moreover the integral is regular in the limit $x_\pA \rightarrow \infty$, and one concludes that (\ref{LimitxB}) is indeed the right expression.

 Note that using the symmetry with respect to the interchange $x \leftrightarrow x_\pA$, one obtains as well the limit of $F_{A,BC}$ as $x_\pA \rightarrow x_\pB$
\begin{multline}  F_{A,BC}  = \frac{  (x_\pA-x_\pB)\hspace{-1mm}\cdot \hspace{-1mm}(x_\pC - x_\pB)}{ |x-x_\pB| |x_\pA-x_\pB| |x_\pB - x_\pC|^3}\\ +  \frac{  (x -x_\pB)\hspace{-1mm}\cdot \hspace{-1mm}(x_\pC - x_\pB)}{ |x -x_\pB|^2|x -x_\pC|  |x_\pB - x_\pC|^2} - \frac{1}{|x  - x_\pC| |x_\pB - x_\pC|^2}+\mathcal{O}(x_\pA-x_\pB) \label{LimitxAB}  \end{multline} 
 
 \subsubsection*{Limit $x \rightarrow x_\pA$}
 The limit $x \rightarrow x_\pA$ is clearly given by the convergent integral 
 \bea  F_{A,BC}(x_\pA) &=& - \int \frac{d^3 y}{2\pi} \frac{ (y-x_\pB)\hspace{-1mm}\cdot \hspace{-1mm}(y - x_\pC)}{ |y-x_{\pA}|^2 |y-x_{\pB}|^3 |y-x_{\pC}|^3} \CR
 &=& - \frac{\partial\, }{\partial x_\pB} \cdot \frac{\partial\, }{\partial x_\pC} \int \frac{d^3 y}{2\pi} \frac{1}{ |y-x_{\pA}|^2 |y-x_{\pB}| |y-x_{\pC}|} \CR
 &=& \left( \frac{\partial\, }{\partial x_\pA} \cdot \frac{\partial\, }{\partial x_\pC}  + \Delta_{x_\pC} \right) \int \frac{d^3 y}{2\pi} \frac{1}{ |y-x_{\pA}|^2 |y-x_{\pB}| |y-x_{\pC}|} \CR
 &=& 2  \int \frac{d^3 y}{2\pi} \frac{ (y-x_\pA)\hspace{-1mm}\cdot \hspace{-1mm}(y - x_\pC)}{ |y-x_{\pB}| |y-x_{\pA}|^4 |y-x_{\pC}|^3} - \frac{2}{|x_\pA-x_\pC|^2|x_\pB-x_\pC|}
 \eea
and therefore 
\be \Delta_{x_\pB}  F_{A,BC}(x_\pA)  = - 4 \frac{ (x_\pB-x_\pA)\hspace{-1mm}\cdot \hspace{-1mm}(x_\pB - x_\pC)}{|x_\pB-x_{\pA}|^4 |x_\pB-x_{\pC}|^3} \ee
We conclude that 
\be F_{A,BC}(x_\pA)  =  - 2 \frac{(x_\pB-x_\pA)\hspace{-1mm}\cdot \hspace{-1mm}(x_\pC - x_\pA)}{|x_\pB-x_\pA|^2 |x_\pC-x_\pA|^2 |x_\pB-x_\pC|} + H_{BC} \ee
for a harmonic function in $x_\pB$, symmetric in the interchange of $x_\pB$ and $x_\pC$. Comparing with the axisymmetric case (\ref{AxiABC}), one obtains that $H_{BC}$ must vanish whenever $x_\pA,x_\pB,x_\pC$ are aligned. Moreover, analysing the asymptotic behaviour of $ F_{A,BC}(x_\pA) $ in the limit $|x_\pB - x_\pA| \rightarrow \infty$, one obtains that $H_{BC} = \mathcal{O}( |x_\pB - x_\pA|^{-3})$. We conclude therefore that $H_{BC}$ must vanish identically and that 
\be  F_{A,BC}(x_\pA)  =  - 2 \frac{(x_\pB-x_\pA)\hspace{-1mm}\cdot \hspace{-1mm}(x_\pC - x_\pA)}{|x_\pB-x_\pA|^2 |x_\pC-x_\pA|^2 |x_\pB-x_\pC|} \ee

\subsubsection*{Limit $x \rightarrow \infty$}
Let us now study the limit $x\rightarrow \infty$. We consider a ball $\mathcal{B}_\Lambda$ centred at $ 0$ of radius $\Lambda >> \sup( |x_\pA|, |x_\pB |, |x_\pC|)$, but such that $|x|>> \Lambda$. We decompose the integral into the integrals over the interior and the exterior of the ball. On the exterior, $y$ is very large everywhere and one can expand the integrand as a Laurent series in $y$
\bea && - \int_{\mathds{R}^3 \setminus {\mathcal{B}_\Lambda}}  \frac{d^3 y}{2\pi} \frac{ (y-x_\pB)\hspace{-1mm}\cdot \hspace{-1mm}(y - x_\pC)}{|y-x| |y-x_{\pA}| |y-x_{\pB}|^3 |y-x_{\pC}|^3}  \CR
&=& - \int_{\mathds{R}^3 \setminus {\mathcal{B}_\Lambda}}  \frac{d^3 y}{2\pi}\biggl(  \frac{1}{|y-x| |y|^5} + \frac{ ( x_\pA + 2 x_\pB + 2 x_\pC)\hspace{-1mm}\cdot \hspace{-1mm} y }{|y-x| |y|^7} + \mathcal{O}(y^{-7}) \biggr)\CR
&=& \frac{1}{3|x|^3} + \frac{ ( x_\pA + 2 x_\pB + 2 x_\pC)\hspace{-1mm}\cdot \hspace{-1mm} x }{5 |x|^5} + \mathcal{O}(x^{-5})\CR
&& \hspace{25mm}  - \frac{1}{\Lambda^2} \biggl( \frac{1}{|x|} +  \frac{ ( x_\pA + 2 x_\pB + 2 x_\pC)\hspace{-1mm}\cdot \hspace{-1mm} x }{3 |x|^3} + \mathcal{O}(x^{-3}) \biggr)  + \mathcal{O}( \Lambda^{-3}) \label{Exter}
  \eea
 On the interior, $x$ is very large compared to $y$ itself, and one can expend the integral in the harmonics centred at $x=0$. 
 \bea && - \int_{ {\mathcal{B}_\Lambda}}  \frac{d^3 y}{2\pi} \frac{ (y-x_\pB)\hspace{-1mm}\cdot \hspace{-1mm}(y - x_\pC)}{|y-x| |y-x_{\pA}| |y-x_{\pB}|^3 |y-x_{\pC}|^3}  \CR
 &=& - \frac{1}{|x|} \int_{ {\mathcal{B}_\Lambda}}  \frac{d^3 y}{2\pi} \frac{ (y-x_\pB)\hspace{-1mm}\cdot \hspace{-1mm}(y - x_\pC)}{ |y-x_{\pA}| |y-x_{\pB}|^3 |y-x_{\pC}|^3} \CR && \hspace{25mm} - \frac{x}{|x|^3} \cdot \int_{ {\mathcal{B}_\Lambda}}  \frac{d^3 y}{2\pi} y \frac{ (y-x_\pB)\hspace{-1mm}\cdot \hspace{-1mm}(y - x_\pC)}{ |y-x_{\pA}| |y-x_{\pB}|^3 |y-x_{\pC}|^3} + \mathcal{O}(x^{-3}) \label{Inter} \eea
 By definition the terms in $\Lambda$ cancel each others, as one can straightforwardly check at this order, by computing the contribution to these integrals in the neighbourhood of the sphere $|y|= \Lambda$.  Defining $\mbox{Lau}_{x\rightarrow \infty}(F)$ as the Laurent series in $x$ of the expansion of the function $F$ at $x\rightarrow \infty$, we conclude that 
 \begin{multline} \mbox {Lau}_{x\rightarrow \infty} \scal{ F_{A,BC} } = - \lim_{\Lambda \rightarrow \infty}  \int_{\mathds{R}^3 \setminus {\mathcal{B}_\Lambda}}  \frac{d^3 y}{2\pi} \frac{1}{|y-x|}  \mbox{Lau}_{y\rightarrow \infty}\left( \frac{ (y-x_\pB)\hspace{-1mm}\cdot \hspace{-1mm}(y - x_\pC)}{ |y-x_{\pA}| |y-x_{\pB}|^3 |y-x_{\pC}|^3} \right) \\
 -  \int_{\mathds{R}^3}   \frac{d^3 y}{2\pi} \mbox{Lau}_{x\rightarrow \infty}\left(  \frac{1}{|y-x|}\right) \frac{ (y-x_\pB)\hspace{-1mm}\cdot \hspace{-1mm}(y - x_\pC)}{ |y-x_{\pA}| |y-x_{\pB}|^3 |y-x_{\pC}|^3} 
  \end{multline} 
Note that the first term is also a Laurent series in $x$, because each term in the Laurent expansion in $y$ will have a definite scaling in $y$, such that the corresponding integral will have a definite scaling in $x$ (after having taken the limit $\Lambda \rightarrow \infty$) .

Applying the Laplace operator, one directly obtains that 
\bea \Delta \mbox {Lau}_{x\rightarrow \infty} \scal{ F_{A,BC} } &=&   \lim_{\Lambda \rightarrow \infty}  \int_{\mathds{R}^3 \setminus {\mathcal{B}_\Lambda}}  d^3 y \delta^\ord{3}(x-y)  \mbox{Lau}_{y\rightarrow \infty}\left( \frac{ 2 (y-x_\pB)\hspace{-1mm}\cdot \hspace{-1mm}(y - x_\pC)}{ |y-x_\pA | |y-x_{\pB}|^3 |y-x_{\pC}|^3} \right)  \CR&=&  \mbox{Lau}_{x \rightarrow \infty}\left(    \frac{ 2 (x-x_\pB)\hspace{-1mm}\cdot \hspace{-1mm}(x - x_\pC)}{ |x-x_\pA| |x-x_{\pB}|^3 |x-x_{\pC}|^3} \right) \eea
because 
\be  \mbox{Lau}_{x\rightarrow \infty}\scal{  \delta^\ord{3}(x-y)} = 0 \ee
and $\mathcal{B}_\Lambda$ does not contain $x$ (before to take the limit) by definition. Therefore it follows that $ \mbox {Lau}_{x\rightarrow \infty} \scal{ F_{A,BC} } $ satisfies the correct Laplace equation as a Laurent series. 

In order to confirm that this formula makes sense, let us also consider the asymptotic behaviour of $F_{A,BC}$ in the limit $x_\pA \rightarrow \infty$.  $F_{A,BC}$ is by definition symmetric with respect to the interchange $x \leftrightarrow x_\pA$, and therefore the same formula must hold in this limit, \ie 
  \begin{multline} \mbox {Lau}_{x_\pA\rightarrow \infty} \scal{ F_{A,BC} } = - \lim_{\Lambda \rightarrow \infty}  \int_{\mathds{R}^3 \setminus {\mathcal{B}_\Lambda}}  \frac{d^3 y}{2\pi} \frac{1}{|y-x_\pA|}  \mbox{Lau}_{y\rightarrow \infty}\left( \frac{  (y-x_\pB)\hspace{-1mm}\cdot \hspace{-1mm}(y - x_\pC)}{ |y-x | |y-x_{\pB}|^3 |y-x_{\pC}|^3} \right) \\
 -  \int_{\mathds{R}^3}   \frac{d^3 y}{2\pi} \mbox{Lau}_{x_\pA\rightarrow \infty}\left(  \frac{1}{|y-x_\pA|}\right) \frac{ (y-x_\pB)\hspace{-1mm}\cdot \hspace{-1mm}(y - x_\pC)}{ |y-x| |y-x_{\pB}|^3 |y-x_{\pC}|^3} 
  \end{multline} 
 Applying the Laplace operator, we obtain the same result, but now only the second term contributes 
 \bea \Delta \mbox {Lau}_{x_\pA\rightarrow \infty} \scal{ F_{A,BC} } &=&    \int_{\mathds{R}^3}   d^3 y   \delta^\ord{3}(x-y) \mbox{Lau}_{x_\pA\rightarrow \infty}\left(  \frac{1}{|y-x_\pA|}\right) \frac{ 2 (y-x_\pB)\hspace{-1mm}\cdot \hspace{-1mm}(y - x_\pC)}{ |y-x_{\pB}|^3 |y-x_{\pC}|^3} \CR
 &=&  \mbox{Lau}_{x_\pA \rightarrow \infty}\left(  \frac{1}{|x-x_\pA|}\right) \frac{ 2 (x-x_\pB)\hspace{-1mm}\cdot \hspace{-1mm}(x - x_\pC)}{ |x-x_{\pB}|^3 |x-x_{\pC}|^3} \eea
It follows that the two terms are necessary for $ \mbox {Lau}_{x\rightarrow \infty} \scal{ F_{A,BC} } $ to satisfy the correct Laplace equation and to be the asymptotic series of a symmetric function in $x$ and $x_\pA$. 
 
 \vskip 4mm

Using the asymptotic expansion (\ref{Exter},\ref{Inter}), one therefore concludes that 
\begin{multline}  F_{A,BC} =- \frac{1}{|x|} \int \frac{d^3 y}{2\pi} \frac{ (y-x_\pB)\hspace{-1mm}\cdot \hspace{-1mm}(y - x_\pC)}{ |y-x_{\pA}| |y-x_{\pB}|^3 |y-x_{\pC}|^3} \\ - \frac{x}{|x|^3} \cdot \int  \frac{d^3 y}{2\pi} y \frac{ (y-x_\pB)\hspace{-1mm}\cdot \hspace{-1mm}(y - x_\pC)}{ |y-x_{\pA}| |y-x_{\pB}|^3 |y-x_{\pC}|^3} + \mathcal{O}(x^{-3}) \end{multline} 
where the integrals are now computed over $\mathds{R}^3$. These integrals are both regular at any value of $x_\pA$, and it follows that they define the associated solutions of the Laplace equations in $x_\pA$ which have no pole in $x_\pA$, as we just discussed in general for the Laurent series. For instance 
\be - \Delta    \int \frac{d^3 y}{2\pi} \frac{ (y-x_\pB)\hspace{-1mm}\cdot \hspace{-1mm}(y - x_\pC)}{ |y-x | |y-x_{\pB}|^3 |y-x_{\pC}|^3}  = \frac{  2 (x-x_\pB)\hspace{-1mm}\cdot \hspace{-1mm}(x - x_\pC)}{ |x-x_{\pB}|^3 |x-x_{\pC}|^3} \ee
and therefore 
\be-   \int \frac{d^3 y}{2\pi} \frac{ (y-x_\pB)\hspace{-1mm}\cdot \hspace{-1mm}(y - x_\pC)}{ |y-x_{\pA}| |y-x_{\pB}|^3 |y-x_{\pC}|^3}   = \frac{|x_\pB - x_\pC|- |x_\pA - x_\pB|-|x_\pA - x_\pC|}{|x_\pB - x_\pC| |x_\pA - x_\pB||x_\pA-x_\pC|}  \ee
And similarly 
\be - \Delta \int  \frac{d^3 y}{2\pi} {\bf y} \frac{ (y-x_\pB)\hspace{-1mm}\cdot \hspace{-1mm}(y - x_\pC)}{ |y-x | |y-x_{\pB}|^3 |y-x_{\pC}|^3} = {\bf x} \frac{  2 (x-x_\pB)\hspace{-1mm}\cdot \hspace{-1mm}(x - x_\pC)}{ |x-x_{\pB}|^3 |x-x_{\pC}|^3} \ee
implies that 
\bea &&-  \int  \frac{d^3 y}{2\pi} {\bf y}  \frac{ (y-x_\pB)\hspace{-1mm}\cdot \hspace{-1mm}(y - x_\pC)}{ |y-x_{\pA}| |y-x_{\pB}|^3 |y-x_{\pC}|^3} \CR
&=& \frac{ - {\bf x}_\pA + {\bf x}_\pB + {\bf x}_\pC}{|x_\pA - x_\pB||x_\pA - x_\pC|} - \frac{1}{|x_\pB- x_\pC|}\biggl( \frac{ {\bf x}_\pB}{|x_\pA-x_\pC|} + \frac{{\bf x}_{\pC}}{|x_\pA-x_{\pB}|}\biggr) \CR
&&\hspace{15mm} + 2 \frac{ |x_\pB-x_\pC|^2 ( {\bf x}_\pA-{\bf x}_\pB) - ( x_\pB - x_{\pC})  \hspace{-1mm}\cdot \hspace{-1mm}(x_\pA- x_\pB) ( {\bf x}_\pB - {\bf x}_\pC)}{ |x_\pB-x_\pC||x_\pA-x_\pB||x_\pA-x_\pC|\scal{  |x_\pA-x_\pB| + |x_\pA-x_{\pC}| + |x_\pB-x_\pC| }}  \eea  
In order to find this solution one first observes that 
\begin{multline}  \Delta  \frac{ - {\bf x}_\pA + {\bf x}_\pB + {\bf x}_\pC}{|x_\pA - x_\pB||x_\pA - x_\pC|} =   {\bf x} \frac{  2 (x-x_\pB)\hspace{-1mm}\cdot \hspace{-1mm}(x - x_\pC)}{ |x-x_{\pB}|^3 |x-x_{\pC}|^3} \\ + 2 \frac{ |x_\pB-x_\pC|^2 ( {\bf x} - {\bf x}_\pB) - ( x_\pB - x_\pC ) \hspace{-1mm}\cdot \hspace{-1mm}(x - x_\pB) ( {\bf x}_\pB - {\bf x}_\pC)}{ |x-x_\pB|^3 |x-x_\pC|^3} \end{multline} 
The second equation 
\bea  \Delta  {\bf G}_{BC} &=& - 2 \frac{ |x_\pB-x_\pC|^2 ( {\bf x} - {\bf x}_\pB) - ( x_\pB - x_\pC ) \hspace{-1mm}\cdot \hspace{-1mm}(x - x_\pB) ( {\bf x}_\pB - {\bf x}_\pC)}{ |x-x_\pB|^3 |x-x_\pC|^3}  \CR
&=& 2   \Scal{ |x_\pB-x_\pC|^2 {\bf \nabla}_{x_\pB}  -  ( {\bf x}_\pB - {\bf x}_\pC) ( x_\pB - x_\pC ) \hspace{-1mm}\cdot \hspace{-1mm}\nabla_{x_\pB} } \frac{1}{|x-x_\pB||x-x_\pC|^3} \eea
can be solved by considering in a first step the equation
\be \Delta F_{BC} = \frac{2}{ |x-x_\pB||x-x_\pC|^3} = \frac{2}{\sqrt{ \rho^2 + (z-z_\pB)^2 } (  \rho^2 + (z-z_\pC)^2 )^\frac{3}{2}} \ee
Expanding the harmonic factor, one computes that a particular solution to this equation is obtained as the integral 
\bea F_{BC} &=& \int_0^1 \frac{d\alpha}{\sqrt{   \rho^2 + (z-z_\pC)^2 } \sqrt{ \rho^2 + ( z - z_\pC  - \alpha ( z_\pB - z_\pC))^2 } }\CR
&=&\frac{\mbox{ln} \biggl( \frac{ z-z_\pB + \sqrt{ \rho^2 + ( z-z_\pB)^2}}{ z-z_\pC + \sqrt{ \rho^2 + ( z-z_\pC)^2} } \biggr) }{(z_\pC-z_\pB) \sqrt{ \rho^2 + ( z-z_\pC)^2} } \CR
&=& \frac{\mbox{ln} \Bigl( \frac{ ( x - x_{\pB})  \hspace{-0mm}\cdot \hspace{-0mm}(x_\pC- x_\pB) + |x - x_{\pB}||x_\pC- x_\pB|}{( x - x_{\pC})  \hspace{-0mm}\cdot \hspace{-0mm}(x_\pC- x_\pB) + |x - x_{\pC}||x_\pC- x_\pB|} \Bigr)}{|x_\pC-x_\pB||x-x_\pC|}
\eea
The corresponding $x_\pB$ derivative only involves the derivative with respect to $x-x_\pB$ because of the projection and gives 
\begin{multline} G^\prime_{BC} = \frac{ |x_\pB-x_\pC|^2 ( {\bf x}-{\bf x}_\pB) - ( x_\pB - x_{\pC})  \hspace{-1mm}\cdot \hspace{-1mm}(x- x_\pB) ( {\bf x}_\pB - {\bf x}_\pC)}{ |x_\pB-x_\pC||x-x_\pB||x-x_\pC|} \\ \times  \left( \frac{1}{ |x-x_\pB| + |x-x_{\pC}| + |x_\pB-x_\pC| } - \frac{1}{ |x-x_\pB| + |x-x_{\pC}| - |x_\pB-x_\pC| }\right)  \end{multline}
However the second factor is singular whenever $x$ goes to the segment $[x_\pB,x_\pC]$, whereas $G_{BC}$ is regular everywhere. We consider therefore the harmonic function \bea H_{BC} &=& \frac{ |x_\pB-x_\pC|^2 ( {\bf x}-{\bf x}_\pB) - ( x_\pB - x_{\pC})  \hspace{-1mm}\cdot \hspace{-1mm}(x- x_\pB) ( {\bf x}_\pB - {\bf x}_\pC)}{ |x_\pB-x_\pC||x-x_\pB||x-x_\pC|} \\ &&  \times  \left( \frac{1}{ |x-x_\pB| + |x-x_{\pC}| + |x_\pB-x_\pC| } + \frac{1}{ |x-x_\pB| + |x-x_{\pC}| - |x_\pB-x_\pC| }\right) \CR
&=& \frac{ |z_\pB - z_\pC| \rho e^{i\varphi} \scal{ \sqrt{ ( z-z_\pB)^2 + \rho^2 } + \sqrt{ ( z-z_\pC)^2 + \rho^2 } }}{ \sqrt{ ( z-z_\pB)^2 + \rho^2 } \sqrt{ ( z-z_\pC)^2 + \rho^2 }({ \scriptstyle  \sqrt{ ( z-z_\pB)^2 + \rho^2 }  \sqrt{ ( z-z_\pC)^2 + \rho^2 } + ( z-z_\pB) ( z-z_\pC) + \rho^2} )} \nn
 \eea
 which interpolates between the singular harmonic function $\nabla \log(\rho^2) = \frac{2 e^{i \varphi}}{\rho} $ near the segment  $[x_\pB,x_\pC]$, and $\frac{\rho e^{i\varphi}}{(z^2 + \rho^2)^\frac{3}{2}}$ in the asymptotic region. Adding this contribution, one obtains the everywhere regular function
 \be G_{BC} =2  \frac{ |x_\pB-x_\pC|^2 ( {\bf x}-{\bf x}_\pB) - ( x_\pB - x_{\pC})  \hspace{-1mm}\cdot \hspace{-1mm}(x- x_\pB) ( {\bf x}_\pB - {\bf x}_\pC)}{ |x_\pB-x_\pC||x-x_\pB||x-x_\pC| \scal{ |x-x_\pB| + |x-x_{\pC}| + |x_\pB-x_\pC| } }   \ee

We have therefore the expansion 
\begin{multline}  F_{A,BC} = \frac{|x_\pB - x_\pC|- |x_\pA - x_\pB|-|x_\pA - x_\pC|}{|x| |x_\pB - x_\pC| |x_\pA - x_\pB||x_\pA-x_\pC|} \\
+ \frac{{\bf x}}{|x|^3} \cdot \left( \frac{ - {\bf x}_\pA + {\bf x}_\pB + {\bf x}_\pC}{|x_\pA - x_\pB||x_\pA - x_\pC|} - \frac{1}{|x_\pB- x_\pC|}\biggl( \frac{ {\bf x}_\pB}{|x_\pA-x_\pC|} + \frac{{\bf x}_{\pC}}{|x_\pA-x_{\pB}|}\biggr)\biggl .  + G_{BC}(x_\pA) \biggr) \right . \\ 
  + \mathcal{O}(|x|^{-3})
\end{multline} 
This result is consistent with the axisymmetric solution, because the third term vanishes identically whenever $x_\pA$ (respectively $x$) is aligned with $x_\pB$ and $x_\pC$. One can also check that in the limit $x_\pA \rightarrow x_\pB$, this expression further reduces to 
\bea  F_{A,BC} &=&\biggl(  \frac{ ( x_\pA - x_{\pB})  \hspace{-1mm}\cdot \hspace{-1mm}(x_\pC- x_\pB)}{ |x_\pA - x_\pB| |x_\pB - x_\pC|^3 }  - \frac{1}{|x_\pB- x_\pC|^2 }\biggr)   \Bigr( \frac{1}{|x|} +   \frac{ x \cdot x_\pB }{|x|^3} \Bigr)  + \mathcal{O}(|x|^{-3},x_\pA - x_\pB) \CR
&=& \frac{ ( x_\pA - x_{\pB})  \hspace{-1mm}\cdot \hspace{-1mm}(x_\pC- x_\pB)}{|x-x_\pB|  |x_\pA - x_\pB| |x_\pB - x_\pC|^3 }  - \frac{1}{|x-x_\pC| |x_\pB- x_\pC|^2 } \CR
&& \hspace{55mm} + \frac{ x  \cdot  ( x_\pC - x_\pB) }{|x|^3 |x_\pB - x_\pC|^3} +  \mathcal{O}(|x|^{-3},x_\pA - x_\pB) 
\eea
which coincides with (\ref{LimitxAB}) in the limit $x\rightarrow \infty$. 

\vskip 4mm

This confirms the asymptotic expansion of $F_{A,BC}$. For the composite black hole solution discussed in the paper, we are in fact interested in the function symmetrised in $A,B$
\bea &&  F_{(A,B)C} + \frac{1}{|x_\pA-x_\pC||x_\pB-x_\pC||x-x_\pC|} \\
&=& \frac{x}{2|x|^3} \cdot \biggl( \frac{ x_\pA - x_\pB}{|x_\pA - x_\pB||x_\pB - x_\pC|} + \frac{ x_\pB - x_\pA}{|x_\pB- x_\pA||x_\pA - x_\pC|} + \frac{ 2 x_\pC - x_\pA - x_\pB}{|x_\pA-x_\pC||x_\pB-x_\pC|} \biggr) \CR
&& + \frac{ |x_\pA - x_\pB|^2 ( x_\pB - x_\pC ) \hspace{-1mm}\cdot \hspace{-1mm} x - ( x_\pA - x_\pB ) \hspace{-1mm}\cdot \hspace{-1mm} ( x_\pB - x_\pC ) ( x_\pA - x_\pB ) \hspace{-1mm}\cdot \hspace{-1mm} x }{|x|^3 |x_\pA-x_\pB||x_\pA-x_\pC||x_\pB-x_\pC|\scal{   |x_\pA-x_\pB| + |x_\pA-x_{\pC}| + |x_\pB-x_\pC| }} + \mathcal{O}(x^{-3})\nn
\eea

\section{Bases for the solvable subalgebras}

In order to deal with nilpotent orbits representatives, it will is convenient to decompose the general charge $\C \in {\bf 42} \cong \e_{6(6)} \ominus \sp(8,\mathds{R})$ according to a Cartan basis of $\sp(8,\IR)$. Up to redefinitions of the harmonic functions that define the solutions, one can restrict oneself to a Cartan basis in the purely non-compact component ${\bf 20}$ according to the decomposition $\sp(8,\IR) \cong \mathfrak{u}(4) \oplus{\bf 20}$, \ie for which only $X_{abc}$ is non-zero in (\ref{Sp}). The most general such quartet of generators is parametrized by $U(4)$, and we will restrict ourselves to a particular case for simplicity. There is Cartan basis which is naturally associated to the $D0$--$D4$ configuration in the $STU$ model, and which is defined as the generators $H_0, H_i$
 for which $X_{abc}$ is defined  as in (\ref{EQP}) with only $Q_0$ and $P^i$ non-zero, and respectively equal to
 \be H_0 : Q_0 = P^i = - \stfrac{1}{2} \ , \qquad H_i : Q_0 = P^i = \stfrac{1}{2} ,\,  P^j = - \stfrac{1}{2} \mbox{ for } j\ne i\ee
 The elements of the coset component decompose with respect to their weight in terms of these four generators. For the charges which sit in the $STU$ truncation, one has 
 \be\begin{split} D0 \ : \   &(1,1,1,1) \\
  D2 \ : \ &  (1,-1,1,1) \\
 &  (1,1,-1,1) \\
 &  (1,1,1,-1) \\
 D4 \ : \ &  (1,1,-1,-1)\\
 &  (1,-1,1,-1) \\
 &  (1,-1,-1,1) \\
 D6 \ : \ &  (1,-1,-1,-1)\\
\overline{D0} \ : \   &(-1,-1,-1,-1) \\
 \overline{D2} \ : \ &  (-1,1,-1,-1) \\
 &  (-1,-1,1,-1) \\
 &  (-1,-1,-1,1) \\
 \overline{D4} \ : \ &  (-1,-1,1,1)\\
 &  (-1,1,-1,1) \\
 &  (-1,1,1,-1) \\
 \overline{D6} \ : \ &  (-1,1,1,1)
 \end{split}\hspace{10mm}\begin{split}
  N &= - \Xi_i = - Q_i = - P^0 \\
  M &= - \Sigma_1 = \Sigma_2 = \Sigma_3 = \stfrac{1}{2} P^1 \\
M &=  \Sigma_1 = -\Sigma_2 = \Sigma_3 = \stfrac{1}{2} P^2 \\
 M &=  \Sigma_1 = \Sigma_2 = -\Sigma_3 = \stfrac{1}{2} P^3 \\
N &= - \Xi_1 = \Xi_2 = \Xi_3 = Q_1 = - Q_2 = - Q_3 = P^0  \\
N &=  \Xi_1 = -\Xi_2 = \Xi_3 = -Q_1 =  Q_2 = - Q_3 = P^0  \\
N &=  \Xi_1 = \Xi_2 = -\Xi_3 = -Q_1 = - Q_2 =  Q_3 = P^0  \\
 M &= - \Sigma_i = \stfrac{1}{2} Q_0 \\
  N &= - \Xi_i = Q_i =  P^0 \\
  M &= - \Sigma_1 = \Sigma_2 = \Sigma_3 = -  \stfrac{1}{2} P^1 \\
M &=  \Sigma_1 = -\Sigma_2 = \Sigma_3 = - \stfrac{1}{2} P^2 \\
 M &=  \Sigma_1 = \Sigma_2 = -\Sigma_3 = - \stfrac{1}{2} P^3 \\
N &= - \Xi_1 = \Xi_2 = \Xi_3 = - Q_1 =  Q_2 =  Q_3 = - P^0  \\
N &=  \Xi_1 = -\Xi_2 = \Xi_3 = Q_1 = - Q_2 =  Q_3 = -P^0  \\
N &=  \Xi_1 = \Xi_2 = -\Xi_3 = Q_1 =  Q_2 =  -Q_3 = -P^0  \\
 M &= - \Sigma_i = - \stfrac{1}{2} Q_0 
 \end{split}\ee
where all the unspecified charges are zero. Out of the the $STU$ truncation one has 
 \be\begin{split}  D2 \ : \ &  (1,1,0,0) \\
 &  (1,0,1,0) \\
 &  (1,0,0,1) \\
 D4  \ : \ &  (1,-1,0,0) \\
 &  (1,0,-1,0) \\
 &  (1,0,0,-1) \\
 Y \ : \ & (0,0,1,1) \\
 & (0,1,0,1) \\
 & (0,1,1,0) \\
 T \ : \ & (0,0,-1,1) \\
 & (0,-1,0,1) \\
 & (0,-1,1,0) \\
 \mbox{flat}  \ : \ &  (0,0,0,0)\\
 \overline{D2} \ : \ &  (-1,-1,0,0) \\
 &  (-1,0,-1,0) \\
 &  (-1,0,0,-1) \\
 \overline{D4}  \ : \ &  (-1,1,0,0) \\
 &  (-1,0,1,0) \\
 &  (-1,0,0,1) \\
 \overline{Y} \ : \ & (0,0,-1,-1) \\
 & (0,-1,0,-1) \\
 & (0,-1,-1,0) \\
 \overline{T} \ : \  & (0,0,1,-1) \\
 & (0,1,0,-1) \\
 & (0,1,-1,0) 
 \end{split}\hspace{10mm}\begin{split}
  - \eta^2 &= \eta^3 = \y = - p_1 \\
    \zeta_1 &= - \zeta_3 = \z = - p_2 \\
  - \xi^1  &= \xi^2 = \x = - p_3 \\
   \eta_0  &= \eta_1 = \bar \y = - q^1 \\
  \zeta^0  &=  \zeta^2 = \bar \z = - q^2 \\
   \xi_0  &= \xi_3 = \bar \x = - q^3 \\
    - \eta_0  &= \eta_1 = \bar \y = q^1 \\
 - \zeta^0  &=  \zeta^2 = \bar \z = q^2 \\
  - \xi_0  &= \xi_3 = \bar \x =  q^3 \\
 \eta^2 &= \eta^3 =- \y =- p_1 \\
    \zeta_1 &= \zeta_3 = \z =  p_2 \\
   \xi^1  &= \xi^2 = -\x =  -p_3 \\
  \varsigma_{\pm} & \\
   - \eta^2 &= \eta^3 = - \y =  p_1 \\
    \zeta_1 &= - \zeta_3 = -\z =  p_2 \\
  - \xi^1  &= \xi^2 =- \x =  p_3 \\
   \eta_0  &= \eta_1 = -\bar \y =  q^1 \\
  \zeta^0  &=  \zeta^2 = -\bar \z =  q^2 \\
   \xi_0  &= \xi_3 = -\bar \x =  q^3 \\
    \eta_0  &= -\eta_1 = \bar \y = q^1 \\
 \zeta^0  &=  -\zeta^2 = \bar \z = q^2 \\
   \xi_0  &= -\xi_3 = \bar \x =  q^3 \\
     \eta^2 &= \eta^3 = \y = p_1 \\
    \zeta_1 &= \zeta_3 = - \z = - p_2 \\
   \xi^1  &= \xi^2 = \x =  p_3 
     \end{split}\ee
Each system of differential equations associated to a given nilpotent orbit is defined by possibly sourced harmonic functions valued in the positive eigenvector of a Cartan generator. For the BPS system, the Cartan generator is simply $H_0$, and the system is generated by all $D0,D2,D4,D6$ charges, which all carry eigenvalue $1$. The single-centred non-BPS system is associated to $\frac{1}{2} ( H_0 + \sum_i H_i)$, and is generated by the charges $\overline{D6},Y,D2$ and $D0$, carrying respectively eigenvalue $1$ and $2$. The almost BPS  system is associated to $2H_0 +  \sum_i H_i$, and is generated by the charges $\overline{D6}, D4$ of eigenvalue $1$, $Y$ of eigenvalue $2$, $D2$ of eigenvalue $3$ and $D0$ of eigenvalue $5$. The non-BPS system is associated to  $\sum_i H_i$, and is generated by the charges $D4,\overline{D2}$ of eigenvalue $1$,   $Y$ of eigenvalue $2$, and  $\overline{D6},D0$ of eigenvalue $3$. 

Another interesting system is the one associated to the maximal nilpotent orbit, for which the Cartan generator is $5 H_0 + H_1 + 2 H_2 + 3 H_3$. It extends the almost BPS system by bringing the charges $T$ into the game, giving rise to a system that includes as many harmonic functions as there are electromagnetic charges
\be {\bf 20} = D0 + 6 \times D2 + 6 \times D4 + \overline{D6} + 3 \times Y + 3 \times T \ee 
Similarly, the maximal locally BPS system is defined by the Cartan generator $H_2 + 2H_3 +4H_0$ and includes
\be  {\bf 20} = D0 + 6 \times D2 + 6 \times D4 + {D6} + 3 \times Y + 3 \times T \ee 
and the maximal composite non-BPS system is defined by the Cartan generator $2H_1 + 3H_2 + 4 H_3$ and includes 
\be {\bf 20} = D0 + 6 \times D2 + 6 \times \overline{D4} + \overline{D6} + 3 \times Y + 3 \times T \ee



\end{document}